\documentclass[10pt]{iopart}

\usepackage[utf8]{inputenc} %
\usepackage[T1]{fontenc}    %
\usepackage{hyperref}       %
\usepackage{url}            %
\usepackage{booktabs}       %
\usepackage{amsfonts}       %
\usepackage{nicefrac}       %
\usepackage{microtype}      %
\usepackage[table, x11names]{xcolor}         %

\usepackage{siunitx} %
\usepackage{fontawesome5}  %

\usepackage{amsmath} %
\usepackage{mathtools}

\usepackage{url} %
\usepackage{float}

\usepackage{caption}
\usepackage{subcaption}

\usepackage{listings} %

\usepackage{booktabs}

\definecolor{amaranth}{rgb}{0.9, 0.17, 0.31}
\colorlet{green}{green!20}
\colorlet{yellow}{yellow!60}
\colorlet{red}{red!30}

\usepackage{setspace}
\usepackage[linesnumbered,ruled,vlined]{algorithm2e}

\SetKwComment{Comment}{\color{green!80!black!190}// }{}

\newcommand{\var}{\texttt}

\SetKwProg{Function}{function}{}{}
\SetKw{Continue}{continue}

\DontPrintSemicolon
\SetAlFnt{\small}
\SetAlgorithmName{Pseudocode}{algorithmautorefname}

\newcommand{\makebrace}[6]{%
    \begin{tikzpicture}[overlay, remember picture]
        \draw [decoration={brace,amplitude=#4},decorate]
        let \p1=(#1), \p2=(#2) in
        ({max(\x1,\x2)}, {\y1+#5+0.8em}) -- node[right=#6] {#3} ({max(\x1,\x2)}, {\y2+#5});
    \end{tikzpicture}
}
\usepackage{tikz}
\usetikzlibrary{decorations.pathreplacing,calc}
\newcommand{\ntikzmark}[2]{#2\thinspace\tikz[overlay,remember picture,baseline=(#1.base)]{\node[inner sep=0pt] (#1) {};}}

\usepackage{enumitem}

\usepackage{siunitx} %

\usepackage{diagbox} %

\usepackage{multirow}  

\usepackage{longtable} %

\usepackage{array}
\newcolumntype{?}[1]{!{\vrule width #1}}
\newcommand{\phantomgraphics}[2][]{%
  \leavevmode\phantom{\includegraphics[#1]{#2}}%
}

\def\showcomments{} %

\ifx\showcomments\undefined
\newcommand{\GD}[1]{}
\newcommand{\YP}[1]{}
\newcommand{\KM}[1]{}
\else
\newcommand{\GD}[1]{\textcolor{red}{[GD: #1]}}
\newcommand{\YP}[1]{\textcolor{purple}{[YP: #1]}}
\newcommand{\KM}[1]{\textcolor{olive}{[KM: #1]}}
\fi
\newcommand{\model}{DIV1D-NN}  %

\usepackage{times}

\begin{document}

\ioptwocolmod
\twocolumn[{
\begin{@twocolumnfalse}
\title[Fast Dynamic 1D Simulation of Divertor Plasmas with Neural PDE Surrogates]{Fast Dynamic 1D Simulation of Divertor Plasmas with Neural PDE Surrogates}
\vspace{-2pt}
\author{Yoeri Poels$^{1,3}$, Gijs Derks$^{2,4}$, Egbert Westerhof$^4$, Koen Minartz$^1$, Sven Wiesen$^5$, Vlado Menkovski$^1$}

\address{$^1$Eindhoven University of Technology, Mathematics and Computer Science, NL-5600MB Eindhoven, The Netherlands}
\address{$^2$Eindhoven University of Technology, Control Systems Technology, NL-5600MB Eindhoven, The Netherlands}
\address{$^3$École Polytechnique Fédérale de Lausanne, Swiss Plasma Center, CH-1015 Lausanne, Switzerland}
\address{$^4$Dutch Institute for Fundamental Energy Research, NL-5612AJ Eindhoven, The Netherlands}
\address{$^5$Forschungszentrum Jülich GmbH, Institut für Energie- und Klimaforschung - Plasmaphysik, DE-52425 Jülich, Germany}
\vspace{-2.6pt}
\ead{y.r.j.poels@tue.nl}
\vspace{1.4pt}
\begin{indented}
\item[]September 2023
\end{indented}
\textbf{Abstract}\\
Managing divertor plasmas is crucial for operating reactor scale tokamak devices due to heat and particle flux constraints on the divertor target. Simulation is an important tool to understand and control these plasmas, however, for real-time applications or exhaustive parameter scans only simple approximations are currently fast enough. We address this lack of fast simulators using \textit{neural PDE surrogates}, data-driven neural network-based surrogate models trained using solutions generated with a classical numerical method. The surrogate approximates a time-stepping operator that evolves the full spatial solution of a reference physics-based model over time. We use DIV1D, a 1D dynamic model of the divertor plasma, as reference model to generate data. DIV1D's domain covers a 1D heat flux tube from the X-point (upstream) to the target. We simulate a realistic TCV divertor plasma with dynamics induced by upstream density ramps and provide an exploratory outlook towards fast transients. State-of-the-art neural PDE surrogates are evaluated in a common framework and extended for properties of the DIV1D data. We evaluate (1) the speed-accuracy trade-off; (2) recreating non-linear behavior; (3) data efficiency; and (4) parameter inter- and extrapolation.  
Once trained, neural PDE surrogates can faithfully approximate DIV1D's divertor plasma dynamics at sub real-time computation speeds: In the proposed configuration, $\text{\SI{2}{\milli\second}}$ of plasma dynamics can be computed in $\approx\text{\SI{0.63}{\milli\second}}$ of wall-clock time, several orders of magnitude faster than DIV1D.
\vspace{.13cm}
\hrule
\vspace{.13cm}
\end{@twocolumnfalse}
}\vspace{-2.9cm}]

\renewcommand*{\thefootnote}{\arabic{footnote}}

\maketitle

\section{Introduction}\label{sec:intro}

Tokamak devices operate in a diverted plasma configuration to decrease the effect of plasma-wall interaction. In this configuration, open field lines rapidly transport particles leaking out of the core plasma to the divertor region. These particle and heat fluxes must be mitigated before they reach the divertor plates, as they can far exceed material limits if left uncontrolled~\cite{kukushkin2011,wiesen2017,wischmeier2015}. However, mitigation techniques such as injecting impurities into the plasma edge have limits as they can degrade or can even be incompatible with core plasma performance~\cite{pacher2015,pitts2019}. 
Consequently, fast and accurate simulation is crucial for understanding and controlling the behavior of the plasma in the divertor region. %
Recent modeling efforts, such as SOLPS-ITER~\cite{wiesen2015}, UEDGE~\cite{rognlien1994}, SD1D~\cite{dudson2019} and DIV1D~\cite{derks2022}, are showing great promise in simulating divertor plasmas on varying levels of fidelity. However, when aiming for real-time applications or for the examination of large parameter spaces, only simple analytical functions such as the two-point models~\cite{stangeby2020} are fast enough, presenting a gap for fast high-fidelity simulation. In general, high-fidelity simulation is the cornerstone of model-based design and control for future reactors~\cite{siccinio2022}.

\begin{figure}[t]
\begin{center}\includegraphics[width=.96\linewidth]{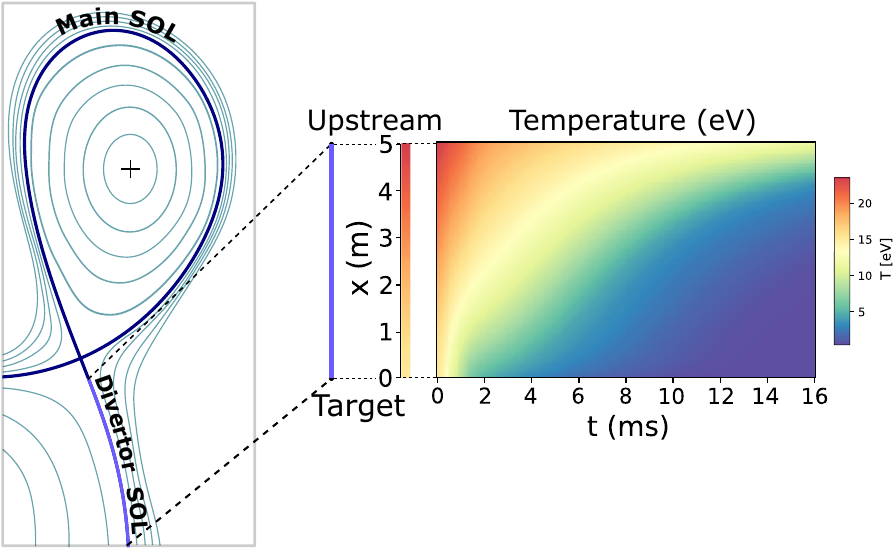}\end{center}
    \caption{Simplified overview of the computational domain of DIV1D, visualized using an illustration of a TCV plasma. DIV1D models a 1D heat flux tube from just below the X-point up to the divertor target. The state of this tube is then evolved over time (illustrated here for the temperature only).}
    \label{fig:1doverview}%
\end{figure}

To enable real-time high-fidelity simulation of complex dynamics, machine learning-based surrogates are showing great potential.
Recently, there has been an uptick in the development of artificial neural network (NN) based surrogate models for Partial Differential Equation (PDE) solvers~\cite{brandstetter2022,cao2021,gupta2022,li2021,stachenfeld2022}. These \textit{neural PDE surrogates} are trained in a data-driven manner: A classical numerical method first generates a dataset of PDE solutions, after which the NN is optimized to approximate the dynamics present in the dataset. At inference time, the NN can generate new solutions of a fidelity similar to the original solver, at a fraction of the computational cost.

Considering the above, we propose to use neural PDE surrogate models for fast and full-fidelity 1D simulation of divertor plasma dynamics. %
In this work we use DIV1D~\cite{derks2022}, a 1D dynamic model of the divertor plasma, as the physics-based model that generates the dataset of simulations. %

The domain covers a 1D heat flux tube spanning from just below the X-point (upstream) to the divertor target; see Figure~\ref{fig:1doverview} for an illustration. While the plasma behavior in DIV1D is simplified compared to higher fidelity codes such as SOLPS-ITER~\cite{wiesen2015}, a recent benchmark shows good agreement~\cite{derks2022}. Moreover, DIV1D can simulate dynamically at a relatively low computational cost, allowing us to generate rich training datasets. The data represents a divertor plasma of the Tokamak à Configuration Variable (TCV), with plasma conditions set around a recent evaluation of DIV1D w.r.t. SOLPS-ITER~\cite{derks2022}. To evaluate the capabilities of a neural PDE surrogate, we vary the upstream parallel heat flux, the upstream plasma density, and the carbon concentration. Dynamics are induced by varying the upstream plasma density over time at various ramp rates. Additionally, we provide an initial investigation in (surrogate) modeling fast transients, simulated through fast spikes in the upstream heat flux and upstream plasma density. %

Using this data we build surrogate models reproducing the full spatiotemporal solutions of DIV1D, specifically for a density ramp and a fast transient dataset. We build upon recent developments in neural PDE surrogates and tackle challenges such as long-rollout stability and efficiency. We evaluate a large number of methods and extend them to account for properties specific to these datasets, such as time-varying boundary conditions (BCs) and large variations in simulation length. We refer to the resulting surrogate model as~\model, and evaluate it with respect to its simulation accuracy and its utility for downstream applications, e.g. detachment studies. For the latter, we evaluate properties and structures such as the approximate emission front~\cite{ravensbergen2021}, the target temperature given upstream conditions, and the reconstruction of phenomena such as a bifurcation in the target temperature~\cite{capes1992} and roll-over in the target ion flux~\cite{loarte1998}. Moreover, we investigate the reproduction of fast transients somewhat resembling Edge-Localized Modes (ELMs)~\cite{zohm1996}. All evaluations are done in relation to the computational cost of the proposed surrogate models.

In general, machine learning-based methods have been used in nuclear fusion research in various settings, for example in disruption prediction~\cite{pau2019,zhu2020}, diagnostic processing \cite{fischer2010,pavone2020} or for accelerating simulation~\cite{ho2021,meneghini2020}; see~\cite{anirudh2022} for an exhaustive overview. Adjacent to our setting, NN-based surrogates have been proposed for accelerating scrape-off layer (SOL) simulation~\cite{dasbach2023,gopakumar2020}. For control purposes,~\cite{lore2023} employ sparse regression techniques (SINDy~\cite{brunton2016}) on SOLPS-ITER simulations to identify reduced models of key boundary plasma quantities. In~\cite{gopakumar2023} an exploration is conducted for data-driven simulation of high-speed camera footage capturing MAST divertor dynamics (among other settings) using the Fourier Neural Operator~\cite{li2021}; in our case, we focus on surrogate modeling for simulators and consider a broader set of surrogates and evaluations. Closest to our setting is a data-driven surrogate for divertor plasmas using UEDGE~\cite{zhu2022}. They propose a fast NN-based surrogate for static prediction of divertor detachment. 
The surrogate maps upstream density, injected power, and carbon concentration to a set of synthesized diagnostics. In contrast, we use DIV1D to simulate the dynamical behavior of the divertor plasma, and we approximate the full spatiotemporal profile of the simulations. %
Modeling the divertor plasma as a dynamical system allows us to capture bifurcations in the solution~\cite{capes1992}, whereas static modeling (i.e., not taking into account previous equilibria) could be ill-suited for these phenomena.

The paper is organized as follows. Section~\ref{sec:bg} formalizes the problem setting. Section~\ref{sec:data} describes the data generation procedure. Section~\ref{sec:method} provides an overview of neural PDE surrogates and describes adaptations we make. The model is evaluated in Section~\ref{sec:results}, and the applications and limitations are discussed. Finally, Section~\ref{sec:conclusions} gives conclusions and an outlook. A summary of our contributions is as follows:
    \begin{itemize}
        \item We generate a dataset of dynamic simulations representing a realistic TCV divertor plasma, containing challenging non-linear phenomena such as the roll-over in target ion flux and bifurcations in the target temperature. In addition, we generate a dataset for the exploratory analysis of reproducing transient events with surrogate models.
        \item We combine and implement a large number of state-of-the-art neural PDE surrogates in a common framework, with the purpose of building a dynamic surrogate model for divertor plasmas. We extend these methods to account for properties specific to the considered dynamics, such as the dependence on time-varying BCs and the wide range of simulation lengths, spanning two orders of magnitude between the shortest and longest simulated time.
        \item We conduct an in-depth evaluation of neural PDE surrogates and data properties, evaluating the following:
        \begin{itemize}
            \item The trade-off between error and computation time.
            \item Recovering higher-level properties and structures arising from non-linear behavior.
            \item The efficiency with respect to the dataset size.
            \item An analysis of inter- and extrapolation in the parameter space.
        \end{itemize}
        \item We show that using neural PDE surrogates we can accurately approximate the dynamic behavior in the divertor plasma with up to 5 orders of magnitude speedup compared to DIV1D. As such, real-time use cases are within reach:~\model~generates $\text{\SI{2}{\milli\second}}$ of plasma dynamics in $\approx\text{\SI{0.63}{\milli\second}}$ of wall-clock time. The surrogate captures high-level structures well, and can be trained using only hundreds of simulations. It shows strong performance within the evaluated parameter space. Global features of high-frequency transients can be reproduced with the proposed architecture.
        
    \end{itemize}

\section{Problem Formulation}\label{sec:bg}
The target domain, the divertor plasma, is described through solutions of DIV1D, a 1D dynamic model of the scrape-off layer~\cite{derks2022}. The DIV1D model solves 4 coupled time-dependent 1D fluid equations for the plasma density, plasma momentum, static plasma pressure and for neutral particles. These are solved along the divertor leg, where the computational domain extends from just below the X-point (upstream) to the target plate. For more details on the DIV1D model we refer to~\cite{derks2022}, with the settings used in this work described in Section~\ref{sec:data}.

The PDE solutions are converted to plasma density ($n$), velocity ($v_\parallel$), temperature ($T$) and the neutral density ($n_\text{n}$). These variables represent the quantities we simulate with a surrogate model. We denote the solution of all quantities with solution function $u(t,x)$, characterizing the solution as follows:
\begin{align}
\begin{split}
       u(t, x) &= \begin{bmatrix*}[r]
           n(t, x) \\
           v_\parallel(t, x) \\
           T(t, x) \\
           n_\text{n}(t, x)
         \end{bmatrix*},
\end{split}
\begin{split}
    t &\in [0, t_{\text{max}}],\\
    x &\in \mathbb{X},
\end{split}
\end{align}
where a simulation runs from time 0 to $t_{max}$ over spatial domain $\mathbb{X}$. Consequently, the solution function ${u \colon [0, t_{\text{max}}] \times \mathbb{X} \to \mathbb{R}^4}$ denotes a mapping from points in time and space to the four quantities.

In practice we operate on discretized solutions. To support neural PDE surrogates that assume fixed domains we use a fixed discretization for both the temporal and spatial domain. 
Spatial domain $\mathbb{X}$ is discretized to an equidistant grid with \textit{Nx} gridpoints over a length of \textit{L} (so $\textit{dx} = \frac{\textit{L}}{\textit{Nx}-1}$), denoted with $\mathbf{x} \in \mathbb{R}^{\textit{Nx}}$. The temporal domain is discretized with fixed timestep \textit{dt} to \textit{Nt} timesteps per simulation, denoted with $\mathbf{t} \in \mathbb{R}^{\textit{Nt}}$. Consequently, a discretized solution %
is denoted as ${\mathbf{u}^{\mathbf{t}, \mathbf{x}} \in \mathbb{R}^{\textit{Nt} \times \textit{Nx} \times 4}}$. %
Note that this discretization pertains to the data as used for the NN surrogate, not the settings used by DIV1D. The DIV1D solutions are downsampled and interpolated to the aforementioned discretization, for details on the numerics used with DIV1D we refer to Section~\ref{sec:data}.

Intuitively, the data-driven surrogate modeling task boils down to mapping the varying conditions to the solution $\mathbf{u}^{\mathbf{t}, \mathbf{x}}$. 
More precisely, we assume varying initial conditions and denote these as $\mathbf{u}^{0, \mathbf{x}} \in \mathbb{R}^{\textit{Nx} \times 4}$. The varied boundary conditions are denoted as $\mathbf{b}_{\text{s}} \in \mathbb{R}^{\textit{Ns}}$ and $\mathbf{b}_{\text{d}}^\mathbf{t} \in \mathbb{R}^{\textit{Nt} \times \textit{Nd}}$ for \textit{Ns} static and \textit{Nd} dynamic boundary conditions, respectively. Additional static conditions are denoted as $\mathbf{c} \in \mathbb{R}^{\textit{Nc}}$, for \textit{Nc} such conditions. In the current work we do not consider additional spatial or time-varying constraints, although the methodology could easily be extended to such a setting. In summary, the goal is to learn the following mapping:%
\begin{align}\label{eq:solmapping}
    f_\theta(
    \mathbf{u}^{0, \mathbf{x}},
    \mathbf{b}_{\text{s}},
    \mathbf{b}_{\text{d}}^\mathbf{t},
    \mathbf{c}) = \mathbf{u}^{\mathbf{t}, \mathbf{x}},
\end{align}
where $f_\theta$ denotes the to-be-learned function. Since the DIV1D equations are autonomous with respect to time, i.e., they do not directly depend on $t$, we can generate solutions by approximating a time-stepping operator with no explicit dependence on $t$. In other words, we parametrize an \textit{autoregressive} model that evolves the state of the system with time \textit{dt}, denoted as follows: %
\begin{align}\label{eq:solautoregressive}
\begin{split}
    \mathbf{u}^{t_{i-1}, \mathbf{x}} + \textit{dt} \cdot f_\theta(&
    \mathbf{u}^{t_{i-1}, \mathbf{x}},
    \mathbf{b}_{\text{s}},
    \mathbf{b}_{\text{d}}^{t_{i}},
    \mathbf{c}) = \mathbf{u}^{t_{i}, \mathbf{x}},
    \\
    &0 < i < \textit{Nt},
\end{split}
\end{align}
where $t_i$ denotes the $i$th element in $\mathbf{t}$. This formula is applied iteratively starting at $t_0=0$ to generate a full solution. Since $f_\theta$ is a neural network, this formulation is much akin learning a neural ordinary differential equation~\cite{chen2018} with an Euler solver with fixed timestep \textit{dt}. By predicting solutions in this form, the invariance to time is built into the model formulation, which should aid the surrogate model's performance. Additionally, generating solutions with an arbitrary number of timesteps is now possible, whereas in Equation~\ref{eq:solmapping} function $f_\theta$ can only predict solutions up to a fixed horizon \textit{Nt}.

\section{Data Generation}\label{sec:data}
The data used in this work is generated using DIV1D. In~\cite{derks2022}, it is shown that the DIV1D model can characterize divertor plasma behavior over a range of upstream plasma densities with a single model setting. As such, the model forms a good starting point to generate data approximating the dynamics of the divertor plasma in TCV. However, we note that DIV1D does not self-consistently solve outside of the SOL, hence the generated data does not cover time-dependent interactions between the SOL and external domains.

The settings of DIV1D equal those in \cite{derks2022}. The parameters that represent physics are set as follows: Connection length $L= \text{\SI{5}{\meter}}$; angle between magnetic field and target $\sin(\theta)=0.06$; effective cross-field heat flux expansion $\varepsilon_{\rm f}=2.3$; neutral cross-field transport $\tau_{\rm n}=\text{\SI{3}{\micro\second}}$.  %

Two scenarios are explored for the data-driven surrogates. We primarily consider the setting of ramps in the upstream density, and explore a parameter space for which DIV1D has been partially validated~\cite{derks2022}. This setting should represent a realistic divertor plasma in TCV; we describe the details in Subsection~\ref{ss:densityramps}. Additionally, we consider a dataset with fast transients as time-dependent boundary conditions. Since these settings have not been physically validated, this exploration is focused on evaluating the capabilities of neural PDE surrogates in more challenging settings that are potentially found in tokamak physics. These settings are described in Subsection~\ref{ss:fasttransients}.

\subsection{Density Ramps}\label{ss:densityramps}
Density ramps are typically used in experiments to investigate the transition of the divertor plasma from attached to detached conditions \cite{loarte1998}. 
To test the neural PDE surrogates' capabilities in forming a surrogate of DIV1D, we simulate ramps in the upstream plasma density with an exponential-like distribution to cover different timescales: ${\dot{n}_{\mathrm{u}}\in \pm \{1.0, 2.5, 5.0, 10.0, 25.0, 50.0, 100.0\}\text{\SI{e20}{\per\meter\cubed\per\second}}}$. The resulting simulations are between $\text{\SI{4}{\milli\second}}$ and $\text{\SI{400}{\milli\second}}$, ramping up and down between $n_{\mathrm{u}} \in [1.0, 5.0]\text{\SI{e19}{\per\meter\cubed}}$. 
The neutral density external to the plasma is changed together with the upstream plasma density as $n_{\mathrm{nb}}=[2.3-1.6n_{\mathrm{u}}\cdot 10^{-19}+1.3(n_{\mathrm{u}}\cdot 10^{-19})^2]10^{17}$ (taken from \cite{derks2022}).
Statically, the following boundary and internal conditions for DIV1D are varied: Upstream heat flux density ${q_{\|\mathrm{u}} \in \{10, 15, 20, 25, 30\}\text{\SI{}{\mega\watt\per\meter\squared}}}$; carbon concentration ${\xi_{\mathrm{C}} \in \{0.01, 0.02, 0.03, 0.04, 0.05\}~\mathrm{ion/electron}}$.

In summary, we create a dataset of $5 \times 5 \times 7 \times 2 = 350$ simulations. Initial conditions for each simulation are found by running DIV1D with static conditions and an initial guess until a steady state is reached. The numerical settings for DIV1D are as follows. The spatial domain is discretized using a finite difference scheme to a non-equidistant grid of 500 cells, with cells becoming smaller the closer they are to the target. The resulting ODE system is evolved with a timestep size of \SI{0.001}{\milli\second} using the DVODE\_F90 solver~\cite{brown1989,dvode}, which internally uses a variable number of timesteps (up to \num{100000}) for each $\text{\SI{0.001}{\milli\second}}$. For more details on the numerical implementation see~\cite{derks2022}.

For the NN surrogate, we use solutions on a much coarser grid than DIV1D uses internally: We linearly interpolate the cells to an equidistant grid with $\textit{Nx} = 100$ points ($dx\approx\text{\SI{0.05}{\meter}}$), and use timesteps of $\textit{dt} = \text{\SI{0.1}{\milli\second}}$. Our simulations span between $\text{\SI{4}{\milli\second}}$ and $\text{\SI{400}{\milli\second}}$, consequently, we have $\textit{Nt} \in [40, 4000]$. Each data channel is standardized before being fed into the NN: The solutions are rescaled to have zero mean and unit variance for each variable (plasma density, temperature etc.) over the entire dataset. 

To verify that we do not lose much information by downsampling solutions we evaluate whether this discretization still represents the dominant frequencies present in the solutions. A set of DIV1D solutions with no downsampling shows that on average more than $95\%$ of the power spectrum can be accounted for with signals below $\text{\SI{2}{\kilo\hertz}}$ in the temporal axis and below 3 cycles per meter in the spatial axis. With the chosen discretizations sampling (more than) 5 times finer in both axes, we ensure the signal is well represented~(following the Nyquist criterion~\cite{shannon1949}). %

The range of values that are used with DIV1D to generate data in this paper extend outside the domain where it was shown that DIV1D provides a realistic representation of the TCV divertor plasma. However, the aim of this work is to test capabilities of machine learning methods in providing fast surrogates for flight simulator and control applications. As such the simulations contain the roll-over of the target ion flux with increasing upstream plasma density~\cite{loarte1998} and the bifurcation of the target temperature as function of upstream plasma density~\cite{capes1992}. Ideally, these non-linear phenomena are reproduced by the machine learning surrogate of DIV1D. Both phenomena are important when the goal of the plasma exhaust is to both maintain low temperature and ion flux on the wall. Moreover, covering such phenomena greatly enhances the scope of applications for the surrogate models.

\subsection{Fast Transients}\label{ss:fasttransients}
\begin{figure}[h]
    \centering
\begin{center}\includegraphics[width=0.7102\linewidth]{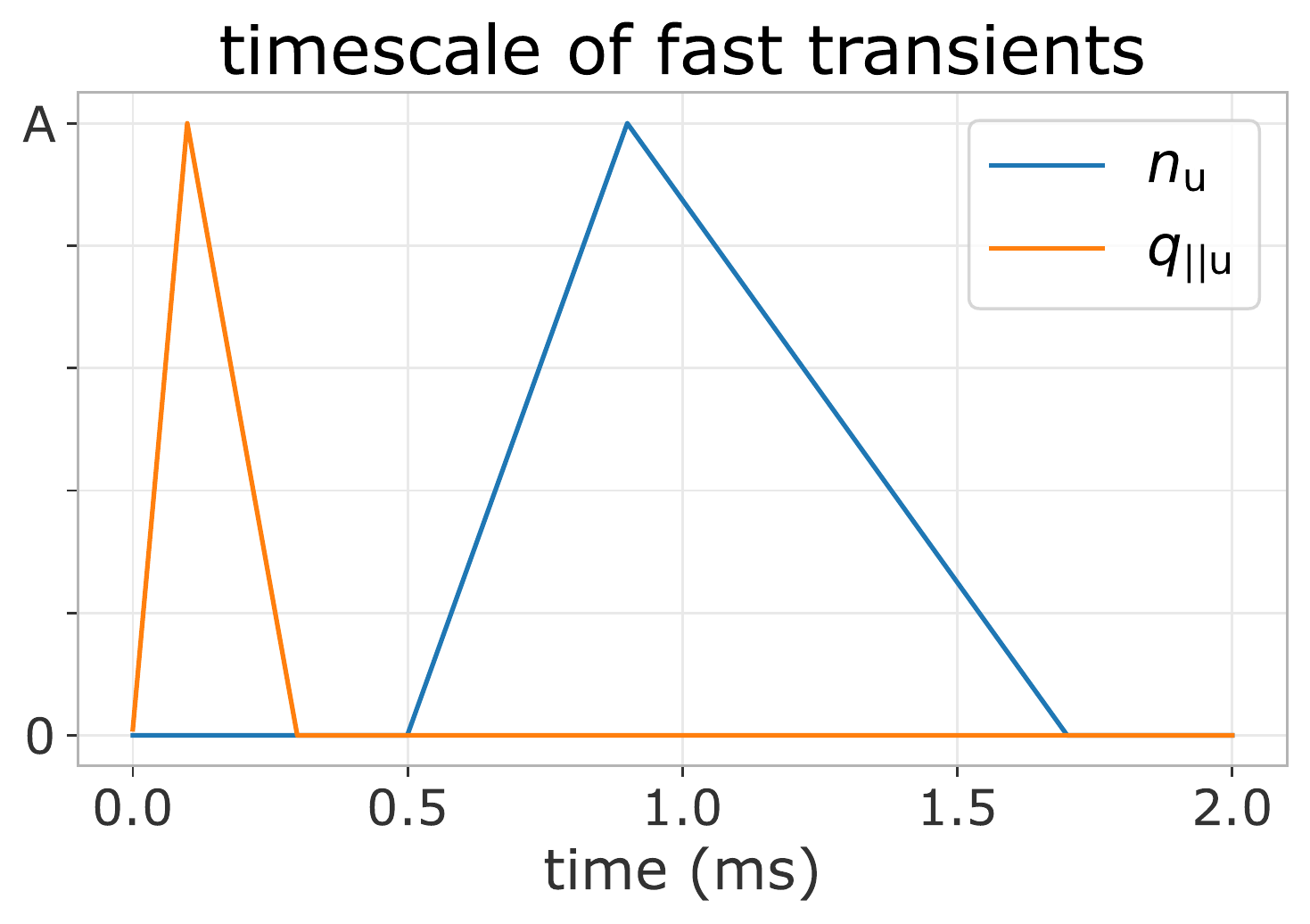}\end{center}
    \caption{The timescale of a single transient event.}
    \label{fig:transient}%
\end{figure}

As a more challenging setting, albeit not necessarily realistic, we create a dataset generated by fast transients happening upstream. We model the transients as a spike in the upstream parallel heat flux $q_{\|\mathrm{u}}$ followed by a spike in the upstream plasma density $n_{\mathrm{u}}$. Note that we do not dynamically couple the external neutral density $n_{\mathrm{nb}}$ to $n_{\mathrm{u}}$ as before but leave it static. The spike in $q_{\|\mathrm{u}}$ takes $\text{\SI{0.3}{\milli\second}}$ and is followed by a spike in $n_{\mathrm{nb}}$ of $\text{\SI{1.2}{\milli\second}}$, see Figure~\ref{fig:transient}. The amplitude of these spikes are chosen as a fraction of the total energy fluence and particle fluence of the incoming plasma over a period $\in$ $[2.5, 20.0]\text{\SI{}{\milli\second}}$, for a fraction $\in$ $[0.05, 0.30]$. For an appropriate range of parameters these transients could resemble ELMs~\cite{frerichs2021divertor,zohm1996}, but the chosen parameter range is not necessarily physically valid for TCV\footnote{In general it has not been assessed if DIV1D is suited to simulate the exact physics of ELMs \cite{derks2022}.}.

\begin{figure}[h]
    \centering
\begin{center}\includegraphics[width=0.8895\linewidth]{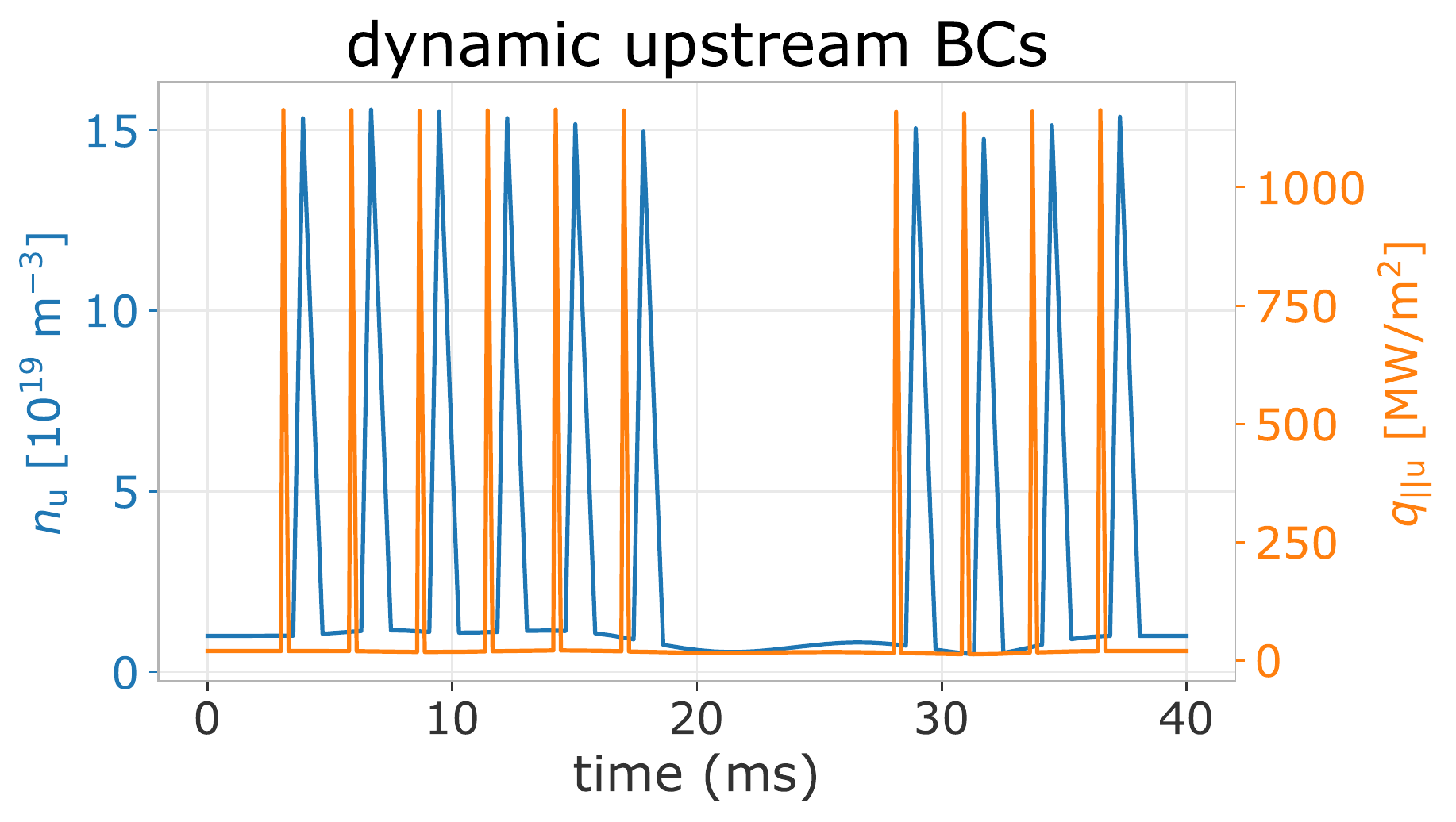}\end{center}
    \caption{Example of dynamic boundary conditions for upstream density $n_{\mathrm{u}}$ and upstream parallel heat flux $q_{\|\mathrm{u}}$.}{\label{fig:upstreambc}}
\end{figure}

We set the base upstream heat flux $q_{\|\mathrm{u}} \in \{10, 20, 30\}\text{\SI{}{\mega\watt\per\meter\squared}}$; base upstream plasma density $n_{\mathrm{u}} \in \{1.0, 3.0, 5.0\}\text{\SI{e19}{\per\meter\cubed}}$; carbon concentration $\xi_{\mathrm{C}} \in \{0.01, 0.02, 0.03, 0.04, 0.05\}~\mathrm{ion/electron}$. Transients are generated with a power fraction $\in \{0.05, 0.10, 0.20, 0.30\}$ and a period of $\{2, 4, 10, 20\}\text{\SI{}{\milli\second}}$, with each simulation covering $\text{\SI{40}{\milli\second}}$ of real time. The simulations in the dataset cover the Cartesian product of these settings. Additionally, we add a set of more general simulations to the dataset. Over the duration of these simulations, we smoothly vary the background level of plasma density $n_{\mathrm{u}}$, the background level of heat flux $q_{\|\mathrm{u}}$, and the transients' periods and amplitudes. An example of such BCs is provided in Figure~\ref{fig:upstreambc}. The complete dataset contains 1130 simulations.

The numerical settings for DIV1D are the same as before. For the resulting dataset's discretization, we again interpolate to an even grid with $\textit{Nx} = 100$ points, and use a finer temporal discretization of $\textit{dt} = \text{\SI{0.01}{\milli\second}}$ (compared to $\textit{dt} = \text{\SI{0.1}{\milli\second}}$ for the density ramp dataset). The fast transients result in high-frequency boundary conditions, which yield high-frequency solutions. To validate the chosen discretization, we again sample a set of solutions. More than 95\% of the power spectrum can be accounted for with spatial frequencies below 3 cycles per meter (as before), and with temporal frequencies below $\text{\SI{10}{\kilo\hertz}}$. Consequently, we opt for the same \textit{dx} as before, and refine the temporal grid by a factor of 10. With this discretization we sample at more than 5 times the highest dominant frequency in both axes, ensuring we still represent the signal properly~\cite{shannon1949}.

\section{Method}\label{sec:method}
\subsection{Method Overview}\label{ss:methodoverview}
\begin{figure}[t]
\begin{center}\includegraphics[width=\linewidth]{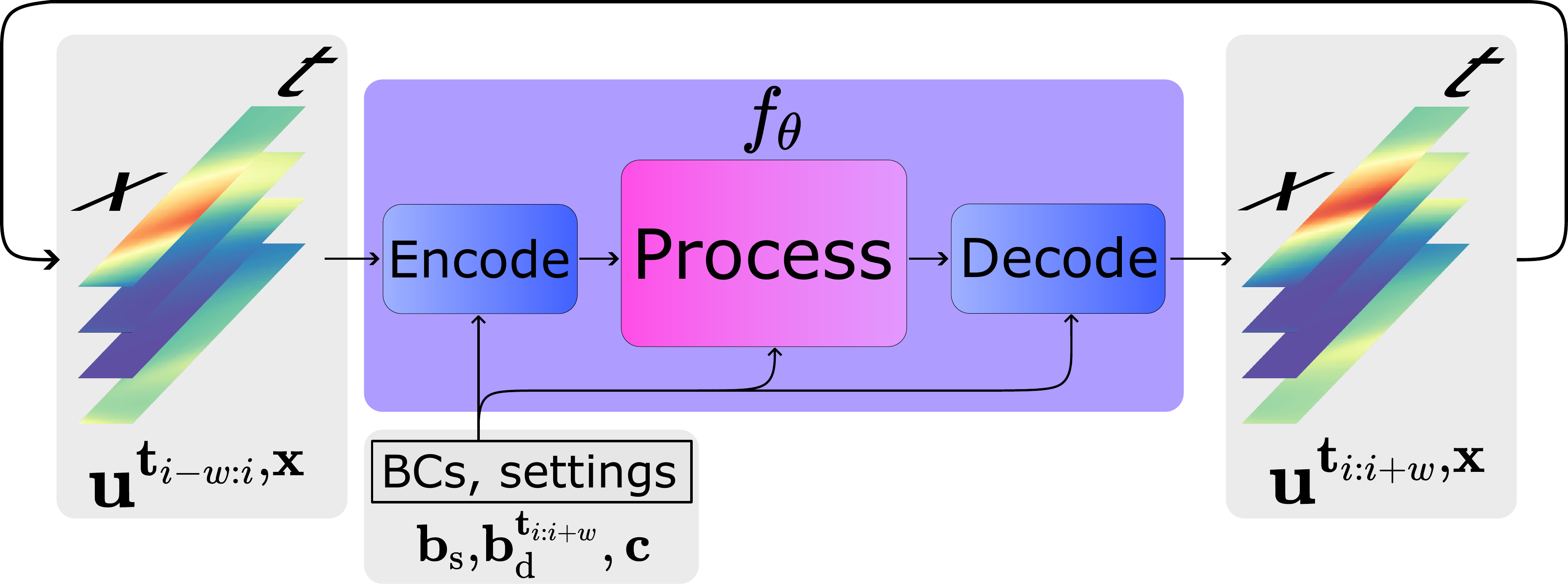}\end{center}
    \caption{Overview of the method. NN function $f_\theta$ takes as input the previous subtrajectory alongside new boundary and internal conditions. Its output is the next subtrajectory, which is inserted as input for the next step. Full solutions are computed by iterating over this procedure.}
    \label{fig:autoregressive}%
\end{figure}

The surrogate model simulates the divertor plasma following the problem formulation described in Section~\ref{sec:bg}. 
We adjust the autoregressive formulation as described in Equation~\ref{eq:solautoregressive} according to developments in neural PDE surrogates. Rather than evolving one timestep at a time, bundling several timesteps together in one neural network forward pass has empirically shown to improve stability and reduce computation time~\cite{brandstetter2022}. As such, we take both as input and output a block of time $\mathbf{t}$ %
rather than a single timestep $t$:
\begin{align}
\begin{split}
   \mathbf{u}^{t_{i-1}, \mathbf{x}} + \mathbf{dt}_{w} \odot f_\theta(
    \mathbf{u}^{\mathbf{t}_{i-w:i}, \mathbf{x}},
    \mathbf{b}_{\text{s}},
    &\mathbf{b}_{\text{d}}^{\mathbf{t}_{i:i+w}},
    \mathbf{c}) \\ & = \mathbf{u}^{\mathbf{t}_{i:i+w}, \mathbf{x}},
    \end{split}\nonumber
\end{align}\vspace{-.7cm}%
\begin{align}\label{eq:soltemporalbundling}
    &\mathbf{t}_{i:i+w} = (t_{i}, t_{i+1}, t_{i+2}, \ldots, t_{i+w-1}), \\
    &\mathbf{dt}_{w} = (\textit{dt}, 2\textit{dt}, 3\textit{dt}, \ldots, (w-1)\textit{dt}),\hspace{-.5cm}\nonumber
\end{align}
where $w$ denotes the time window, the number of timesteps in each input and output block. We predict the delta of future times $t_i$ to $t_{i+w-1}$ with respect to the last known state $\mathbf{u}^{t_{i-1}, \mathbf{x}}$. Intuitively, we can see Equation~\ref{eq:soltemporalbundling} as a vectorized version of Equation~\ref{eq:solautoregressive}; rather than one timestep at a time, we compute $w$ timesteps in parallel. %
Handling time together in blocks is referred to as \textit{temporal bundling}~\cite{brandstetter2022}. Since the model no longer depends on only the current state, but on the past \textit{w} states, we now use a short initial trajectory rather than only initial conditions for predicting full solutions. A simplified overview of the model is depicted in Figure~\ref{fig:autoregressive}.

\let\oldlabelitemi\labelitemi
\let\oldtabcolsep\tabcolsep
\setlength{\tabcolsep}{0em}
\renewcommand\labelitemi{$\vcenter{\hbox{\tiny$\bullet$}}$}
\newlength \figwidth
\setlength \figwidth {0.8\textwidth}
\newcolumntype{P}[1]{>{\centering\arraybackslash}p{#1}}
\begin{figure*}[h]
\centering
\begin{tabular}{m{.318\figwidth}m{.36398\figwidth}m{.318\figwidth}}
\multicolumn{1}{c}{\begin{minipage}{.294\figwidth}
    \centering
    \textbf{Encoder} \\
    \vspace{0.2em}
    \hrule height \heavyrulewidth
    
\begin{itemize}[leftmargin=1.3em]
    \item Linear convolution
    \item Point-wise transform~\cite{brandstetter2022}
\end{itemize}
\end{minipage}}
& 
\multicolumn{1}{c}{\begin{minipage}{.35298\figwidth}
\leftskip=1.3em
\begin{minipage}{\linewidth}
    \centering
    \textbf{Processor} \\
    \vspace{0.2em}
    \hrule height \heavyrulewidth
   
\vspace{-0.2em}
\begin{itemize}[leftmargin=0.85em]
    \item \ntikzmark{FNO}{FNO~\cite{li2021}}
    \item \ntikzmark{UNET}{UNet~\cite{gupta2022}}
    \item \ntikzmark{DRN}{DRN~\cite{stachenfeld2022}}
    \item \ntikzmark{MPPDE}{MP-PDE~\cite{brandstetter2022}}
    \item \ntikzmark{FT}{FT~\cite{cao2021}}
\end{itemize}
\makebrace{FNO}{DRN}{\textit{Convolution}}{0.5em}{-0.1em}{0.5em}
\makebrace{MPPDE}{MPPDE}{\textit{Message passing}}{0.25em}{-0.1em}{0.3em}
\makebrace{FT}{FT}{\textit{Self attention}}{0.25em}{-0.1em}{0.3em}
\end{minipage}
\end{minipage}}
& 
\multicolumn{1}{c}{\begin{minipage}{.3085\figwidth}
\leftskip=0.8em
\begin{minipage}{\linewidth}
    \centering
    \textbf{Decoder} \\
    \vspace{0.2em}
    \hrule height \heavyrulewidth
\begin{itemize}[leftmargin=1.3em,rightmargin=0.4em]
    \item Linear convolution
    \item Channel-wise convolution to time axis~\cite{brandstetter2022}
\end{itemize}
\end{minipage}
\end{minipage}}
\\
\addlinespace[-.23\figwidth]
\multicolumn{3}{c}{
\includegraphics[width=1 \figwidth]{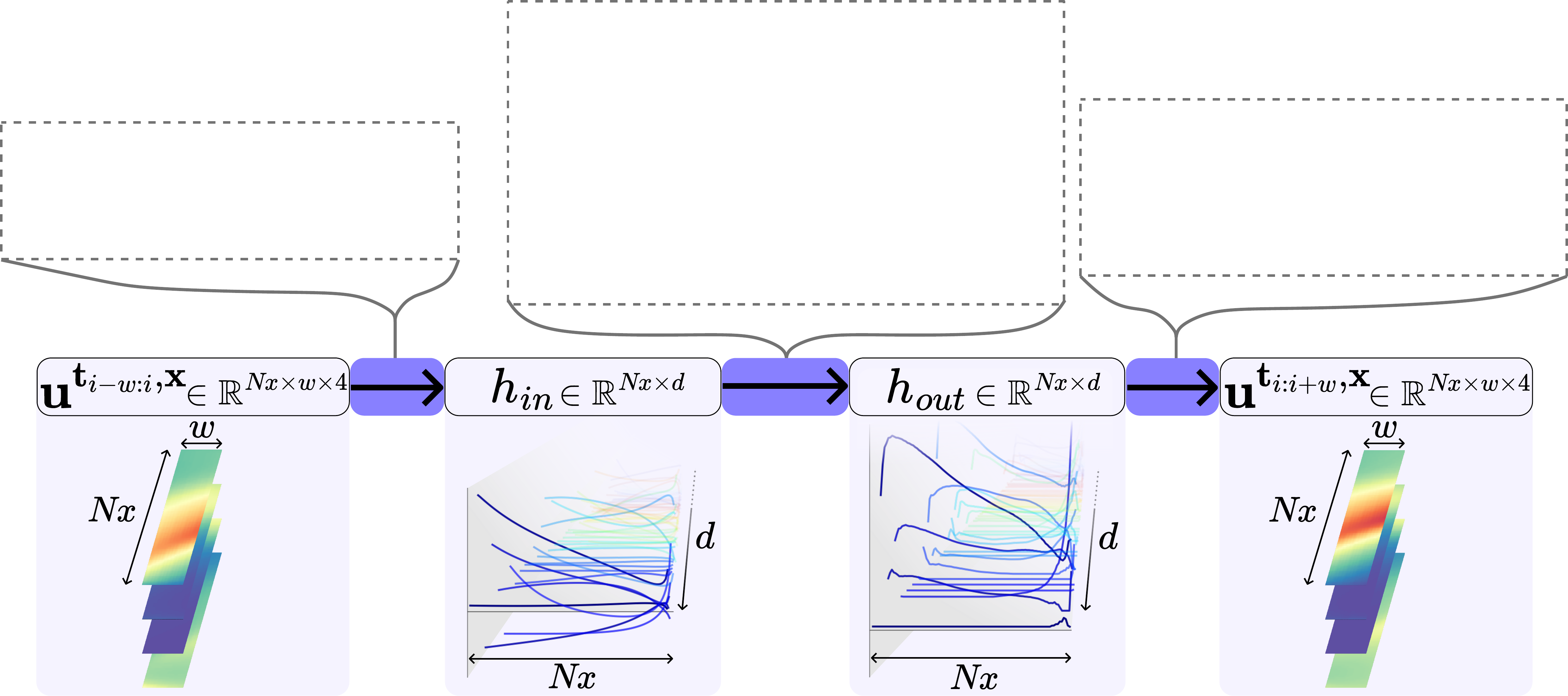}}\\
\addlinespace[-2.32pt]
\end{tabular}
\caption{Overview of the encode-process-decode framework. The encoder lifts signals from input time block $\mathbf{u}\mathbf{b}^{i-w:i}$ to abstract representation $h_{in}$. This representation is processed (evolved) into the representation of the next state, $h_{out}$. The decoder maps this signal back to the observed space as $\mathbf{u}\mathbf{b}^{i:i+w}$. For each component we evaluate several architectures.}
\label{fig:encprocdecbig}
\end{figure*}
\renewcommand\labelitemi{\oldlabelitemi}
\setlength{\tabcolsep}{\oldtabcolsep}
\setlength\tabcolsep{6pt}
\subsection{Model Training}\label{ss:modeltraining}
The objective function for optimizing parameters $\theta$ of our neural network function $f_\theta$ considers the prediction error with respect to DIV1D solutions. To simplify notation we denote the LHS of Equation~\ref{eq:soltemporalbundling} as $\mathcal{M}_\theta(\mathbf{ub}^{i-w:i}, \cdot)$, that is, model $\mathcal{M}_\theta$ predicting new states given a solution block from time $t_{i-w}$ to $t_{i-1}$ (alongside corresponding conditions, omitted for brevity). The RHS is referred to as $\mathbf{ub}^{i:i+w}$, which represents the ground truth values for the solution from time $t_{i}$ to $t_{i+w-1}$. The optimization target in its simplest form minimizes the one-step errors:
\begin{align}\label{eq:onestep}
    \hat{\theta} = \text{argmin}_{{\theta}}\ \mathcal{L}(\mathcal{M}_\theta(\mathbf{ub}^{i-w:i}, \cdot),\mathbf{ub}^{i:i+w}),
\end{align}
for all simulations and timesteps in the dataset, with appropriate loss function $\mathcal{L}$. However, by only minimizing single-step errors surrogate model $\mathcal{M}_\theta$ will likely suffer from instabilities when applied iteratively. Small errors accumulate on each solver step, which will lead to the input gradually falling off the training data manifold, i.e., there is a \textit{distribution shift}. Since the model will likely not generalize to data far out of its training distribution, the prediction quality will suffer. To combat this issue we also optimize with noisy model predictions as inputs, rather than only using clean DIV1D solution blocks. Consequently, the model can learn to correct its own error to stay on the data manifold. This method is referred to as the \textit{pushforward trick}~\cite{brandstetter2022}. We first apply the model $n$ times to gather perturbed prediction $\widetilde{\mathbf{u}}\mathbf{b}^{i-w:i}$, and use this noisy prediction as input. For example, for $n=2$, we get the following optimization target:
\begin{align}
\begin{split}
  \widetilde{\mathbf{u}}\mathbf{b}^{i-w:i} = \mathcal{M}_\theta(\mathcal{M}_\theta(\mathbf{ub}^{i-3w:i-2w}, \cdot), \cdot),
\\
  \hat{\theta} = \text{argmin}_{{\theta}}\ \mathcal{L}(\mathcal{M}_\theta(\widetilde{\mathbf{u}}\mathbf{b}^{i-w:i}, \cdot),\mathbf{ub}^{i:i+w}).
\end{split}
\end{align}
Parameters $\theta$ are optimized with mini-batches of data using standard gradient-based optimization techniques. %
Note the separation between computing $\widetilde{\mathbf{u}}\mathbf{b}^{i-w:i}$ and the loss calculation: The parameter gradients are only computed with respect to the final prediction step. %

\subsection{Model Architectures}\label{ss:modelarch}

The formulation of the surrogate model $\mathcal{M}_\theta$ is based around $f_\theta$, a function approximation using a neural network with parameters $\theta$. The architecture of this NN, which defines the function space of $f_\theta$, is key to finding a good model. Ideally, the architecture captures properties, for example, translational symmetries in the spatial domain, that fit with DIV1D solutions. However, the optimization procedure (model training) depends on this form, but this relation can be highly unpredictable: Selecting one architecture a priori is not sufficient. Consequently, determining the best architecture is primarily an empirical effort. In this work, we consider a representative set of architectures spanning various NN types, that show some of the best results in the field of neural PDE surrogate modeling.

\newcommand{\rulesep}{\hfill}
\begin{figure*}[t]
    \centering
    \begin{subfigure}[b]{0.118\textwidth}
    \begin{center}\includegraphics[width=\textwidth]{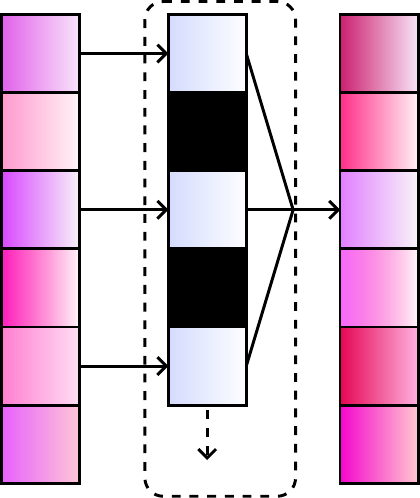}\end{center}
    \caption{\textbf{DRN}}{\label{fig:minidrn}}\vspace{.02cm}
    \end{subfigure}
    \rulesep
    \begin{subfigure}[b]{0.21\textwidth}
\begin{center}\includegraphics[width=0.93\textwidth]{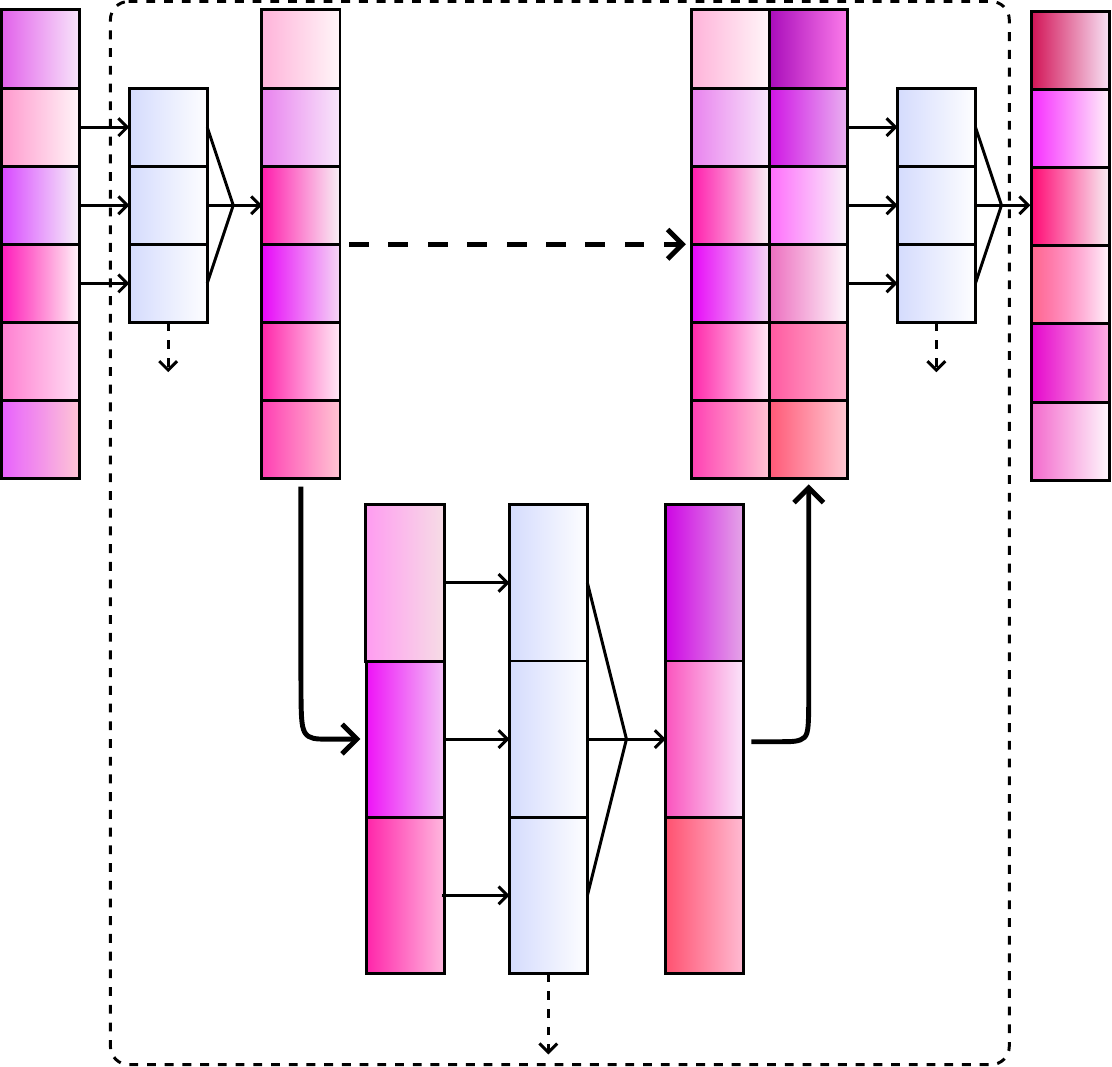}\end{center}
    \caption{\textbf{UNet}}{\label{fig:miniunet}}
    \end{subfigure}
    \rulesep
    \begin{subfigure}[b]{0.18\textwidth}
    \begin{center}\includegraphics[width=\textwidth]{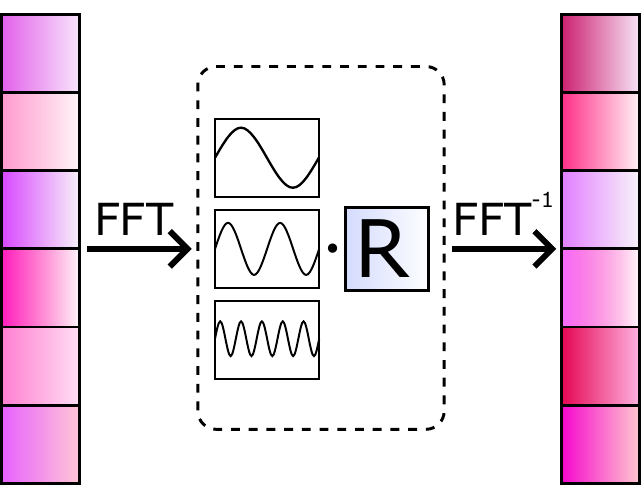}\end{center}
    \caption{\textbf{FNO}}{\label{fig:minifno}}
    \end{subfigure}
    \rulesep
    \begin{subfigure}[b]{0.12\textwidth}
    \begin{center}\includegraphics[width=\textwidth]{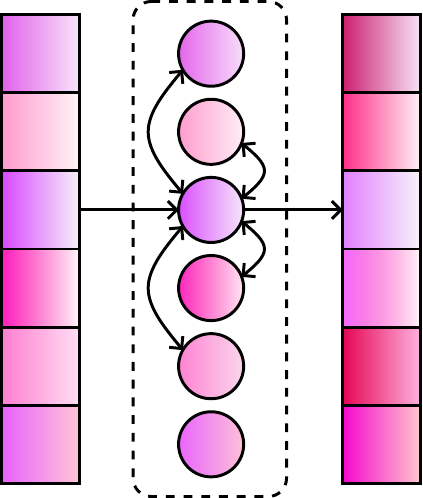}\end{center}
    \caption{\textbf{MP-PDE}}{\label{fig:minimppde}}
    \end{subfigure}
    \rulesep
    \begin{subfigure}[b]{0.19\textwidth}
    \begin{center}\includegraphics[width=\textwidth]{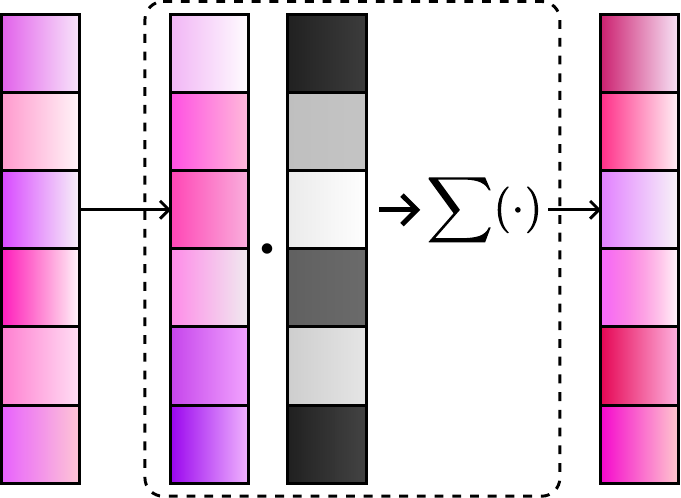}\end{center}
    \caption{\textbf{FT}}{\label{fig:minift}}
    \end{subfigure}
    \caption{Simplified illustrations of different methods for transforming the hidden representation's spatial grid (vertical cells), used in the model mentioned in the caption. All methods besides the UNet are stacked sequentially to form a deep network; UNet depicts a simplified overview of the entire network architecture. Best viewed zoomed in.}%
    \label{fig:uform}%
\end{figure*}

As a whole, the architecture follows the encode-process-decode structure as often used in neural PDE surrogates~\cite{brandstetter2022,li2021,stachenfeld2022}. The \textit{encoder} takes an observed signal and produces an abstract representation (encoding). This representation is then evolved into an abstract representation of the next state by the \textit{processor}. The \textit{decoder} maps this new abstract representation back into the observed space, delivering the model output. Throughout this process the spatial geometry of the solution is maintained. For each of the components we evaluate several architectures, see Figure~\ref{fig:encprocdecbig} for an overview. We first summarize the encoder and decoder architectures, and subsequently describe the considered processors.

The encoder and decoder map between the observed space and the abstract space. For the encoder, input block $\mathbf{ub}^{i-w:i} \in \mathbb{R}^{\textit{Nx}\times w \times 4}$ is flattened to a signal $\in \mathbb{R}^{\textit{Nx}\times 4w}$, a 1D grid of \textit{Nx} points with $4w$ channels. We evaluate two encoder architectures, a linear convolution over the spatial domain and point-wise non-linear transformations~\cite{brandstetter2022}. Both map the input grid to input hidden state $h_{in} \in \mathbb{R}^{\textit{Nx} \times d}$. The decoder maps the output hidden state $h_{out} \in \mathbb{R}^{\textit{Nx} \times d}$ to model prediction $\widetilde{\mathbf{u}}\mathbf{b}^{i:i+w} \in \mathbb{R}^{\textit{Nx} \times w \times 4}$. As architectures we consider a linear convolution over the spatial domain (mapping to $4w$ channels, which are reshaped to the 4 variables over $w$ timesteps), and non-linear convolutions over hidden channels mapping to the time axis~\cite{brandstetter2022}.

The considered processor architectures are visualized in Figure~\ref{fig:uform} (see Appendix~\ref{ap:arch_figs} for more detailed illustrations). We categorize them according to their main underlying inductive bias (convolution-based, message passing-based and self attention-based), and summarize them in subsequent paragraphs. We implement and extend all methods in a common framework to investigate which methods work best for approximating DIV1D dynamics. %

\textbf{Convolution-based networks}. Methods based on convolutional layers~\cite{lecun1989} involve learning filters/kernels that slide over a grid-based representation, making the learned transformation equivariant to shifts in its input domain. These kernels locally detect features on the grid, and by stacking these convolutional layers larger-scale behavior can be modeled. Since the kernels necessarily depend on the relative positions of grid cells, convolutional layers are tied to the grid discretization used in training. 

An architecture showing much success with PDE modeling using convolutional layers is the \textit{Dilated Residual Network}~\cite{stachenfeld2022}, abbreviated as \textbf{DRN}. Dilated convolutions~\cite{he2016} transform grid points not with their direct neighbors but with cells more than 1 step away, an example for 2 steps is given in Figure~\ref{fig:minidrn}. Stacking dilated convolutions with varying dilation rates allows for communication on large spatial scales while preserving local structure. These layers are implemented as a residual network~\cite{yu2016}: For hidden representation $h_{k}$ as output of the $k$th layer $l_{k}$, rather than transforming as $h_{k} = l_{k}(h_{k-1})$, we learn the residual $h_{k} = h_{k-1} + l_{k}(h_{k-1})$.

Another architecture primarily making use of convolutional layers is the \textbf{UNet} architecture~\cite{ronneberger2015}. Here, the representation is first downsampled and then again upsampled in the spatial domain, whilst connecting representations of the same resolution in the down- and upsample pass; see Figure~\ref{fig:miniunet} for an illustration. Intuitively, a UNet transforms the state on multiple spatial scales, resembling multigrid approaches. Strided convolutions are used to downsample the grid: Only one out of every $s$ grid points is processed to collect the next grid, for stride $s>1$. Transposed convolutions %
are used to upsample the grid. Normalization schemes, residual connections and self-attention are also often used in modern UNet implementations~\cite{gupta2022, ho2020, vaswani2017}; we describe self-attention in more detail below. We evaluate modern implementations of the UNet architecture as they have shown state-of-the-art performance in PDE modeling tasks~\cite{gupta2022}.

Orthogonal to the aforementioned approaches, the \textit{Fourier Neural Operator}~\cite{li2021}, abbreviated as \textbf{FNO}, aims to learn a convolution operator in Fourier space. Rather than explicitely parametrizing convolution kernels, the spatial domain is transformed to the frequency domain by the Fast Fourier Transform (FFT)~\cite{cooley1965}. In this frequency representation of our hidden dimensions, a truncated $m$ spectral coefficients are multiplied by a learned weight matrix. The result is transformed back by the inverse FFT, and it is summed to a point-wise transformation of the input grid. Transforming $h_{k-1}$ to $h_{k}$ with such an FNO layer can be formulated as follows:
\begin{align}
    h_k = \sigma\big(\text{FFT}^{-1}(\mathbf{R}_k\text{FFT}(h_{k-1})) + \mathbf{W}_k h_{k-1}\big),
\end{align}
for learned weight matrices $\mathbf{R}_k \in \mathbb{R}^{d \times d \times m}$ and $\mathbf{W}_k \in \mathbb{R}^{d \times d}$ ($d$ hidden dimensions; $m$ fourier modes), and non-linear activation function $\sigma$. A simplified illustration of the principle behind the FNO is given in Figure~\ref{fig:minifno}. One reason for the FNO's power stems from the combination of linear, global integral operators and non-linear, local activation functions. A significant benefit of this formulation is the invariance to the spatial discretization: There is no direct dependence on the grid size as the hidden representation is transformed in the frequency domain (and point wise). %

\textbf{Message passing-based networks}. By representing data as a graph consisting of nodes and edges, message passing neural networks~\cite{gilmer2017} transform a representation by updating individual nodes using a function of the node and its neighbors. Generally, a message passing step updates node $x_k^{i}$, where $i$ denotes the node index and $k$ the layer index, with the following formulation:
\begin{align}
    x_k^{i}=\gamma_{k}\Big(x^i_{k-1}, \cup_{j\in \mathcal{N}(i)} \phi_{k}(x^i_{k-1}, x^j_{k-1}, e^{j,i})\Big),
\end{align}
where $\phi$ denotes the \textit{edge transfer function}, $\cup$ denotes the \textit{aggregation function}, $\gamma$ denotes the \textit{node update function}, $\mathcal{N}(i)$ denotes the indices of the neighbors of $x^i$, and $e^{j,i}$ denotes the data associated with the edge between nodes $x^i$ and $x^j$. Functions $\phi$ and $\gamma$ are parametrized by small NNs, whereas aggregation $\cup$ is usually a simple function such as the mean or sum of the edge embeddings. As long as $\cup$ is invariant to the order of the edges, the network as a whole will be equivariant with respect to permutations of the input graph.

In the case of PDE modeling we represent the grid as a geometric graph, i.e., nodes are defined by the features on the gridpoint and the grid coordinates (consequently, the network is no longer equivariant to permutations of node features). Nodes are connected according to relative distances on the grid, for example by adding edges between points that lie within 2 cells from each other. We use graph neural networks (GNNs) to benefit from their expressive power: One can consider a message passing step as a generalization of a convolution. Additionally, since GNNs only operate on relations between nodes through their edges, GNNs are in principle capable of using arbitrary spatial discretizations. A notable implementation of GNNs for PDE modeling we evaluate is the \textit{Message Passing PDE} solver, abbreviated as \textbf{MP-PDE}~\cite{brandstetter2022}, which implements message passing according to differences in features and positions of nodes. %
A simplified illustration of message passing is provided in Figure~\ref{fig:minimppde}.

\textbf{Self attention-based networks}. The self attention mechanism, most prominently used in the transformer architecture~\cite{vaswani2017}, operates on sequences by transforming each element according to all other elements in the sequence. A dynamically weighted attention score is computed for all pairs of elements in the sequence, which is used in tandem with another transformation of the input sequence to compute the output sequence. Self attention is equivariant to permutations in the sequence as no information regarding the position of elements is used in this computation. One strength of self attention is its expressiveness, as it merges information over the entire sequence with dynamically computed weightings.

In more detail, elements are transformed by three learned matrices: Query matrix $\mathbf{W_Q} \in \mathbb{R}^{d \times d}$, key matrix $\mathbf{W_K}\in \mathbb{R}^{d \times d}$ and value matrix $\mathbf{W_V} \in \mathbb{R}^{d \times d}$, for elements with $d$ dimensions. We denote the transformed sequences as $\mathbf{Q} \in \mathbb{R}^{n \times d}$, $\mathbf{K} \in \mathbb{R}^{n \times d}$ and $\mathbf{V} \in \mathbb{R}^{n \times d}$, for sequences of length $n$. The attention score is computed as the softmax over the product of sequences $\mathbf{Q}$ and $\mathbf{K}^\top$ normalized by the number of hidden dimensions. This score is then multiplied by $\mathbf{V}$ to generate the final output\footnote{For simplicity we describe the case of only a single attention head in so-called multi-head attention. In multi-head attention, feature dimensions $d$ of each sequence element are split over multiple attention heads, i.e., each self-attention computation transforms a subset of the data's dimensions.}:
\begin{align}
    Attention(\mathbf{Q}, \mathbf{K}, \mathbf{V}) = softmax\Big(\frac{\mathbf{Q}\mathbf{K}^\top}{\sqrt{d}}\Big)\mathbf{V}.
\end{align}
An alternative interpretation of the self attention mechanism is provided by~\cite{cao2021}, where rather than considering rows of embedded sequences $\mathbf{Q}$, $\mathbf{K}$ and $\mathbf{V}$ as point-wise feature embeddings, we can consider the columns as vector representations of learned basis functions of the representation. Motivated by this interpretation they provide alternative formulations of the self-attention mechanism for PDE modeling, of which we use the \textit{Transformer with Fourier-type Attention}. We refer to this model as the Fourier Transformer, abbreviated as \textbf{FT}. Fourier-type attention is formulated as follows:
\begin{align}
    Attention_{FT}(\mathbf{Q}, \mathbf{K}, \mathbf{V}) = (\widetilde{\mathbf{Q}}\widetilde{\mathbf{K}}^\top)\mathbf{V}/n,
\end{align}
where $\widetilde{\diamond}$ denotes layer normalization~\cite{ba2016} and $n$ the number of elements in the sequence. Since our sequence represents the spatial grid, $n = \textit{Nx}$. This attention formulation alongside residual connections and a point-wise NN form the basis of the (Fourier) Transformer. Positional information is concatenated with the element features to let the model use positional relations of grid points; consequently, it is no longer equivariant to permutations of the input sequence. Additionally, since the positional information is used only as input feature, a transformer is in principle capable of working with arbitrary spatial discretizations. A simplified illustration of self-attention is provided in Figure~\ref{fig:minift}. We also evaluate the use of FT layers followed by FNO layers (as proposed in the FT paper~\cite{cao2021}), denoted as \textbf{FT-FNO}.

\subsection{Adaptations for DIV1D Data}\label{ss:model_div1d}
Besides implementing various existing methods in a unified framework for training autoregressive encode-process-decode models with temporal bundling and the pushforward trick, we also make adaptations tailored towards the DIV1D data. Specifically, we address two properties: A wide variation in simulation length \textit{Nt} (as we model density ramp dynamics spanning multiple timescales) and the dependence of solutions on (time-varying) conditions. To address the former, we propose an adapted strategy for sampling batches while training. To address the latter, we describe a simple conditioning method that we implement across all architectures described in Subsection~\ref{ss:modelarch}.

\textbf{Sampling with scheduled unrolling}. As described in Section~\ref{ss:densityramps}, simulations in the density ramp dataset range from 40 to 4000 timesteps. We found that using many unrolling steps with the pushforward trick (Subsection~\ref{ss:modeltraining};~\cite{brandstetter2022}) is key to training models that are stable over long timeframes. However, since training is done using minibatches of randomly sampled datapoints (to exploit the parallel computation of GPUs), simulations with large differences in length are often batched together. Due to these differences in length, long unrollings during training require tricks to deal with simulations shorter than the unrolling window, such as padding and adapting the model input. These tricks come at a cost: For example, when processing a simulation of length 40 in a batch where we unroll 10 times with $w=20$, the forward computation is done on 200 timesteps, whereas only 20 timesteps can be used for the pushforward loss computation. The redundant computations amount to a significant part of the total cost, making training unnecessarily long and expensive\footnote{While the gradient computation, which on itself is most expensive, is done only once at the final prediction step, we found that when unrolling for many steps the majority of time in each optimization step is spend in the unrolling procedure.}.

To combat this issue we propose sampling batches by first sampling an unrolling window and then sampling items that fit in this window. Sample probabilities are adjusted such that all items are sampled uniformly in expectation, as we do not want to bias the learning algorithm towards dynamics occurring in longer simulations. We sample unrolling length $t$ from a chosen distribution $p(t)$ and sample items $x$ from distribution $p(x|t)$ which is constrained as follows:
\begin{align}\label{eq:pxsampling}
\mathbb{E}_{t \sim p(t)}\big[p(x_i|t)\big] = \frac{1}{N},\hspace{.5cm} 0 \leq i < N,
\end{align}
where $N$ denotes the total number of simulations in the dataset. We refer to Appendix~\ref{ap:probs} for details on the computation of $p(x|t)$, an evaluation of multiple unrolling distributions, and a comparison with baselines.

\textbf{Incorporating (time-varying) conditions}. Our function approximation $f_\theta$, and by extension surrogate model $\mathcal{M}_\theta$, not only depends on the previous state of the system but also on (time-varying) boundary conditions and internal conditions. These conditions are inserted into the model's internal representations as follows. We process conditions $(\mathbf{b}_{\text{s}}, \mathbf{b}_{\text{d}}^{\mathbf{t}_{i:i+w}}, \mathbf{c})$ to a conditioning vector $\mathbf{cond}$, and concatenate this vector throughout the internal representations. In more detail, $\mathbf{cond}$ is created by first processing the time-varying conditions $\mathbf{b}_\text{d}^{\mathbf{t}_{i:i+w}}$ with a small NN and concatenating this output with $\mathbf{b}_{\text{s}}$ and $\mathbf{c}$. This vector is repeated for each grid point, making our conditioning vector $\mathbf{cond} \in \mathbb{R}^{\textit{Nx} \times dc}$ for $dc$ conditioning features. To insert this information into the model, we implement the same strategy for all architectures: $\mathbf{cond}$ is concatenated feature-wise to all spatial hidden representations $h \in \mathbb{R}^{\textit{Nx} \times d}$, such that the model learns mappings from ${\mathbb{R}^{\textit{Nx} \times (d + dc)} \to \mathbb{R}^{\textit{Nx} \times d}}$, i.e., the model can make use of the simulation conditions for each internal transformation. Additionally, we correct the model output at each step: For the plasma density, plasma temperature and neutral density we clamp outputs to a minimum value of 0.1, and we set the upstream plasma density values to the time-varying BCs.

\section{Experiments and Results}\label{sec:results}
In this section we evaluate all methods and adaptations with the purpose of constructing a fast and accurate high-fidelity surrogate model of divertor plasmas, based on data from DIV1D. We primarily focus on the density ramp dataset, as this data represents a realistic divertor plasma; the objective of the fast transient dataset is to provide an exploratory outlook towards modeling higher frequency dynamics.

We start with a short summary of the training procedure and settings in Subsection~\ref{ss:recaphyperparam}. In Subsections~\ref{ss:wp} and~\ref{ss:div1dnn} we evaluate all methods and propose the configuration for surrogate model DIV1D-NN using the density ramp dataset. In Subsection~\ref{ss:casestudies} we evaluate DIV1D-NN in more depth through a set of case studies using the density ramp data, and also evaluate its ability to scale to more difficult datasets and dynamics with the fast transient data. Finally, in Subsections~\ref{ss:dataef} and~\ref{ss:interp}, we investigate properties of the surrogate with respect to the datasets, focusing on the number of simulations in the training data and the surrogate's inter- and extrapolation capabilities.

\subsection{Method Recap and Hyperparameters}\label{ss:recaphyperparam}
    The two datasets are split as follows. For the density ramp dataset (350 simulations), we use 300 for training, 25 for validation (model selection) and 25 for testing (final results). The split is randomly sampled for the most part -- the only intervention is that we ensure the test set contains some simulations with identical parameters both as a density ramp up and ramp down. The split is kept fixed throughout the experiments unless stated otherwise. For the fast transient dataset (1130 simulations), we use 904 for training, 113 for validation and 113 for testing. The test split is selected to contain all 60 simulations with transients where the average energy fluence is between $\text{\SI{15}{\kilo\joule\per\square\meter}}$ and $\text{\SI{20}{\kilo\joule\per\square\meter}}$, along with 53 randomly sampled simulations (covering 10\% of the dataset in total). The remaining simulations are randomly split between the train and validation set. 

For training we use a batch size of 16. The procedure consists of first sampling a batch of simulations and the unrolling time (Subection~\ref{ss:model_div1d}). For all simulations in the batch we pick a random starting point, unroll the model predictions, compute and backpropagate the loss and update the model parameters (Subsection~\ref{ss:modeltraining}). This process is repeated for $\lceil \frac{300}{16} \rceil = 19$ batches per epoch for the density ramp dataset and $\lceil \frac{904}{16} \rceil = 57$ batches per epoch for the transients dataset. That is, one epoch is one full pass over the dataset, and we repeat this procedure for \num{20000} epochs for both datasets. Following standard practice, we select the model parameters for which the validation set error is minimal, and report results on the test set.

For the density ramp dataset, we use the root of the squared error (i.e., the $L_2$ distance) as loss function: 
\begin{align}
    \mathcal{L}(\mathbf{ub}, \widetilde{\mathbf{u}}\mathbf{b}) = \sqrt{\textstyle\sum_{i=1}^n|\mathbf{ub}-\widetilde{\mathbf{u}}\mathbf{b}|^2},
\end{align}
for all $n$ points in the batch, where $\mathbf{ub}$ denotes a time block of DIV1D reference solutions and $\widetilde{\mathbf{u}}\mathbf{b}$ denotes the corresponding model predictions. For the fast transient dataset, we average the squared error and absolute error and take the square root:
\begin{align}
\begin{split}
\mathcal{L}&(\mathbf{ub}, \widetilde{\mathbf{u}}\mathbf{b}) = \\
&\sqrt{\textstyle\sum_{i=1}^n \big( 0.5 \cdot |\mathbf{ub}-\widetilde{\mathbf{u}}\mathbf{b}| + 0.5 \cdot |\mathbf{ub}-\widetilde{\mathbf{u}}\mathbf{b}|^2\big)},
\end{split}
\end{align}
with the same variables as before. We use both terms as we empirically found that the absolute error term helped stabilize training in the presence of large spikes in the solutions. We use the Adam optimizer~\cite{kingma2015} with an initial learning rate of $10^{-4}$ and decay the learning rate by 0.4 at epochs 500, 2500, 5000 and 7500. 

All models compute the solution in blocks of $\text{\SI{2}{\milli\second}}$, which corresponds to $w=20$ timesteps for the density ramp dataset (where $dt=\text{\SI{0.1}{\milli\second}}$) and $w=200$ timesteps for the fast transient dataset (where $dt=\text{\SI{0.01}{\milli\second}}$). All models are implemented using PyTorch~\cite{paszke2019}, and are trained and evaluated using an NVIDIA A100 40GB GPU and Intel Xeon Platinum 8360Y CPU unless stated otherwise.

\renewcommand{\rulesep}{\hfill}
\begin{figure*}[h]
    \centering
    \begin{subfigure}[h]{0.49\textwidth}
    \begin{center}\includegraphics[width=\textwidth]{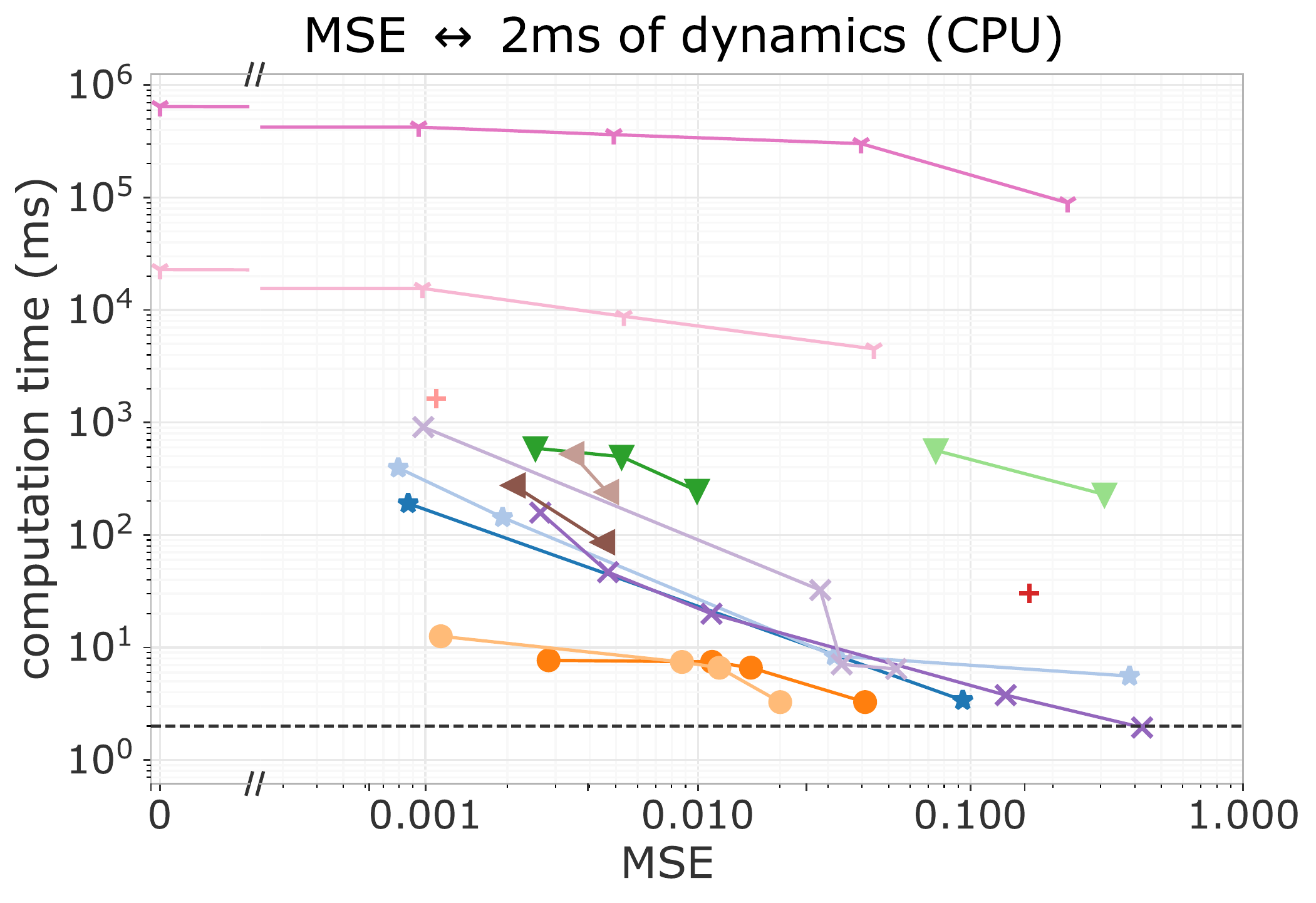}\end{center}
    \includegraphics[width=2\textwidth]{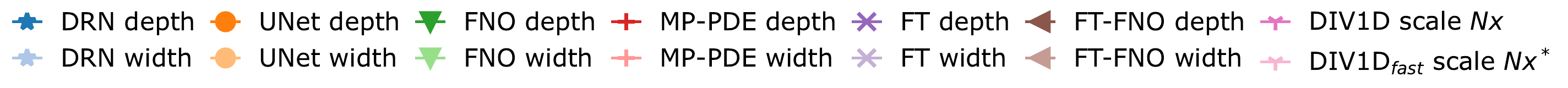}
    \caption{AMD Rome 7H12 CPU, 1 core}{\label{fig:wpdiv1dcpu}}
    \end{subfigure}
    \begin{subfigure}[h]{0.49\textwidth}
    \begin{center}\includegraphics[width=\textwidth]
    {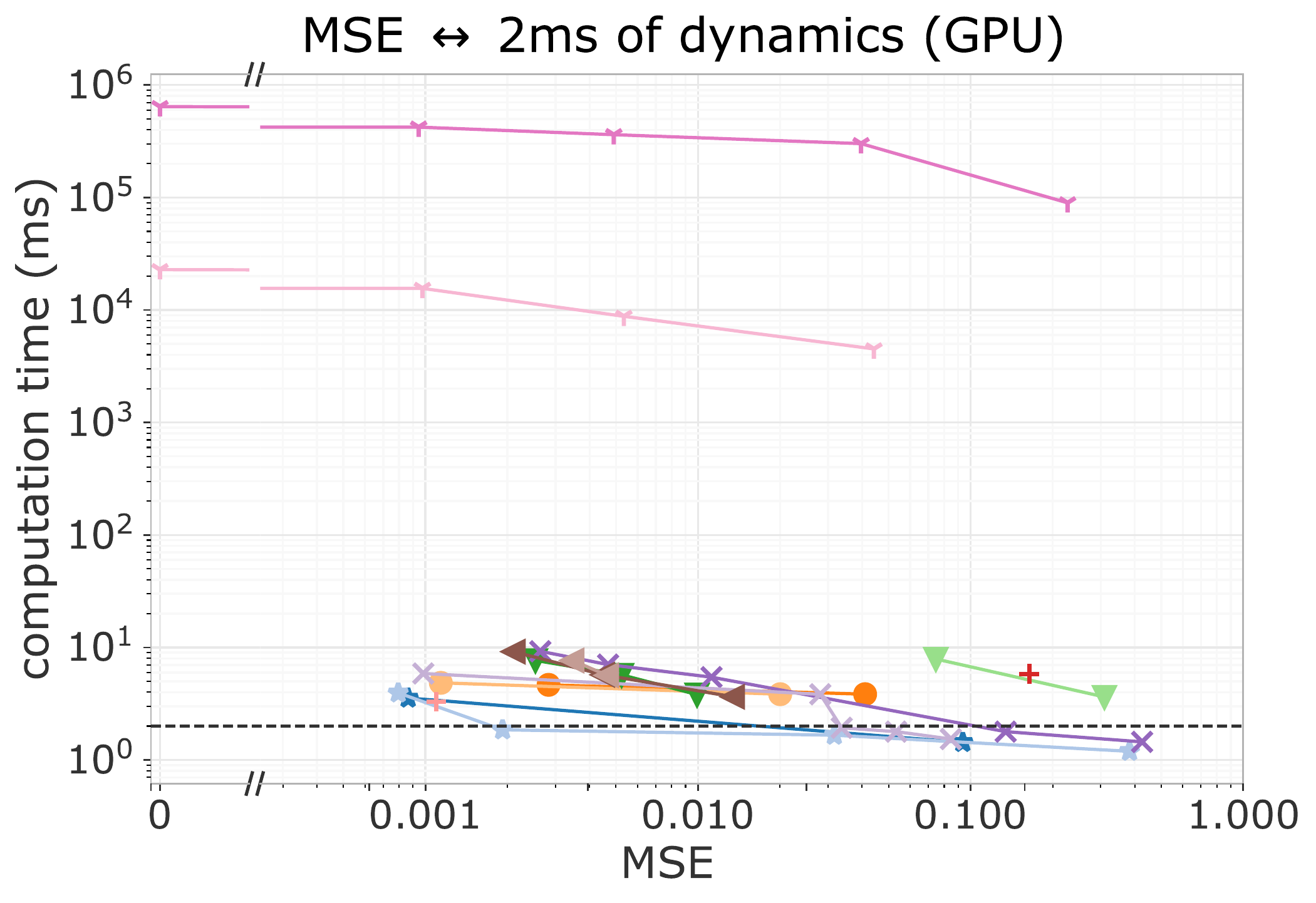}\end{center}
    \phantomgraphics[width=2\textwidth]{figures/wp/legend_all_DIV1D_nobg.pdf}
    \caption{NVIDIA A100 GPU}{\label{fig:wpdiv1dgpu}}
    \end{subfigure}
    \caption{Work-precision diagrams of the MSE for full solution rollouts versus the computation time of 1 block ($\text{\SI{2}{\milli\second}}$ of plasma simulation). We compare all neural PDE surrogates with default DIV1D: Neural PDE surrogates are scaled by their architecture, DIV1D is scaled by decreasing the size of the computational grid. We benchmark using (a) equal hardware and (b) GPUs, if applicable. The dashed line represents the real-time barrier. Note the disconnect in the x-axis: The computation cost of reference solutions is also visualized, but since by definition they have an MSE of 0, they cannot be plotted on a log-scale. $^*$For DIV1D$_{\textit{fast}}$, the MSE is calculated w.r.t. its own solutions at $\textit{Nx}=500$ rather than the DIV1D reference solutions.}
    \label{fig:wpdiv1d}%
\end{figure*}

\subsection{The Trade-Off Between Error and Computation Time}\label{ss:wp}
We investigate different architectures for the model alongside different hyperparameters for each architecture, with the aim of finding a fast and accurate surrogate; the selected configuration is described in Section~\ref{ss:div1dnn}. These evaluations are done using the density ramp dataset, aiming to simulate density ramps in a realistic TCV divertor plasma.

For the encoder and decoder we evaluate two architectures, and for the processors we evaluate the DRN, UNet, FNO, MP-PDE, FT, and FT-FNO (Section~\ref{ss:modelarch}). Since the emphasis of the surrogate model lies on fast computation, we evaluate configurations for a range of inference speeds. For one block ($\text{\SI{2}{\milli\second}}$ of real time) we search configurations with a computation time of $\approx$\{0.5, 1, 2, 4, 6, 8, 10\}$\text{\SI{}{\milli\second}}$. We search for parameters aiming at `wide' and `deep' networks, that is, putting the emphasis on a network with many features per block, or on stacking many blocks. For each processor we define a maximum width and maximum depth configuration and iteratively reduce it until it reaches the desired computation time. For details on these configurations and the identified settings we refer to Appendix~\ref{ap:results}.

We compare predictions on simulations from the test set through work-precision diagrams. Quality is measured through the Mean Squared Error (MSE) of the solutions, starting after the input time block of $w$ timesteps: 
\begin{align}\label{eq:mse}
\begin{split}
    \text{MSE}&\big(\mathbf{u}^{\mathbf{t}, \mathbf{x}}, \widetilde{\mathbf{u}}^{\mathbf{t}, \mathbf{x}}\big) =
    \\    
    &\frac{1}{(\textit{Nt}-w) \cdot \textit{Nx}}\textstyle\sum_{t=t_w}^{t_\textit{Nt}}\textstyle\sum_{x=x_0}^{x_\textit{Nx}} \big(\mathbf{u}^{t, x} - \widetilde{\mathbf{u}}^{t, x}\big)^2,
\end{split}
\end{align}
for the discretized DIV1D reference solution $\mathbf{u}$ and the corresponding prediction $\widetilde{\mathbf{u}}$. We measure with variable-wise standardized solutions so all variables are on the same scale. As a reference, since the test set will have a mean of approximately 0 and a variance of approximately 1, a naïve baseline of predicting all zeros will result in an MSE of approximately 1. Speed is measured as the time to compute $\text{\SI{2}{\milli\second}}$ worth of solutions, which corresponds to 1 output block. In other words, we compute the latency of the model with a batch size of 1 (the total number of solutions per second can be increased by using bigger batch sizes, which better exploits the parallel computation of GPUs). %

\begin{figure*}[h]
    \centering
    \begin{subfigure}[h]{0.315\textwidth}
\begin{center}\includegraphics[width=\textwidth]{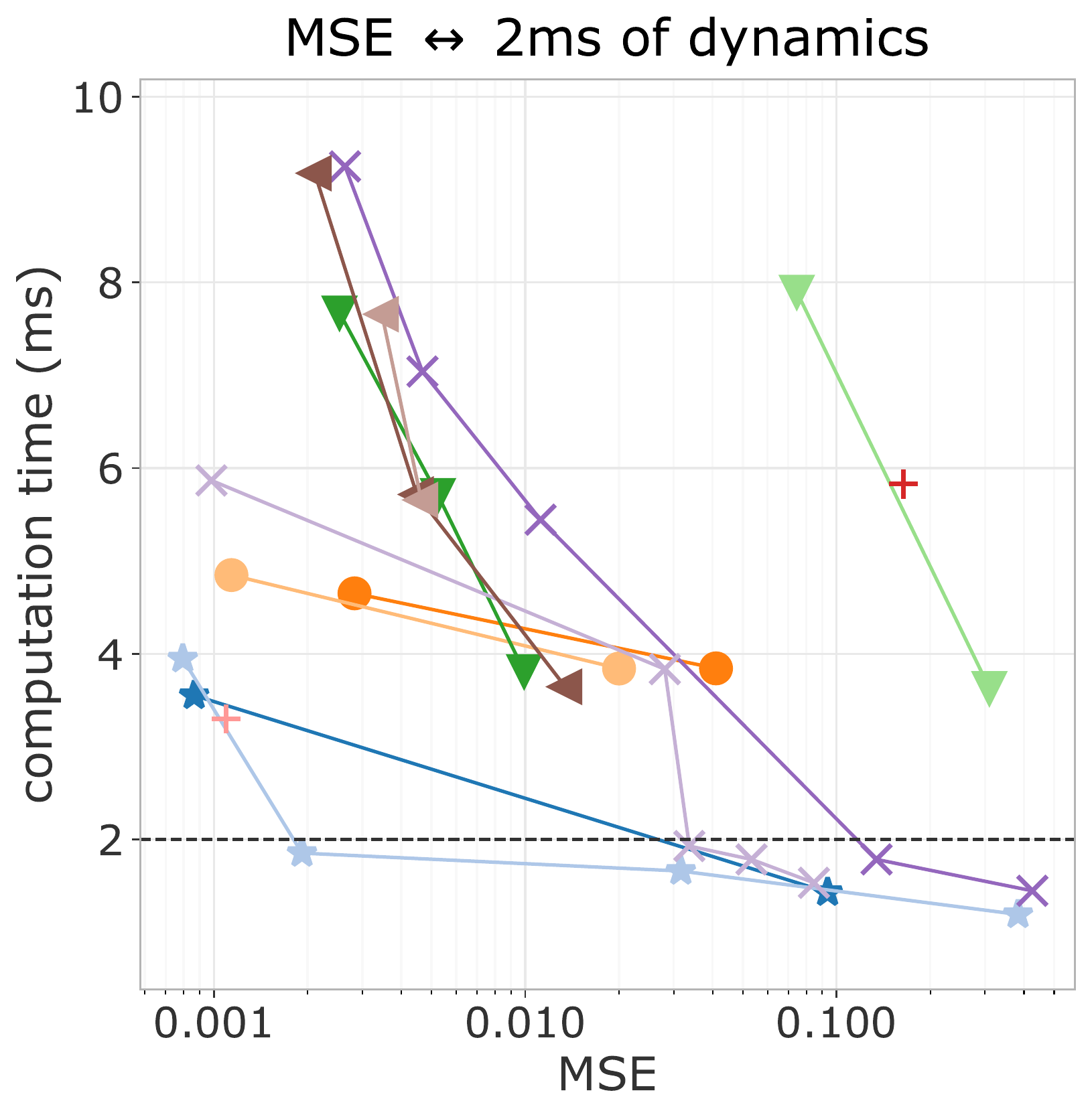}\end{center}
    \includegraphics[width=3.1\textwidth]{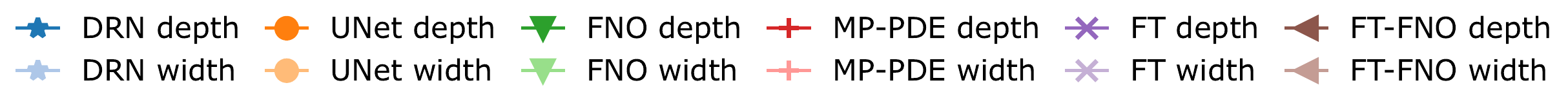}
    \caption{Full simulations}{\label{fig:wpall}}
    \end{subfigure}
    \begin{subfigure}[h]{0.315\textwidth}
\begin{center}\includegraphics[width=\textwidth]{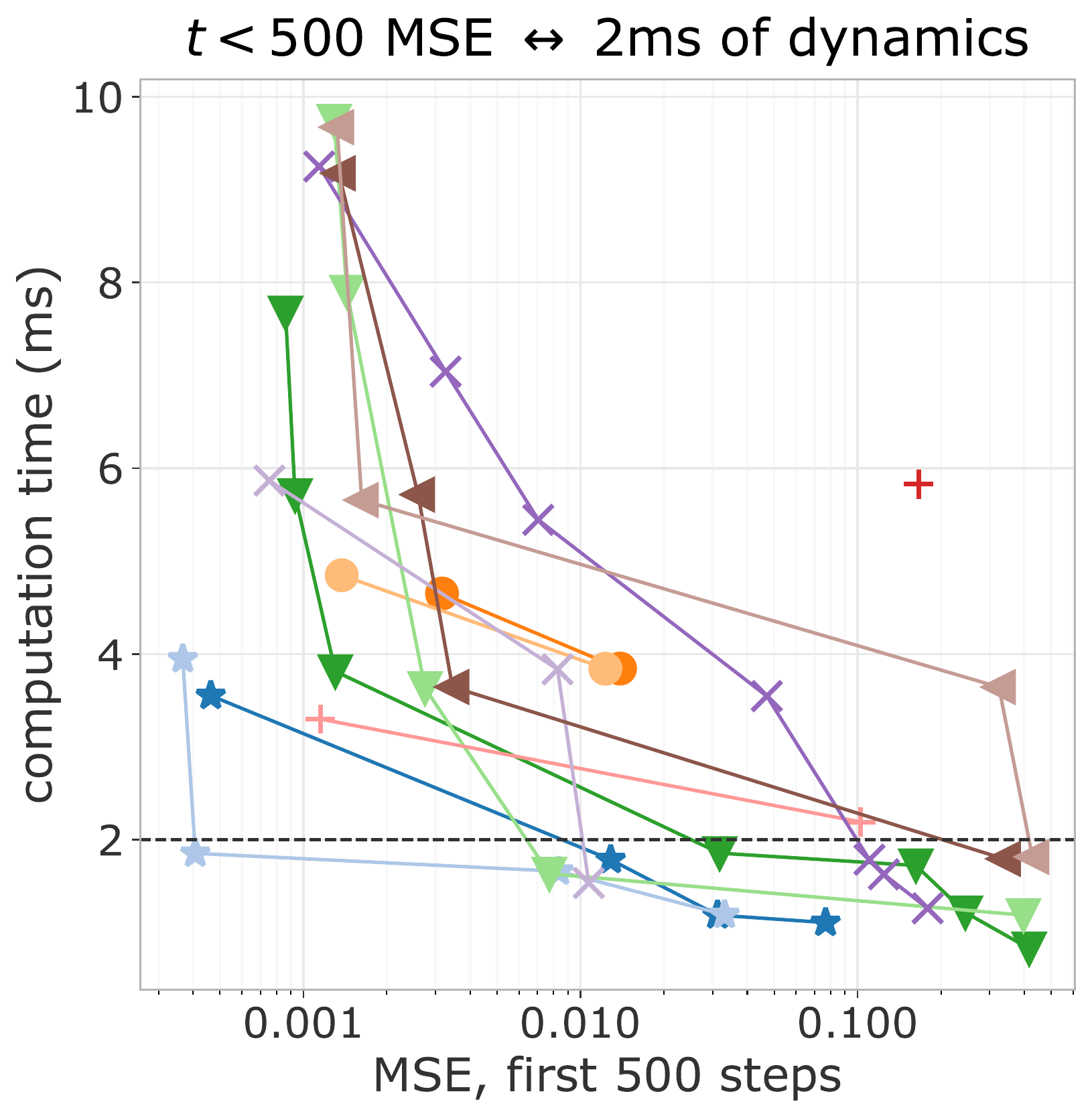}\end{center}
\phantomgraphics[width=3.1\textwidth]{figures/wp/legend_all_nobg.pdf}
    \caption{First 500 steps}{\label{fig:wp500}}
    \end{subfigure}
        \begin{subfigure}[h]{0.317\textwidth}
\begin{center}\includegraphics[width=\textwidth]{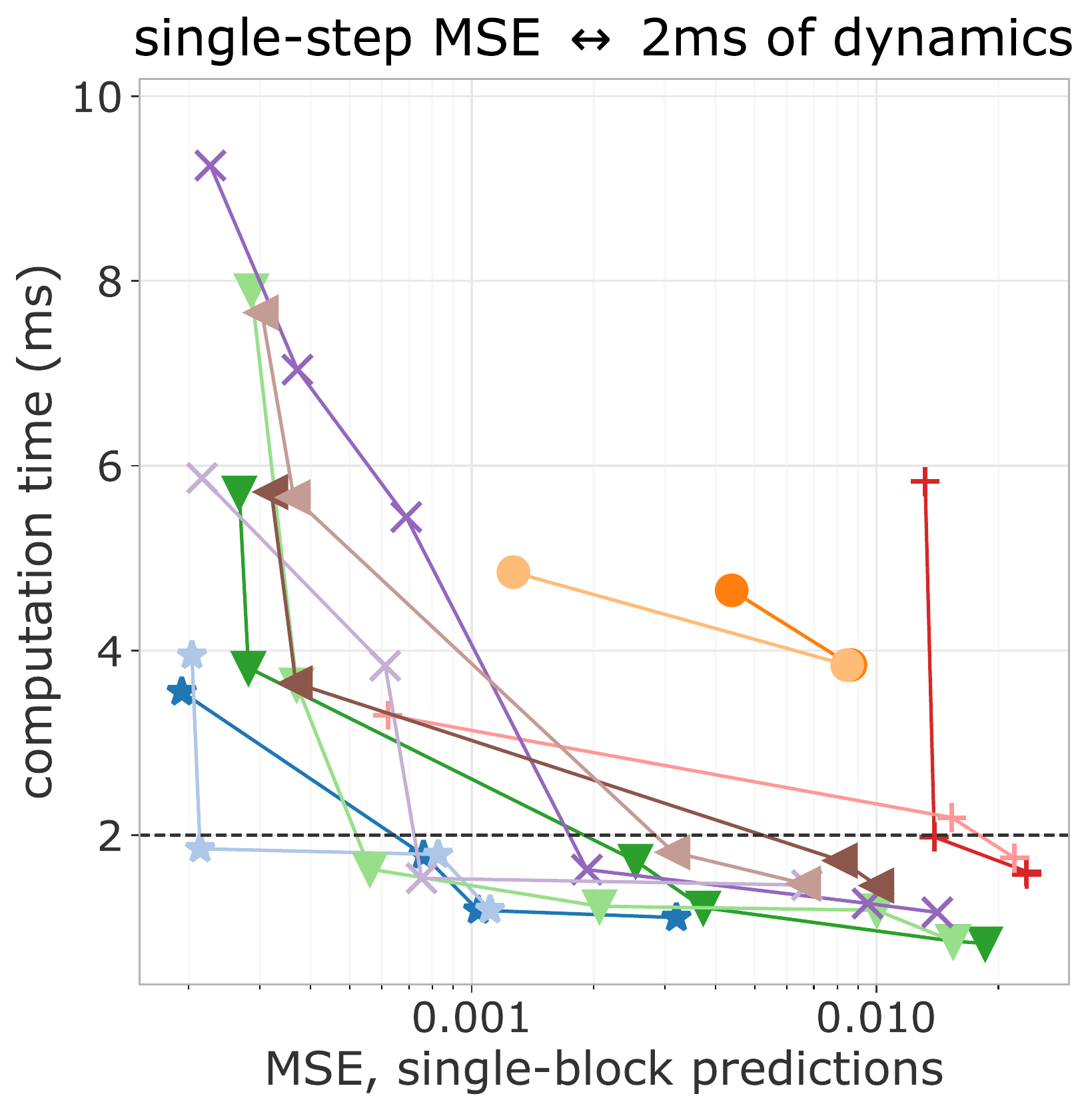}\end{center}
\phantomgraphics[width=3.1\textwidth]{figures/wp/legend_all_nobg.pdf}
    \caption{Single-block predictions}{\label{fig:wpss}}
    \end{subfigure}
    \caption{Work-precision diagrams for the neural PDE surrogates for: (a) full solution rollouts; (b) solutions of 500 steps; and (c) predicting 20 timesteps (one time block) from arbitrary starting points. Computation time is measured on an NVIDIA A100 GPU. The DRN dominates the Pareto front in most cases, although this advantage is smaller for shorter time windows.}
    \label{fig:wpnn}%
\end{figure*}

In the work-precision diagrams (Figures~\ref{fig:wpdiv1d} and~\ref{fig:wpnn}), each processor architecture and parameter-search strategy (deep or wide networks) is shown as one line: Each point represents a different configuration plotted at its speed and error. We plot the Pareto front of each such category to keep the plots readable, and omit all points with an MSE of more than 0.5. For a complete overview of experimental results we refer to Appendix~\ref{ap:results}.

\renewcommand{\rulesep}{\hfill}

\renewcommand{\rulesep}{\hfill}
\begin{figure*}[h]
    \centering
    \begin{subfigure}[b]{0.425\textwidth}
\begin{center}\includegraphics[width=\textwidth]{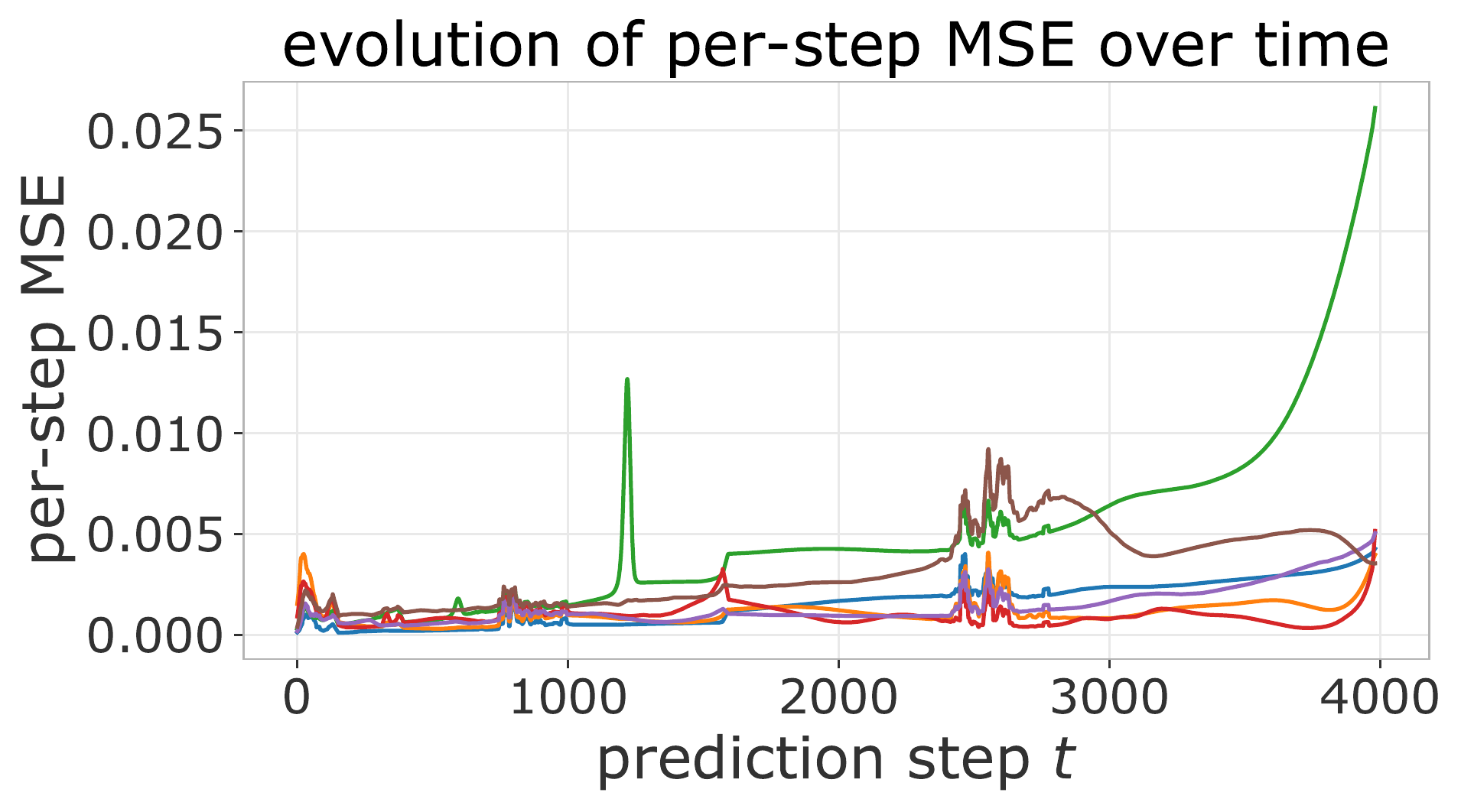}\end{center}
    \caption{Per-step MSE over time}{\label{fig:timeps}}
    \end{subfigure}
\includegraphics[width=.12\textwidth]{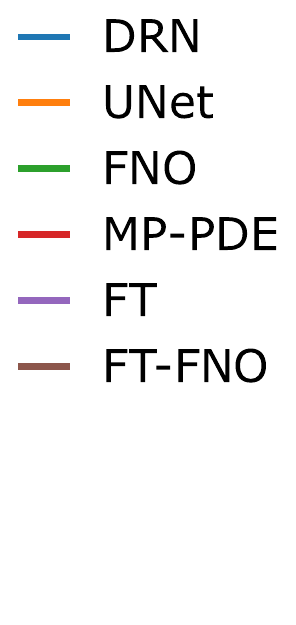}
        \begin{subfigure}[b]{0.425\textwidth}
\begin{center}\includegraphics[width=\textwidth]{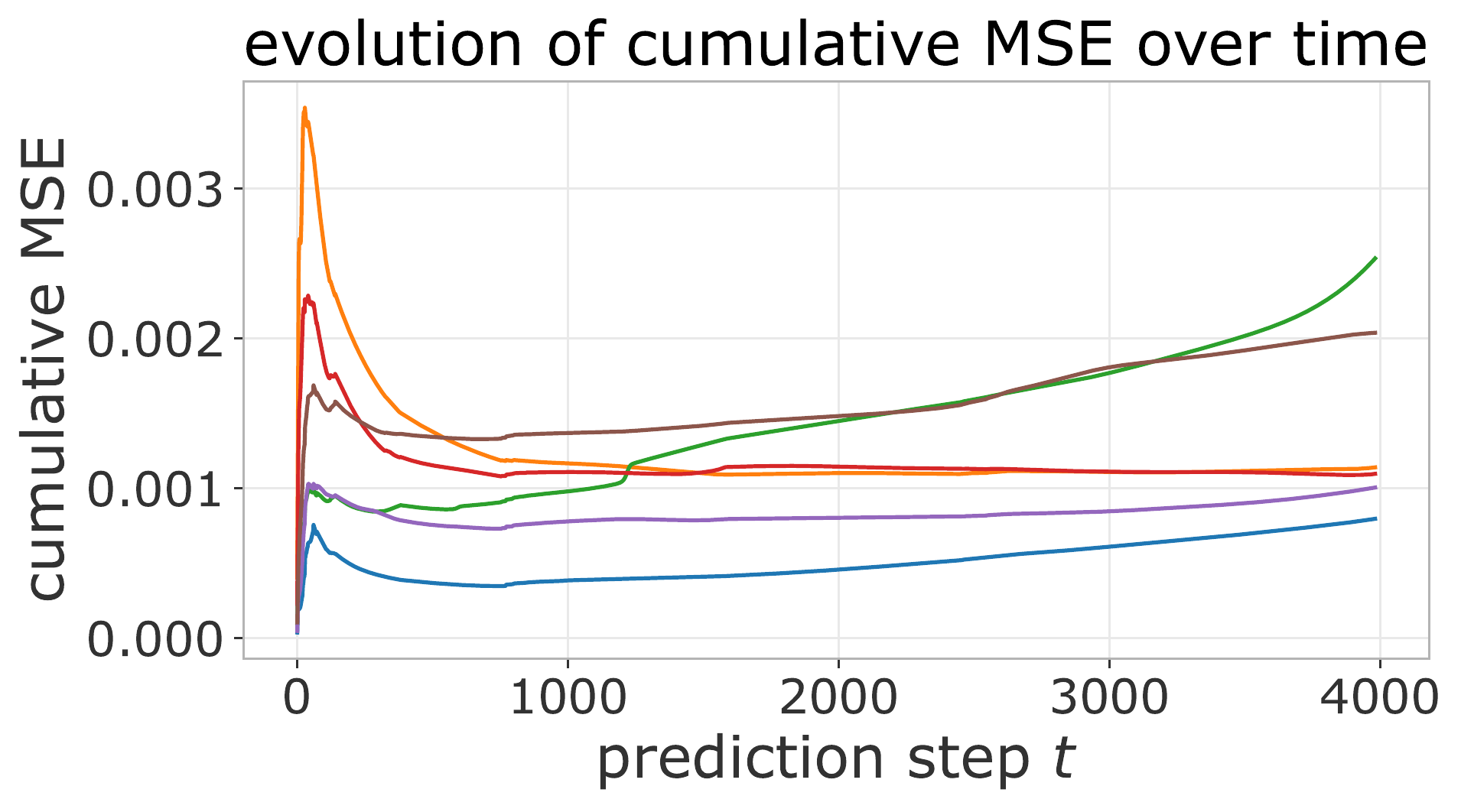}\end{center}
    \caption{Cumulative MSE over time}{\label{fig:timecumul}}
    \end{subfigure}
    \caption{Plots of test simulation error as a function of time. The per-step error is displayed in (a), whereas (b) shows the cumulative error. The common spike at the start in (b) can be explained due to the impact of short simulations: These dynamics are faster and more challenging, and only affect the error computation for their short time span.}
    \label{fig:time}%
\end{figure*}

\textbf{Comparison with DIV1D}. We start by comparing full-length simulations. To place the surrogates in context, we compare them to DIV1D's speed-accuracy trade-off when coarsening its spatial grid. To evaluate the computation time of the neural PDE surrogates with a reasonably optimized code, an effort was made to accelerate the DIV1D code. By considering that neutrals have finite energy in the calculation of charge exchange energy losses \cite[{Eq.~11a}] {derks2022}, the original DIV1D implementation is accelerated considerably. We denote this faster version as DIV1D$_\textit{fast}$. While the solutions are qualitatively similar this deviation in the computation results in notable differences on the gridpoint level. Therefore, we compute all DIV1D/DIV1D$_\textit{fast}$ errors w.r.t. their own best-quality solutions at $\textit{Nx}=500$.

In Figure~\ref{fig:wpdiv1dcpu} the comparisons are plotted for equal hardware (AMD Rome 7H12 CPU, 1 core). The left-most points for DIV1D and DIV1D$_\textit{fast}$ correspond to the reference solutions (hence 0 error and the `gap' in the log-scale x-axis). On equal hardware, neural PDE surrogates already show significant speed-ups compared to DIV1D while keeping high accuracy: The surrogates' best accuracy is comparable to the first step down for DIV1D and DIV1D$_\textit{fast}$, which corresponds to running them using 450 grid points, down from 500 grid points for the reference solutions.

Since NN methods benefit from parallel computation and the use of dedicated accelerators such as GPUs, we evaluate the surrogates using a fast GPU; these results are plotted in Figure~\ref{fig:wpdiv1dgpu}. While this comparison is not even, it is highly non-trivial to exploit this hardware with existing CPU-based numerical codes such as DIV1D. As such, we primarily focus on the fastest possible computation times. Figure~\ref{fig:wpdiv1dgpu} shows that the neural PDE surrogates can generate accurate solutions several orders of magnitude faster.

\begin{figure*}[t]
\begin{center}\includegraphics[width=.99\textwidth]{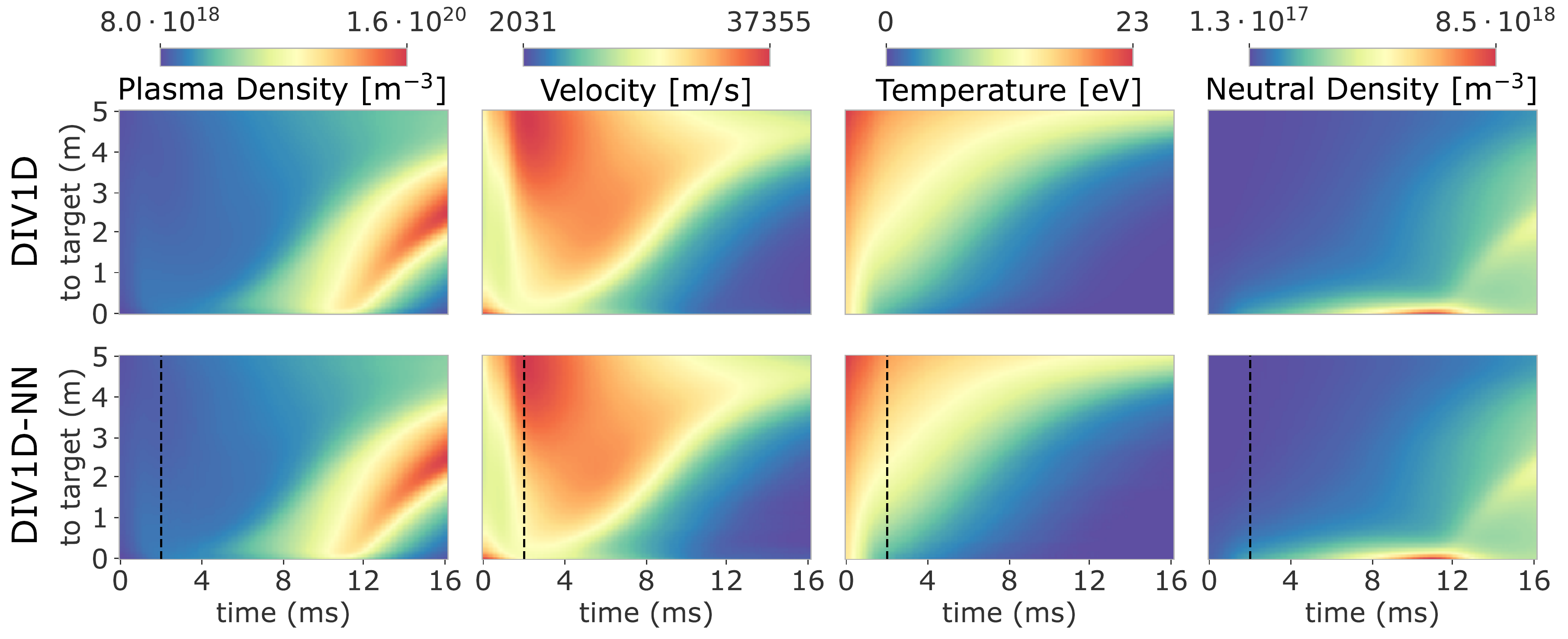}\end{center}\vspace{-0.4cm}
    \caption{Visualization of a simulation from the test set: A ramp up of $n_{\mathrm{u}} = [1.0 - 5.0]\text{\SI{e19}{\per\meter\cubed}}$ over $\text{\SI{16}{\milli\second}}$, $q_{\|\mathrm{u}} = \text{\SI{15}{\mega\watt\per\meter\squared}}$, $\xi_{\mathrm{C}} = 0.04~\mathrm{ion/electron}$. The top row depicts the refence DIV1D simulation, the bottom row the DIV1D-NN simulation. The dashed line indicates the end of the first block, the input for DIV1D-NN.}
    \label{fig:sims}%
\end{figure*}

\begin{figure}[t]
    \centering
    \begin{center}\includegraphics[width=1.\linewidth]{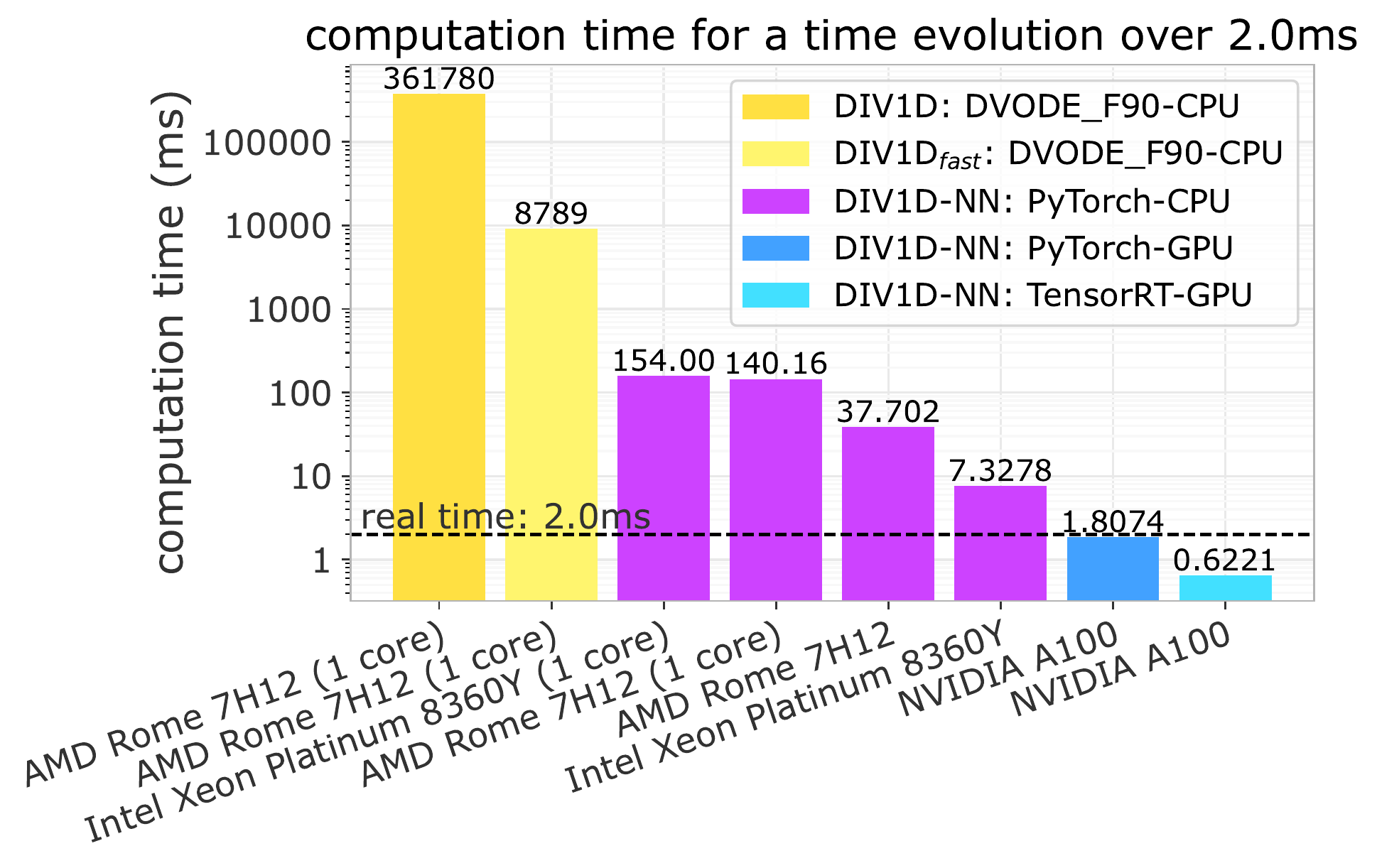}\end{center}
    \caption{Computation times of DIV1D-NN with varying hardware and inference engines, along with DIV1D for comparison. DIV1D-NN's speed is computed as the time of one model forward pass, generating a block of 20 timesteps spanning $\text{\SI{2}{\milli\second}}$ of dynamics. DIV1D's speed is the average time to generate $\text{\SI{2}{\milli\second}}$ of data from the test set. Since DIV1D-NN incurs some error, we use the first setting for DIV1D and DIV1D$_{\textit{fast}}$ that resulted in a higher error than DIV1D-NN ($\textit{Nx}=400$ for both, third points from the left in Figure~\ref{fig:wpdiv1d}).} %
    \label{fig:computationtimes}%
\end{figure}

\textbf{Comparison of neural PDE surrogates}. To compare surrogate methods more precisely we zoom in on the surrogates only, see Figure~\ref{fig:wpall} for full-simulation errors for all methods on the GPU. The DRN shows strong performance, both for computing maximum accuracy solutions at any cost and for computing accurate solutions fast. Notably, the UNet's speed-accuracy trade-off is very favorable when using a single CPU core (Figure~\ref{fig:wpdiv1dcpu}), but this advantage fades when evaluating on the GPU. Likely, the relatively deep and complex architecture of a UNet is not as favorable for the parallelized GPU computations. 

Since the solutions can cover many timesteps--the longest simulations being 4000 timesteps--error accumulation becomes a major factor. Methods could be viable for short-term predictions but show bad performance over long rollouts. To evaluate the latter, we check the error for the MSE for the first 500 steps ($\text{\SI{50}{\milli\second}}$) of all simulations, and errors of single block predictions ($\text{\SI{2}{\milli\second}}$) given arbitrary starting points in the simulation. These are plotted in Figures~\ref{fig:wp500} and~\ref{fig:wpss}, respectively. The DRN still performs best, but the lead is less pronounced.

To further evaluate the influence of error accumulation we plot the error of simulations over time. For each processor architecture we select the model with the lowest MSE and plot this error as a function of time. In Figure~\ref{fig:timeps} the error is plotted at each individual timestep, whereas Figure~\ref{fig:timecumul} displays the cumulative MSE evolving over time (with the final timestep being the total MSE as used in Figure~\ref{fig:wpall}). While most models have a similar distribution over time, the FNO stands out: At short timeframes its error is comparable to the DRN and FT, but it accumulates more error over time. The opposite holds for the MP-PDE and UNet, which show more stability w.r.t. time.

\subsection{Selecting the NN Surrogate Architecture for DIV1D}\label{ss:div1dnn}
For the final divertor plasma surrogate model, we select the configuration we deem to have the best speed-accuracy trade-off. This model is the DRN with an inference speed of $\approx\text{\SI{1.807}{\milli\second}}$ per block and a standardized MSE of 0.001918. We dub this configuration as \textit{DIV1D-NN}. To push the computation time as low as possible we further optimize this implementation and compile the model with the NVIDIA TensorRT SDK~\cite{tensorrt} for fast inference. These optimizations result in an inference speed of $\approx\text{\SI{0.6221}{\milli\second}}$ per block, about 3 times as fast as the simulated plasma dynamics of $\text{\SI{2}{\milli\second}}$. Figure~\ref{fig:computationtimes} depicts a comparison of compute times with varying hardware and with DIV1D. For context, training DIV1D-NN (the offline cost) took just under 4 hours.

As qualitative comparison we plot a DIV1D-NN simulation alongside the reference DIV1D simulation in Figure~\ref{fig:sims}, of dynamics induced by a ramp up. The top row depicts the reference simulation, whereas the bottom row depicts the DIV1D-NN simulation. %
Qualitatively, these simulations align closely. %

\subsection{Case Studies: Recovering Properties and Structures}\label{ss:casestudies}
To assess the utility of DIV1D-NN for downstream tasks we evaluate its performance on recovering a set of relevant properties and structures. In particular, we consider the ability to reconstruct two non-linear phenomena: A roll-over of the target ion flux with increasing upstream plasma density~\cite{loarte1998} and a bifurcation of the target temperature as a function of upstream plasma density~\cite{capes1992}. These phenomena are important when aiming for divertor plasmas that maintain both a low temperature and ion flux on the target. Additionally, we evaluate the reconstruction of the approximate emission front, a useful proxy for detachment control~\cite{ravensbergen2021}. We repeat a subset of these evaluations for fast transient behavior by retraining DIV1D-NN on the fast transients dataset and re-evaluating the results.

Key observation is that we do not explicitly train DIV1D-NN for any of these properties, but rather evaluate whether it is sufficiently accurate w.r.t. DIV1D such that the surrogate can fill various roles of the source model without building a surrogate for each role. An additional benefit relative to building individual surrogates is that we can still evaluate the full trajectories related to these predictions. If a trajectory can be identified as non-physical we can discard the prediction, making our surrogate modeling strategy less of a black box compared to methods that directly map input parameters to target quantities.

\renewcommand{\rulesep}{\hfill}
\begin{figure}[t]
\begin{center}\includegraphics[width=.9\linewidth]{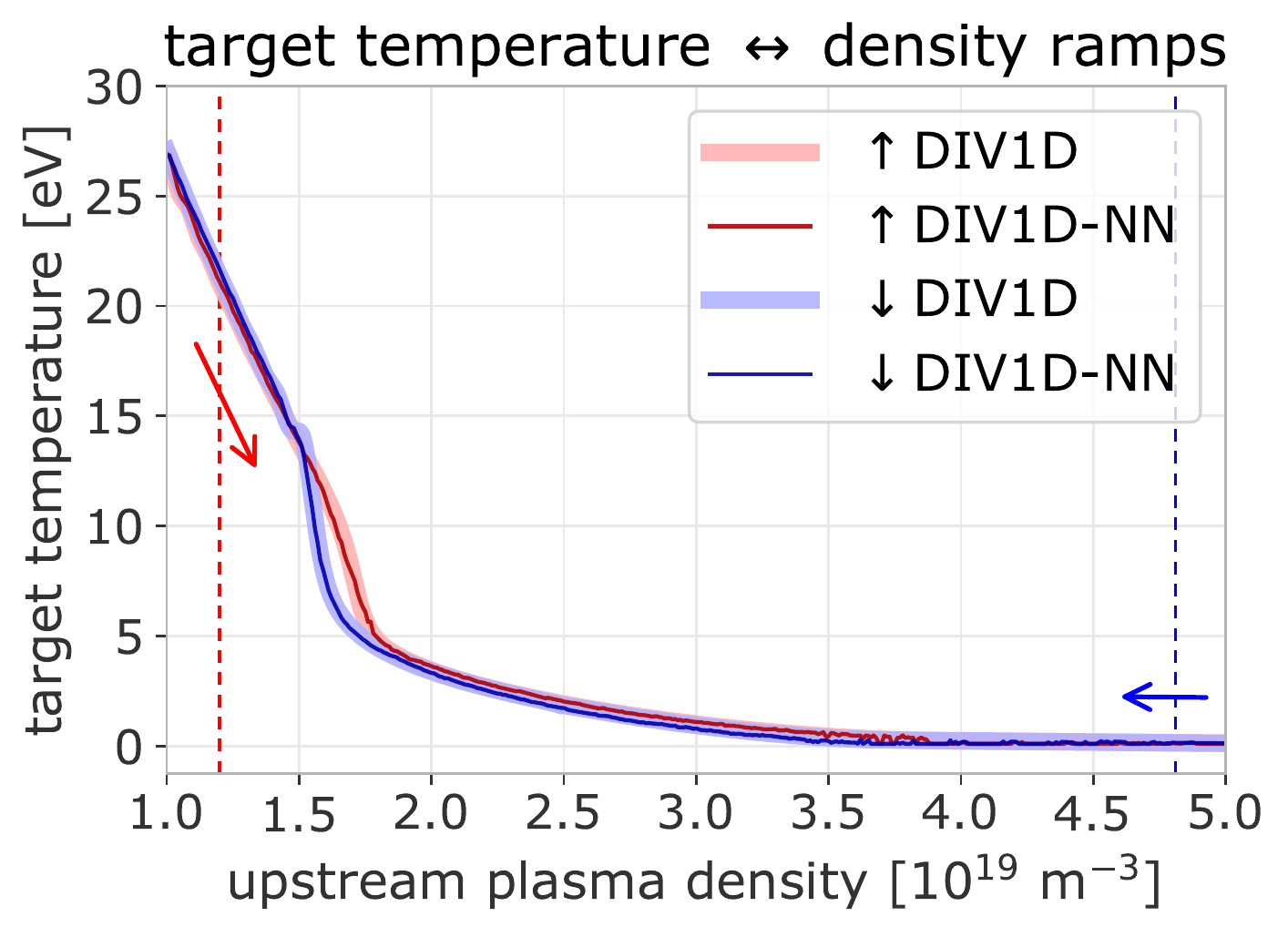}\end{center}
    \caption{Target temperature as function of upstream plasma density, comparing DIV1D and DIV1D-NN for $\text{\SI{40}{\milli\second}}$ ramps with $q_{\|\mathrm{u}} = \text{\SI{25}{\mega\watt\per\meter\squared}}$ and $\xi_{\mathrm{C}} = 0.05$. The plot depicts both a ramp up and ramp down from the test set, with red indicating the ramp up and blue indicating the ramp down. DIV1D-NN matches the bifurcation captured by DIV1D well, with minor artifacts.}
    \label{fig:bifs}%
\end{figure}

\renewcommand{\rulesep}{\hfill}
\begin{figure}[t]
\begin{center}\includegraphics[width=.9\linewidth]{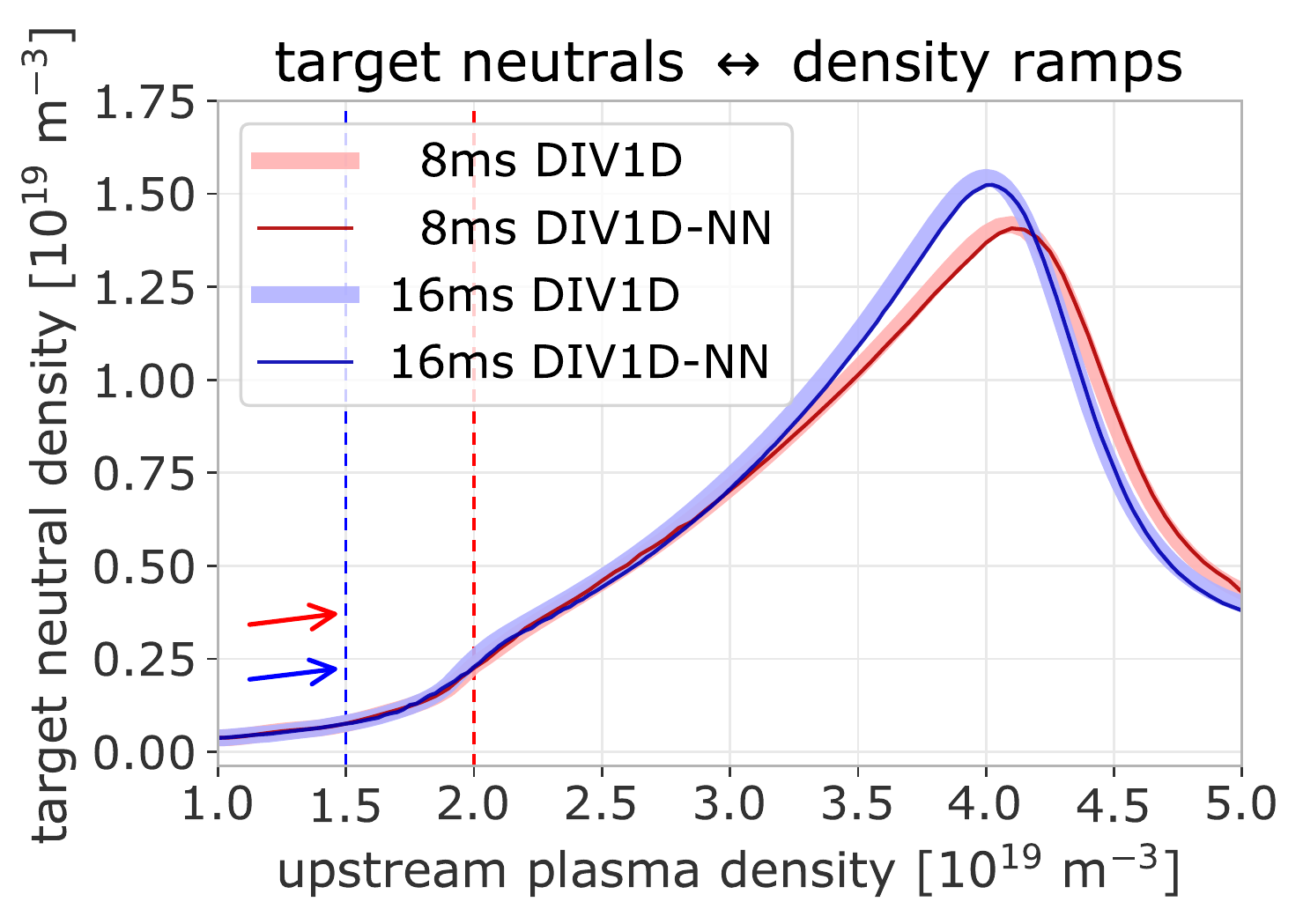}\end{center}
    \caption{Target neutral density as function of upstream plasma density, comparing DIV1D and DIV1D-NN for an $\text{\SI{8}{\milli\second}}$ and $\text{\SI{16}{\milli\second}}$ ramp up with ${q_{\|\mathrm{u}} = \text{\SI{30}{\mega\watt\per\meter\squared}}}$ and ${\xi_{\mathrm{C}} = 0.05}$. The roll-over of target ion flux results in a similar roll-over in the target neutrals; this phenomenon is captured by DIV1D, and recovered by DIV1D-NN.}
    \label{fig:neutrals}%
\end{figure}

\textbf{Bifurcation in target temperature}. First, we evaluate whether DIV1D-NN accurately captures the bifurcations (hystereses) in the target temperature. That is, in certain otherwise identical conditions, the target temperature varies depending on whether the upstream density goes from low to high or vice versa~\cite{capes1992}. We illustrate this bifurcation, with DIV1D-NN compared to DIV1D, in Figure~\ref{fig:bifs}. We find a close match, with only small deviations in the DIV1D-NN simulation. Quantitatively, the standardized MSE of the target temperature is 0.001007 over all test simulations. Rescaled to the physical scale, the average absolute error between predicted and real target temperatures is $\text{\SI{0.1714}{\electronvolt}}$.

\textbf{Target ion flux roll-over}. Next, we consider whether DIV1D-NN accurately captures the roll-over in the ion flux~\cite{loarte1998}, as well as its dependence on the ramp speed. This roll-over can be demonstrated by looking at the target neutrals, due to recombination at the target being proportional to the target ion flux. In Figure~\ref{fig:neutrals} the neutral density at the target is plotted as function of upstream density for varying density rates. The roll-over is reconstructed accurately, with the influence of time dynamics clearly illustrated. 
Quantitatively, the standardized MSE of the target neutral density is 0.08385 over all test simulations. In the physical scale, this error corresponds to an average absolute error of $\text{\SI{2.2873e17}{\per\meter\cubed}}$.

\renewcommand{\rulesep}{\hfill}
\begin{figure}[t]
\begin{center}\includegraphics[width=.9\linewidth]{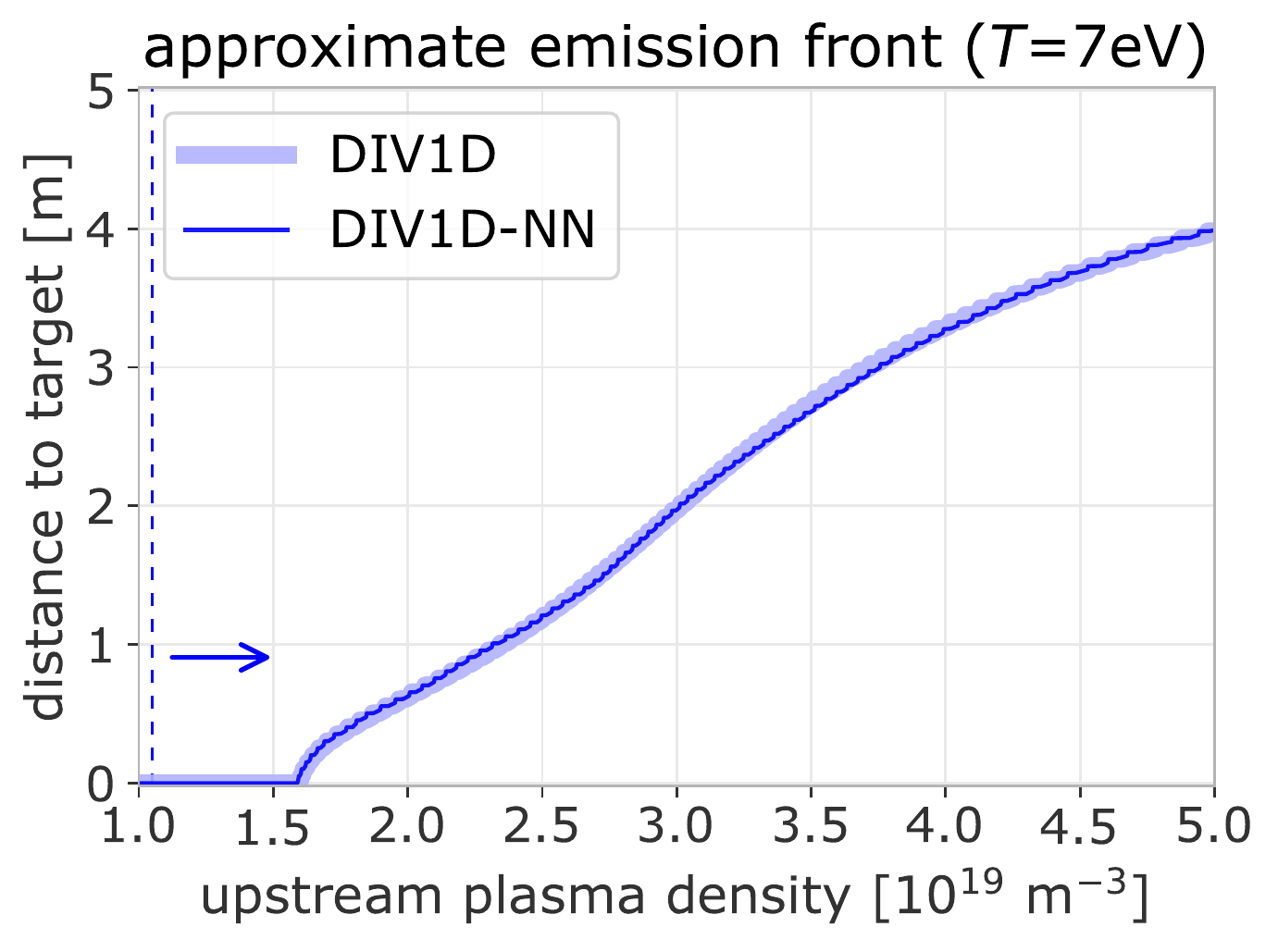}\end{center}
    \caption{Approximate emission front as function of upstream plasma density, comparing DIV1D and DIV1D-NN for a $\text{\SI{160}{\milli\second}}$ ramp up with {$q_{\|\mathrm{u}} = \text{\SI{20}{\mega\watt\per\meter\squared}}$} and {$\xi_{\mathrm{C}} = 0.02$}.}
    \label{fig:radfronts}%
\end{figure}

\textbf{Emission front position}. As final property we consider the approximate location of the carbon impurity emission front, defined as the position along the flux tube where the plasma temperature is equal to $\text{\SI{7}{\electronvolt}}$. The $\text{\SI{7}{\electronvolt}}$ temperature corresponds to the peak of the cooling rate function from~\cite{post1977} used by DIV1D. Ideally, this temperature corresponds to the temperature of the CIII impurity front position as used in~\cite{koenders2022,ravensbergen2021}. In reality the exact temperature of the front position depends on the plasma scenario and is hard to infer (see~\cite{smolders2020,theiler2017} and references therein). As definition of detachment we follow~\cite{stangeby2018}, who define it as the moment where the target temperature is below $\text{\SI{10}{\electronvolt}}$, but we use the $\text{\SI{7}{\electronvolt}}$ temperature as it conveniently corresponds to our definition of the carbon impurity emission front.
The position is detected by scanning the generated temperature profiles, filtering them such that they are monotonically decreasing from upstream towards the target, and interpolating between gridcells to find the location at $\text{\SI{7}{\electronvolt}}$. We scan from upstream towards the target; if no point below $\text{\SI{7}{\electronvolt}}$ is found, we assume the front to lie at the target, corresponding to an attached divertor plasma.

\renewcommand{\rulesep}{\hfill}

In Figure~\ref{fig:radfronts} we overlay the predicted emission front on top of the estimation for a reference DIV1D simulation for a ramp up. Even over the long simulation timeframe of $\text{\SI{160}{\milli\second}}$, the DIV1D-NN prediction still closely aligns with DIV1D. The average absolute error of the predicted emission front location over the entire test set is $\text{\SI{0.01570}{\meter}}$. Additionally, we can use the predicted emission front location to estimate detachment in the simulated plasma: We assume detached if the temperature at the target is smaller than $\text{\SI{7}{\electronvolt}}$, and attached otherwise. A confusion matrix for this evaluation is provided in Table~\ref{tab:attachmatrix}. In most cases DIV1D-NN accurately captures the state modeled by DIV1D, with an accuracy of $\approx$ 99.87\% over the entire test set (as a baseline, $\approx$ 78\% accuracy could be achieved by always predicting detachment).

\begin{table}[h]
\centering
\begin{tabular}{cccc}
& & \multicolumn{2}{c}{DIV1D} \\
& \multicolumn{1}{c?{\heavyrulewidth}}{} & Detached & Attached \\
\addlinespace[-2.32pt]
\cmidrule[\heavyrulewidth]{2-4}
\addlinespace[-\belowrulesep]
\multirow{2}{*}{DIV1D-NN} &\multicolumn{1}{c?{\heavyrulewidth}}{Detached} & \cellcolor[RGB]{151,198,223} 9557 & \cellcolor[RGB]{255,255,255} 7 \\
&\multicolumn{1}{c?{\heavyrulewidth}}{Attached} & \cellcolor[RGB]{255,255,255} 9 & \cellcolor[RGB]{218,232,245} 2612 \\
\addlinespace[-2.32pt]
\cmidrule[\heavyrulewidth]{2-4}
\end{tabular}
\caption{Confusion matrix of attachment predictions, where attachment is defined as the target temperature being below $\text{\SI{7}{\electronvolt}}$. The accuracy of predicting attachment is 99.87\% over the entire test set.}
\label{tab:attachmatrix}
\end{table}

\begin{figure}[b]
    \centering
\begin{center}\includegraphics[width=.995\linewidth]{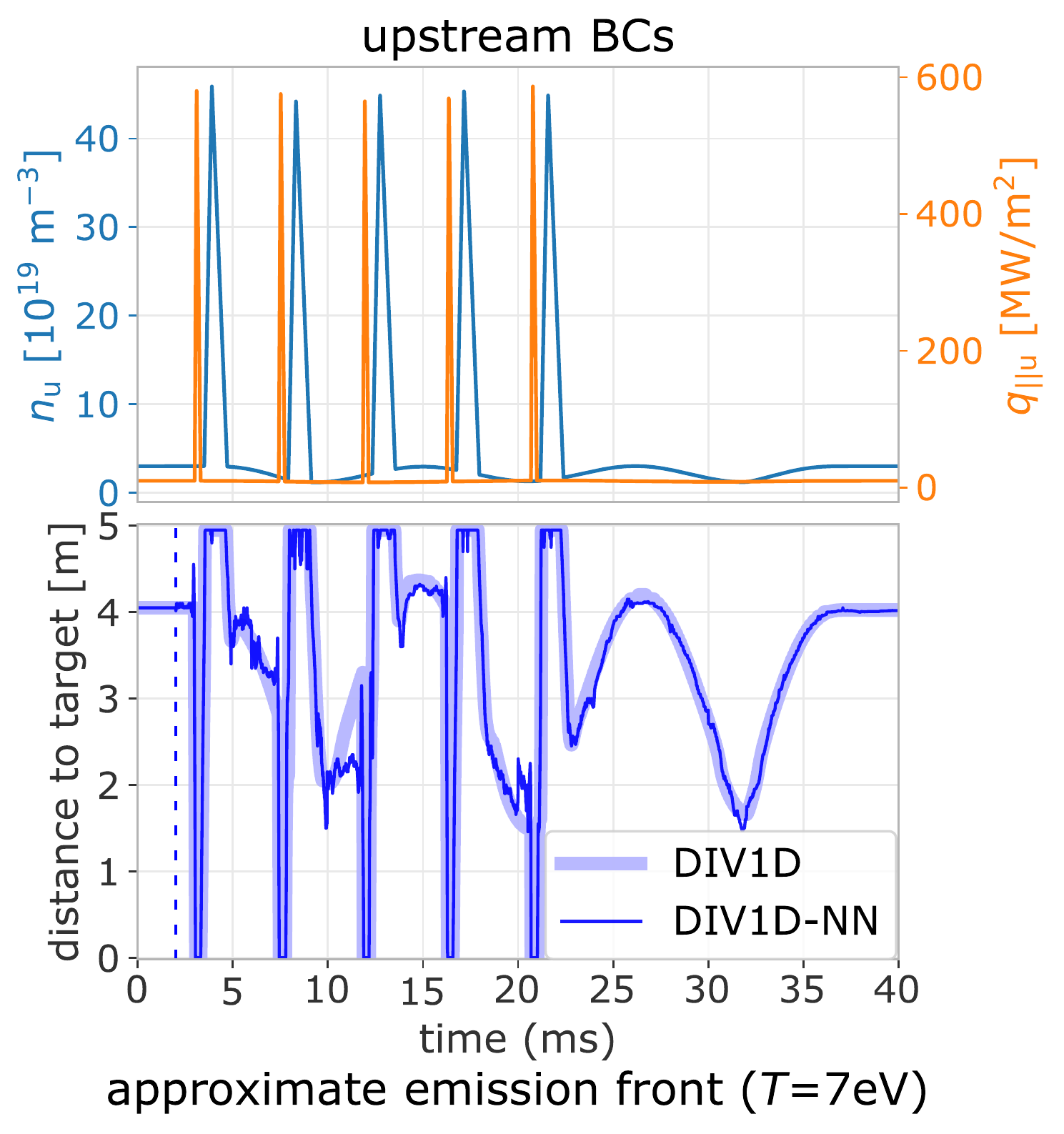}\end{center}
    \caption{Comparing DIV1D and DIV1D-NN for tracking the emission front over a period with fast transients. The corresponding BCs, the upstream density $n_{\mathrm{u}}$ and upstream parallel heat flux $q_{\|\mathrm{u}}$, are depicted on top.}{\label{fig:elmdetachment}}
\end{figure}

\begin{figure*}[h]
\begin{subfigure}[b]{.976\textwidth}
\begin{center}\includegraphics[width=\textwidth]{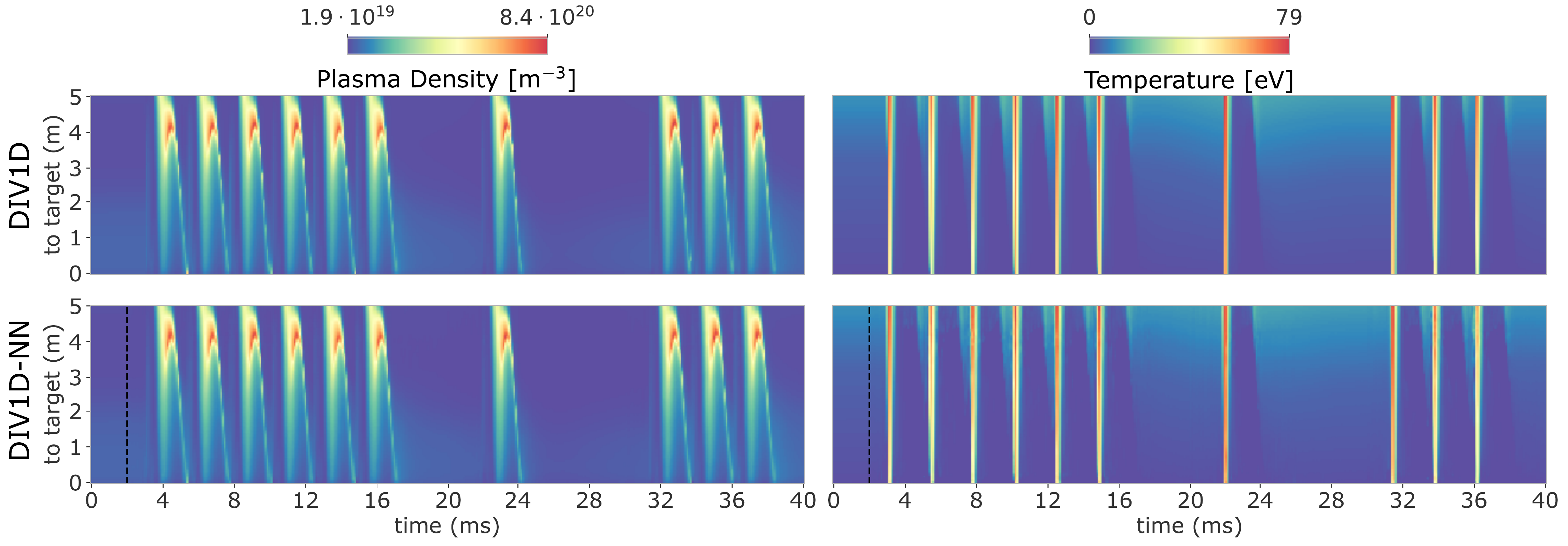}\end{center}\vspace{-0.4cm}
    \caption{A test-set simulation from the fast transient dataset, here showing only the plasma density and temperature. Corresponding dynamic upstream BCs are depicted in (c), the impurity fraction $\xi_{\mathrm{C}}$ is set to 0.03 $\mathrm{ion/electron}$ (static).}
    \label{fig:simelm1}%
\end{subfigure}

\begin{subfigure}[b]{.483\textwidth}
    \centering
\begin{center}\includegraphics[width=\linewidth]{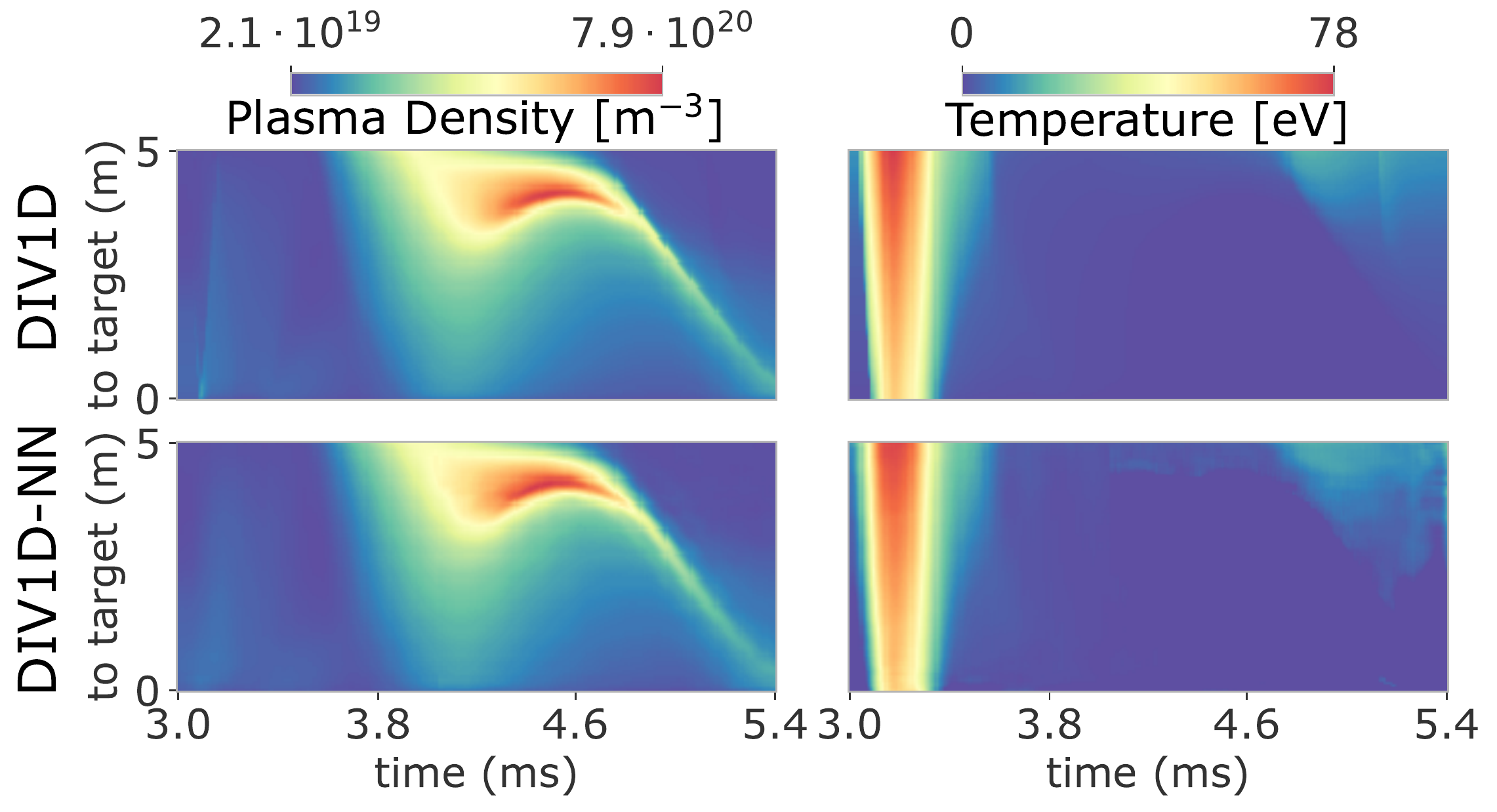}\end{center}
    \caption{Zoomed-in view of the first transient. DIV1D-NN captures the overall structure well, but there are noticeable artifacts and missing details in the surrogate's solution.}{\label{fig:elmshort1}}
\end{subfigure}
\hfill
\begin{subfigure}[b]{.49\textwidth}
\begin{center}\includegraphics[width=\linewidth]{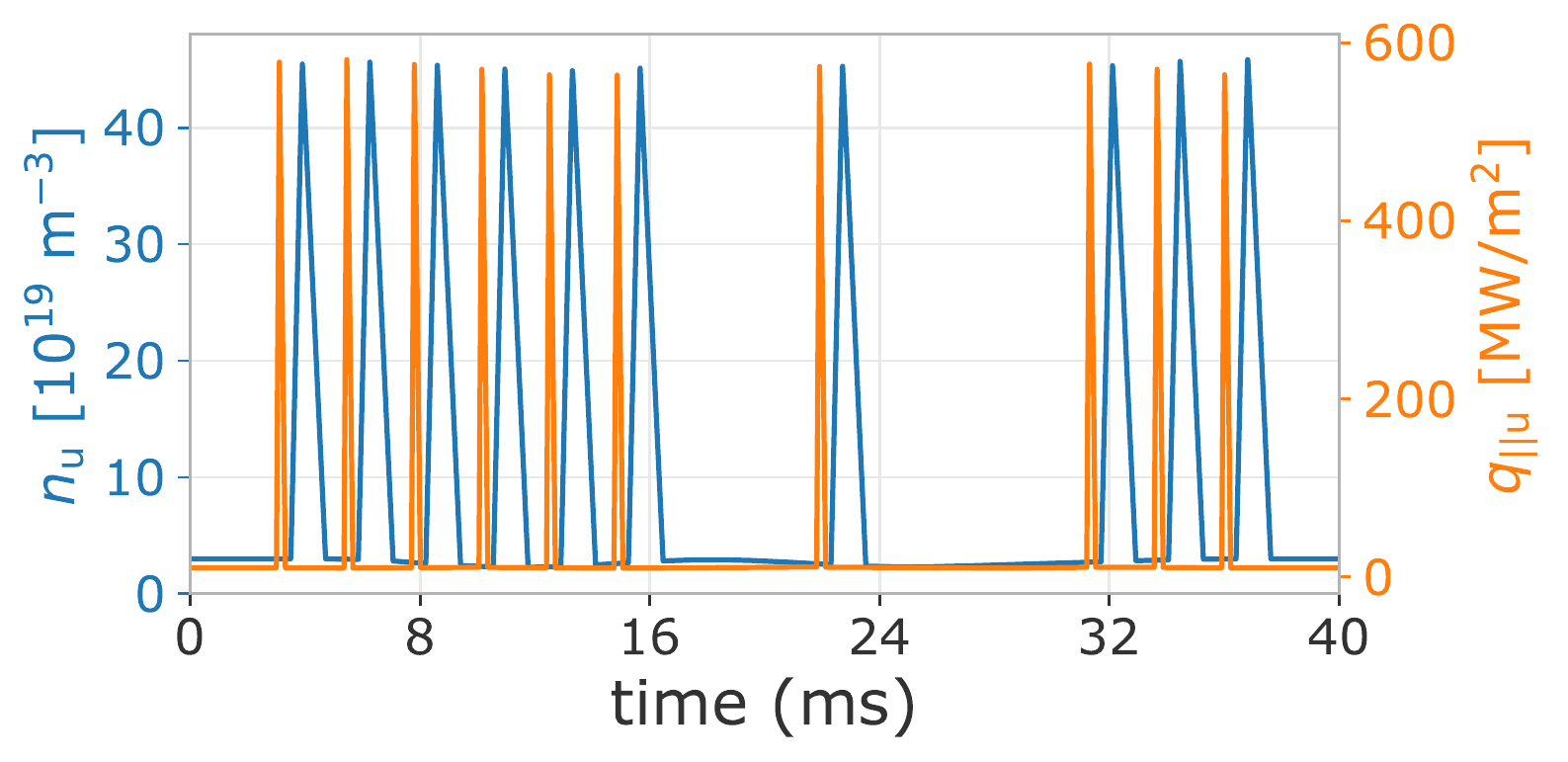}\end{center}\vspace{-0.4cm}
    \vspace{0.2cm}
    \caption{Dynamic boundary conditions for the upstream density $n_{\mathrm{u}}$ and upstream parallel heat flux $q_{\|\mathrm{u}}$.}
    \vspace{0.4cm} %
    \label{fig:simelm1bc}%
\end{subfigure}
\caption{A test-set simulation from the fast transient dataset: (a) evolution of the plasma density and temperature; (b) zoomed-in view of a single transient event; and (c) the dynamic BCs. The impurity fraction $\xi_{\mathrm{C}}$ is static at 0.03 $\mathrm{ion/electron}$.}\label{fig:simelmtotal}
\end{figure*}

\textbf{Fast transients.} To investigate the capabilities of the DIV1D-NN surrogate architecture in a more challenging setting, we retrain this architecture from scratch using the fast transients dataset. These transients could resemble phenomena like ELMs, representing much smaller timescales with much larger gradients than found in the density ramp dataset; the corresponding solutions are much less smooth than those trained and tested with before. Training DIV1D-NN on the fast transient data took about 12 hours of wall-clock time.

Since we predict in blocks of $\text{\SI{2}{\milli\second}}$, we now use window size $w=200$ instead of $w=20$ to account for the shift in time discretization $dt$ from $\text{\SI{0.1}{\milli\second}}$ for the density ramp data to $\text{\SI{0.01}{\milli\second}}$ for the fast transient data. Note that for the transient data, we always start simulations with a steady-state, i.e., all 200 timesteps in the first block are identical.

The architecture of the model's processor is identical to DIV1D-NN as used before, the change in input and output dimension is accounted for in the encoder and decoder respectively. As a result, model inference is slightly slower, taking $\approx\text{\SI{0.6788}{\milli\second}}$ per block to compute $\text{\SI{2}{\milli\second}}$ of plasma dynamics ($\approx\text{\SI{0.6221}{\milli\second}}$ before). Even though we compute at a 10 times finer temporal resolution the effect on computation time is somewhat limited, because in the processor the dimensionality was already scaled up beyond the time discretization, and this dimensionality remains unchanged between the two datasets' models. Additionally, we now directly predict solution values, rather than the delta as written in Equation~\ref{eq:soltemporalbundling}. We found that models struggled to train when predicting the delta, as the correct output values then heavily depend on whether the input block ended on top of a spike in the solution or not.

The MSE on standardized data using DIV1D-NN is 0.02298, which is comparable to running DIV1D with just below 400 grid points, instead of the 500 grid points used for the reference solutions (see Appendix~\ref{ap:transients} for DIV1D's scaling on the fast transient data). The NN surrogate does not scale as well as on the density ramp data, stressing the challenge for data-driven surrogates when modeling high-frequency dynamics. However, large-scale features are still recovered well, see Figure~\ref{fig:simelmtotal} for an example solution.

Of interest is whether aggregate structures can be captured well in the fast transient setting. To evaluate this question, we compare the estimated emission front locations for fast transient simulations. The average absolute error of the front location is $\text{\SI{0.05374}{\meter}}$, with the detachment prediction having an accuracy of 98.61\%; see Appendix~\ref{ap:transients} for the confusion matrix. The model is still quite accurate, but there is a drop in quality compared to results on the smoother density ramp dynamics. An example is provided in Figure~\ref{fig:elmdetachment}. The position is tracked well on a global scale, but on the small scale, such as the regions between the spikes, there are noticeable artifacts.

\renewcommand{\rulesep}{\hfill}

\begin{figure*}[h]
    \centering
    \begin{subfigure}[b]{0.42\textwidth}
\begin{center}\includegraphics[width=\textwidth]{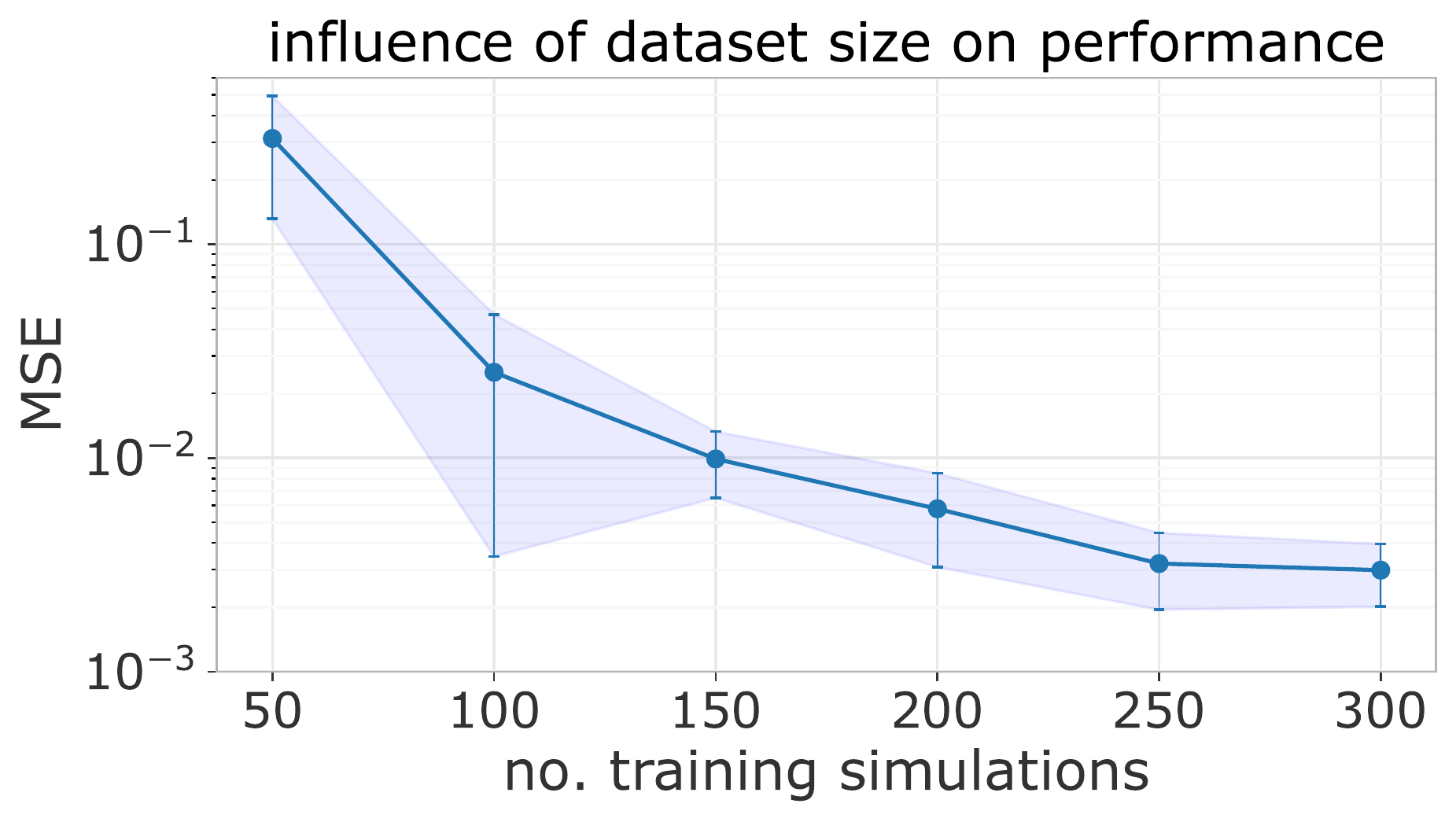}\end{center}
    \caption{Density ramp dataset}{\label{fig:datasetramp}}
    \end{subfigure}
    \hspace{.08\textwidth}
        \begin{subfigure}[b]{0.42\textwidth}
\begin{center}\includegraphics[width=\textwidth]{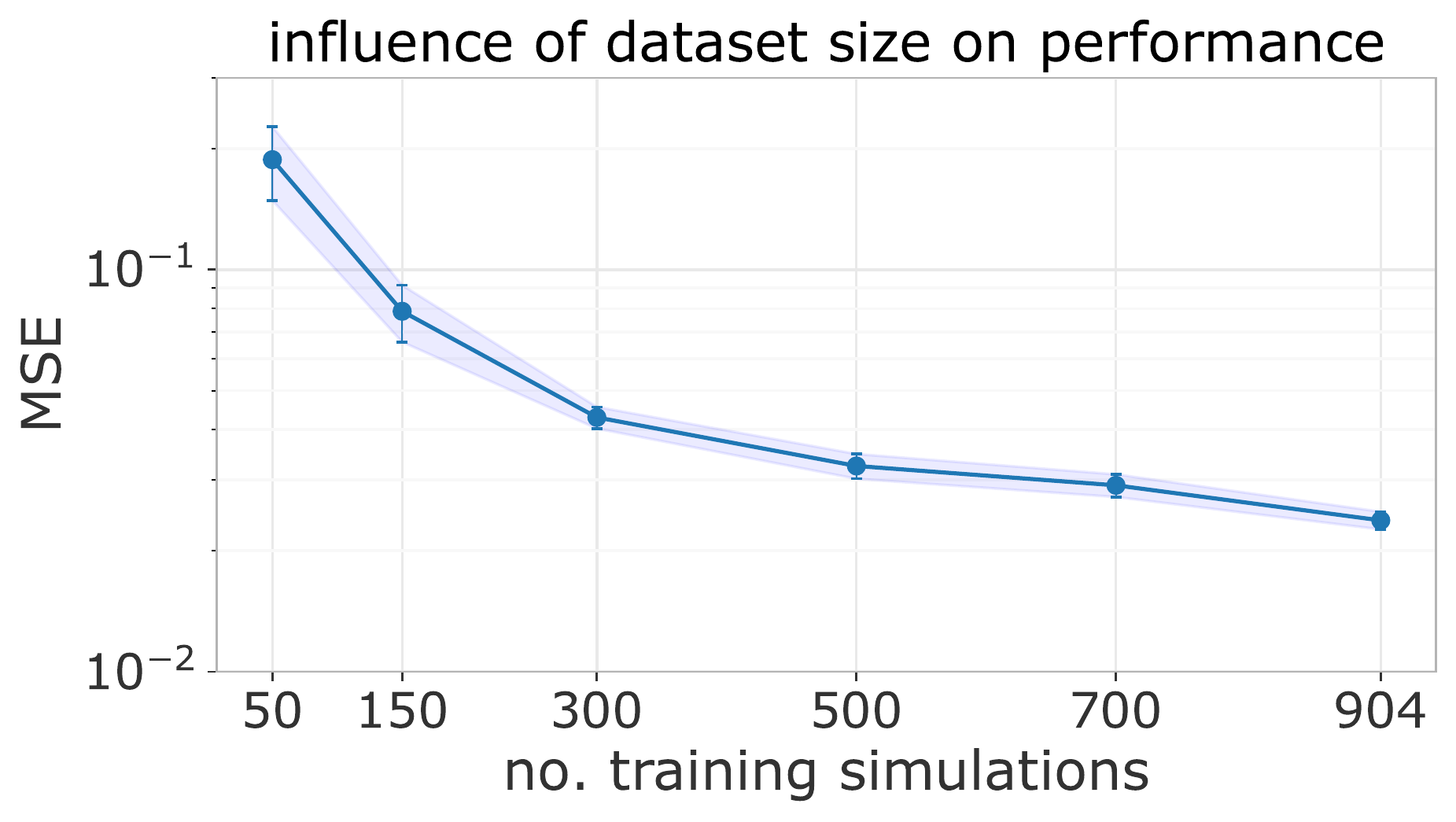}\end{center}
    \caption{Fast transients dataset}{\label{fig:datasetelm}}
    \end{subfigure}
    \caption{MSE of DIV1D-NN as function of dataset size. We use the same validation and test splits as before, but vary the number of train simulations. Each size is evaluated for five different splits; for the full sizes (300 for density ramps and 904 for fast transients) the difference stems from different model initializations. Results are shown as mean $\pm$ standard deviation.}
    \label{fig:dataset}%
\end{figure*}

\renewcommand{\rulesep}{\hfill}
\begin{figure*}[h]
\begin{subfigure}[h]{0.35\textwidth}
\begin{center}\includegraphics[width=\textwidth]{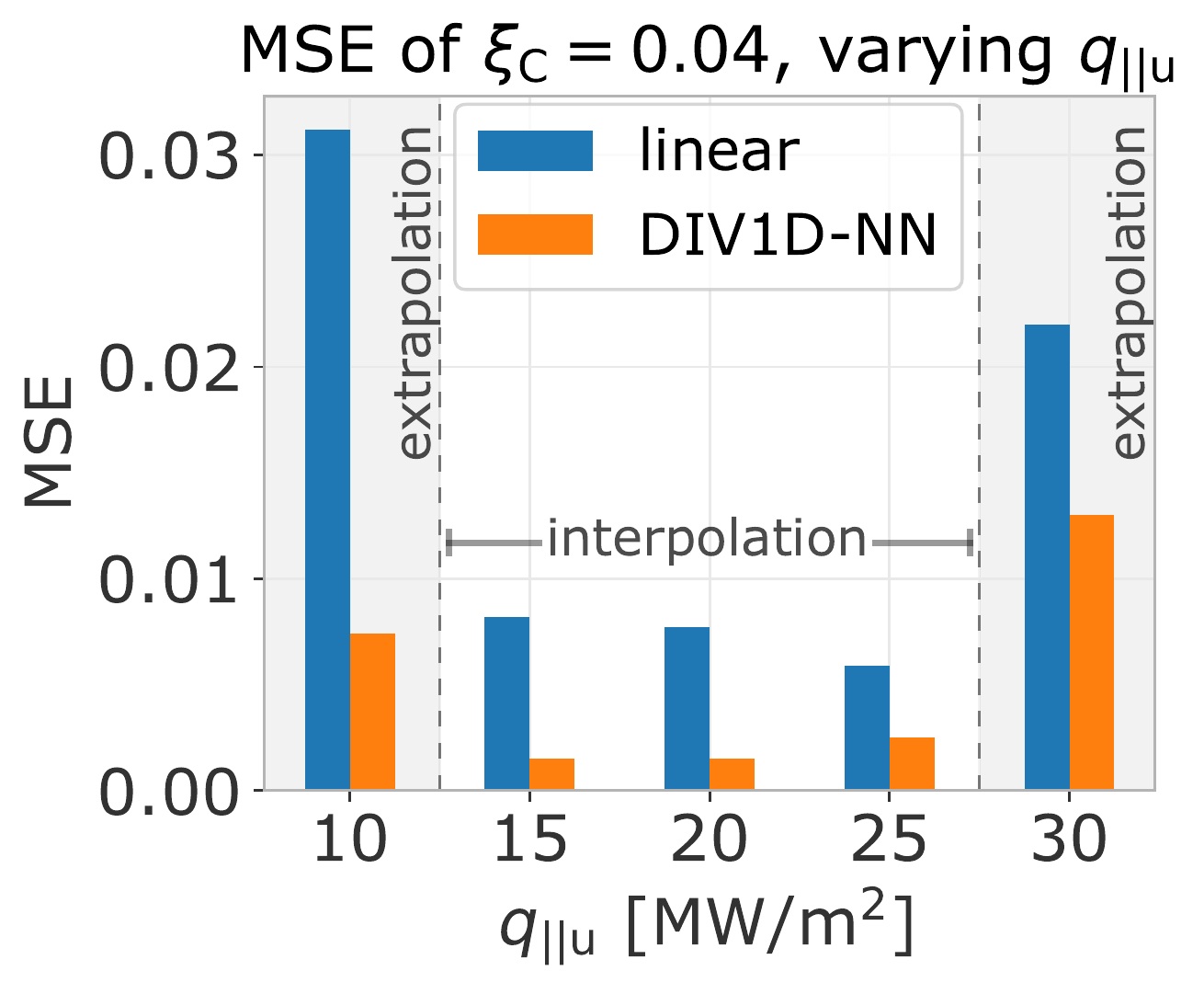}\end{center}
\caption{Evaluation of error when inter- or extrapolating in the parameter space, for both DIV1D-NN and when linearly interpolating surrounding solutions. We use simulations that do not contain either $\xi_{\mathrm{C}}=0.04$ or $q_{\|\mathrm{u}} \in \{10, 15, 20, 25, 30\}$ to estimate solutions with these settings.}{\label{fig:interpmse}}
    \end{subfigure}
    \rulesep
\begin{subfigure}[h]{0.263\textwidth}
\begin{center}\includegraphics[width=\textwidth]{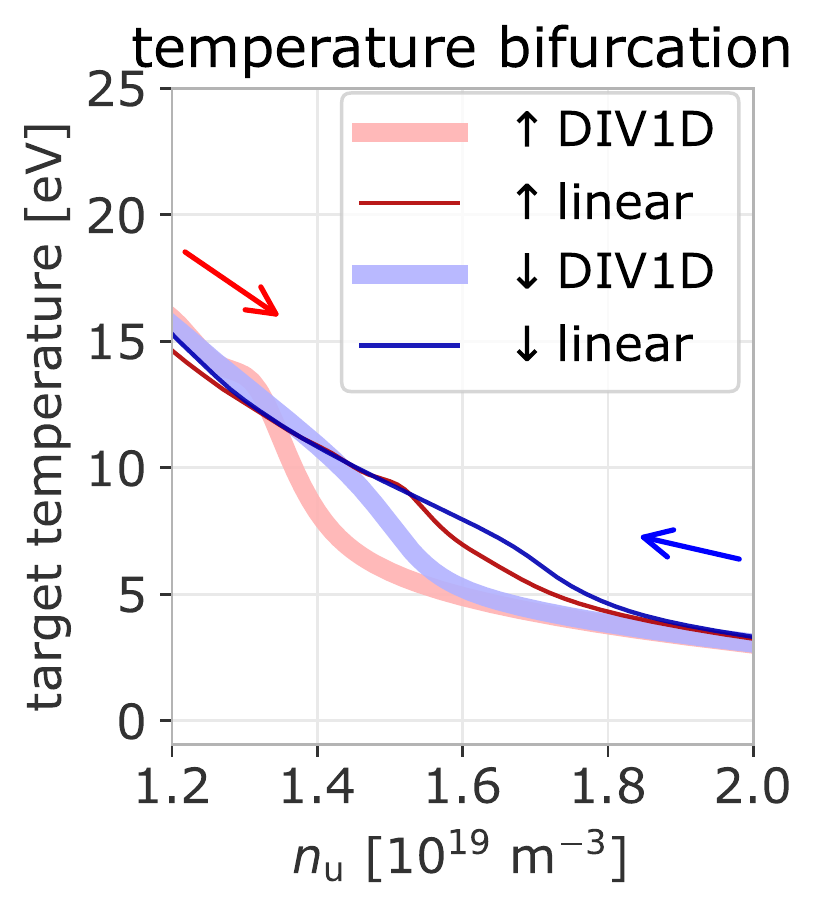}\end{center}
\caption{When linearly interpolating surrounding solutions we cannot capture the bifurcating dynamics correctly.}{\label{fig:bfinterp}}
    \end{subfigure}
    \rulesep
\begin{subfigure}[h]{0.278\textwidth}
\begin{center}\includegraphics[width=\textwidth]{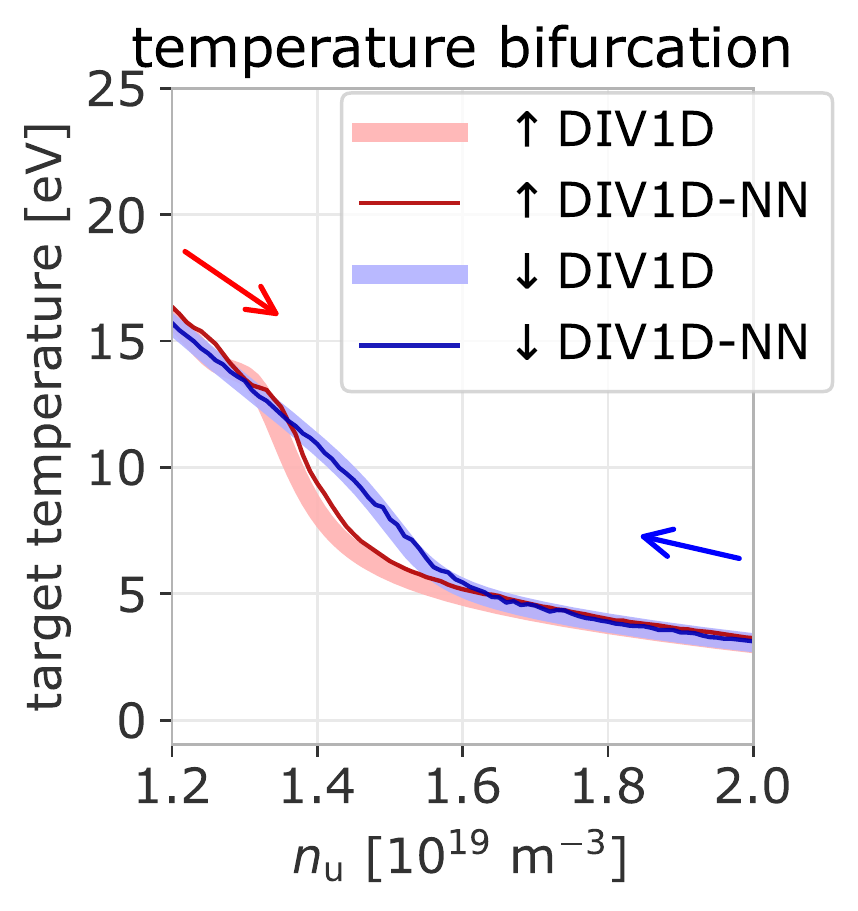}\end{center}
    \caption{NN-based interpolation, using proposed surrogate DIV1D-NN, captures the bifurcating dynamics (up to small artifacts).}\label{fig:bfnn}
    \end{subfigure}
    \caption{Evaluation of the inter- and extrapolation capability of DIV1D-NN, and a comparison with linear interpolation of surrounding solutions. Errors are given for $\xi_{\mathrm{C}} = 0.04$ and varying $q_{\|\mathrm{u}}$ in (a), where a white background denotes interpolation in the parameter space, and a gray background extrapolation. DIV1D-NN performs best when interpolating in the parameter space and consistently shows benefits over linearly interpolating surrounding solutions. This advantage is illustrated using the bifurcation in the target temperature in (b) and (c), for $\text{\SI{40}{\milli\second}}$ ramps with $q_{\|\mathrm{u}} = \text{\SI{20}{\mega\watt\per\meter\squared}}$ and $\xi_{\mathrm{C}} = 0.04$.}
    \label{fig:bifnn}%
\end{figure*}

\subsection{Data Efficiency}\label{ss:dataef}
When creating a data-driven surrogate model, the quality of the resulting model is highly dependent on the size of the dataset. Generating a rich training dataset can be a severe bottleneck in the surrogate creation process\footnote{Another avenue for improving data efficiency considers identifying which simulations are most informative to the surrogate, i.e., \textit{Active Learning}~\cite{settles2009}.}. To investigate the dependence on dataset size, we retrain DIV1D-NN on both datasets with varying levels of training samples, using five different subsets for each `number of samples' setting.

The results for the density ramp data and the fast transient data are plotted in Figures~\ref{fig:datasetramp} and~\ref{fig:datasetelm}, respectively. There is little drop in quality when taking one or two steps down for both datasets, the surrogates can reach satisfactory performance with relatively few simulations; in the order of 200-250 for the density ramps and about 500 for the fast transients. In general, we hypothesize that the surrogates can accurately match DIV1D with relatively few simulations because we exploit the full spatiotemporal signal of the solutions. For example, for 300 density ramp simulations with 100 gridpoints and an average of $\approx$1000 timesteps, there are $300 \times 100 \times 1000 \times 4 = \num{120000000}$ individual points the model uses to find correlations found in DIV1D solutions, a much more substantial sounding dataset.

\subsection{Evaluation of Inter- and Extrapolation}\label{ss:interp}
The utility of a surrogate lies in its ability to provide \textit{new} solutions fast. Consequently, it is crucial to evaluate the capabilities of the proposed surrogate's inter- and extrapolation capability w.r.t. the parameter space; the extent to which DIV1D-NN can accurately generate solutions in relation to the provided training data.

We evaluate the model performance on the density ramp data by leaving out all solutions with a given upstream parallel heat flux $q_{\|\mathrm{u}}$ \textit{or} impurity fraction $\xi_{\mathrm{C}}$. We retrain DIV1D-NN with most of the remaining simulations, leaving out a few as validation data. The resulting model is tested on the simulations with the left-out parameters for $q_{\|\mathrm{u}}$ and $\xi_{\mathrm{C}}$. 

Additionally, we consider linear interpolation between existing solutions as a baseline. New solutions are formed by interpolating between solutions for the surrounding values of $q_{\|\mathrm{u}}$ and $\xi_{\mathrm{C}}$ with identical ramp speed $\dot{n}_{\mathrm{u}}$, or extrapolating from the two closest values if we cannot interpolate. In practice we are lenient towards this method: Not all of these simulations are necessarily available and used as DIV1D-NN's training data, but we assume they are always available for the linear interpolation baseline.

Results for $\xi_{\mathrm{C}} = 0.04~\mathrm{ion/electron}$ and all values of $q_{\|\mathrm{u}}$ are provided in Figure~\ref{fig:interpmse}. As expected, the results are much better when interpolating within the parameter space, compared to extrapolating outside of this range. In general, these surrogate modeling techniques are ill-suited for extrapolating (far) beyond the training data. DIV1D-NN outperforms linear interpolation in all cases by a wide margin, although we note that in some parameter extrapolation settings (not depicted here) linear extrapolation outperforms DIV1D-NN; for details we refer to Appendix~\ref{ap:interp}. Evaluating non-linear dynamics, Figures~\ref{fig:bfinterp} and~\ref{fig:bfnn} depict the bifurcation in the target temperature~\cite{capes1992} for linear interpolation and for DIV1D-NN, respectively. The complete mismatch for linear interpolation shows a clear benefit of using neural PDE surrogates.

\section{Conclusions and Discussion}\label{sec:conclusions}
We have presented the application and extension of neural PDE surrogate techniques for building a fast dynamic 1D surrogate model of divertor plasmas. We demonstrated the application of an autoregressive neural network-based model, which approximates solutions by learning a time-stepping operator that evolves the state of the system, following the structure proposed in~\cite{brandstetter2022}. Within this framework, we investigated and extended five state-of-the-art neural network architectures to approximate said operator. The evaluation showed that many methods can accurately approximate solutions, most notably the Dilated Residual Network~\cite{stachenfeld2022}. We proposed surrogate architecture DIV1D-NN, which showed the best trade-off between error and computation time. DIV1D-NN can simulate density ramp dynamics faster than real-time, simulating a time evolution of $\text{\SI{2}{\milli\second}}$ in $\approx\text{\SI{0.63}{\milli\second}}$ of wall-clock time. 

Due to explicitly taking into account the dynamics of the plasma, we are able to recover non-linear time-dependent phenomena, demonstrated through the bifurcation in the target temperature. %
Further investigation also showed the recovery of properties and structures such as the roll-over in the target ion flux %
and the location of the emission front. %
An introductory evaluation of fast transient dynamics shows that the surrogate can reproduce higher-frequency dynamics, although there are improvements to be made in this area.

The aforementioned surrogate models could be trained with relatively few simulations, in the order of hundreds for density ramps and about 500 for fast transients. Additionally, we have evaluated the inter- and extrapolation capability of DIV1D-NN, showing its suitability for non-linearly interpolating within the parameter space of the training set.

One limitation of this work is that the dynamics in the divertor (along the magnetic field line) generally are very fast. In the density ramp dataset, the slowest ramp rates result in quasi-stationary profiles as a function of upstream conditions. %
With very fast perturbations, as found in the fast transient dataset, the simulated divertor plasma returned to steady-state conditions in the order of milliseconds. As such, many time dependencies are captured in only a few solution blocks (of $\text{\SI{2}{\milli\second}}$ each), the ability to model long-range time dependencies cannot be inferred from the conducted evaluations.

\subsection{Future Work}
The presented DIV1D-NN surrogate represents a real-time model of realistic TCV divertor plasmas. Of interest is the application in real-world use cases. For example, one could explore coupling a fast neural PDE surrogate in a flight simulator-setting~\cite{fable2021,felici2018,romanelli2014}, replacing lower-fidelity approximations. Another promising setting is real-time control, for example, to exploit real-time high-fidelity estimates of the plasma evolution in exhaust control schemes~\cite{ravensbergen2021} or in advanced control algorithms that require a model to be evaluated in real-time~\cite{bosman2022}. To improve a surrogate's utility in this setting one could explore constraining its dynamics, for example by learning a coordinate transform in which dynamics are linear~\cite{gin2020,lusch2018}. Furthermore, one can envision combining diagnostic measurements with fast state estimates in a Kalman filter-like setting~\cite{kalman1960}; especially with potentially limited diagnostics in future reactors~\cite{biel2019} fast high-fidelity plasma state estimates could become vital.

To improve the approximation quality of the surrogate model, one can explore a diverse set of directions. Different methods of conditioning the NN could be explored, for example, attention-based methods have shown strong performance in different domains~\cite{rebain2022}. Another angle is to use alternative approaches to deal with the distribution shift problem (Section~\ref{ss:modeltraining}), one being to inject noise into the training inputs~\cite{stachenfeld2022}. 
To better exploit dynamics found in data, one can structure networks better suited for geometrical transformations found in dynamical PDE solutions with stronger geometric priors~\cite{ruhe2023}. One can consider exploiting additional information following from physics laws~\cite{karniadakis2021,richterpowell2022}, as opposed to the primarily data-driven approach presented here. Stepping away from full surrogate models, one can explore hybrid techniques that combine machine learning and classical numerical methods~\cite{kochkov2021} to accelerate existing physics-based codes. Finally, the proposed techniques make deterministic single-point predictions not taking into account any uncertainties. In settings with turbulent dynamics or with experimental measurements it is crucial to explicitly account for uncertainty, for example, due to unresolved turbulence scales or measurement uncertainty~\cite{lakhlili2020}.

In a broader context, one can apply the discussed techniques to different domains and modalities. Numerical simulation of tokamak plasmas has proven vital to the development and operation of tokamak devices in various plasma regions and on many levels of fidelity~\cite{casali2018,derks2022,felici2011,hoelzl2021,pereverzev2002,raj2020,rognlien1994,wiesen2015}, and has shown reasonable agreement with experimental observations in many cases~\cite{fietz2013,pamela2017,teplukhina2017,wensing2021}. However, in a significant number of codes, the computational cost is prohibitively expensive. If a dataset with a sufficient number of simulations could be generated, the presented surrogate modeling techniques could be applied in order to generate many new simulations at a fraction of the cost or to simulate in time-sensitive control contexts. Since the presented techniques have shown good results using only hundreds of simulations, one could potentially apply them to much more expensive codes for which this number of simulations is obtainable (see e.g.~\cite{park2020}). In a similar vein, one could further explore the presented methods for creating data-driven simulators directly based on real-world physics, using the vast amount of experimental data that has been collected for a wide array of tokamaks.

\section*{Acknowledgements}
This work has been carried out within the framework of the EUROfusion Consortium, funded by the European Union via the Euratom Research and Training Programme (Grant Agreement No 101052200 — EUROfusion). Views and opinions expressed are however those of the author(s) only and do not necessarily reflect those of the European Union or the European Commission. Neither the European Union nor the European Commission can be held responsible for them. 
This work made use of the Dutch national e-infrastructure with the support of the SURF Cooperative using grant no. EINF-3557.

\section*{References}
\bibliographystyle{plainurl_abrev}
\bibliography{main}

\clearpage
\appendix
\onecolumn
\setcounter{footnote}{1}
\counterwithin{figure}{section}
\counterwithin{lstlisting}{section}

\section{Architecture Figures}\label{ap:arch_figs}
\begin{figure}[H]
    \centering
    \includegraphics[width=0.9\linewidth]{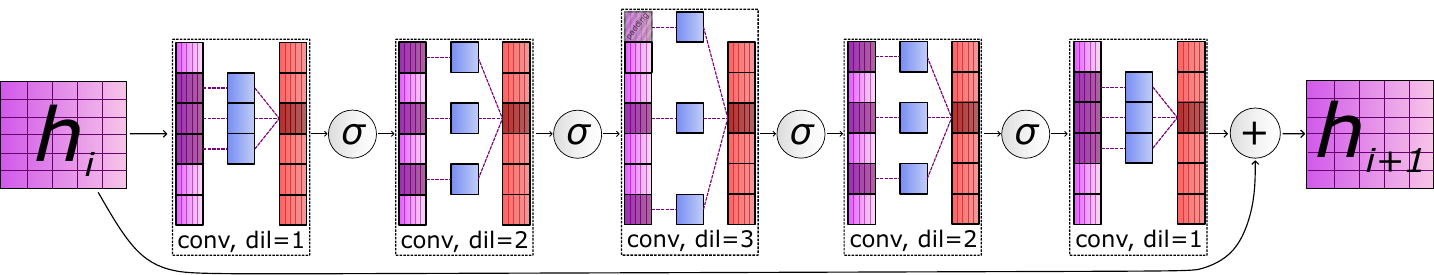}%
    \caption{Dilated Residual Network~\cite{stachenfeld2022}: The \textbf{DRN} applies a sequence of dilated convolutions with varying dilation rates.}
    \label{fig:drn}
\end{figure}

\begin{figure}[H]
    \centering
    \includegraphics[width=0.9\linewidth]{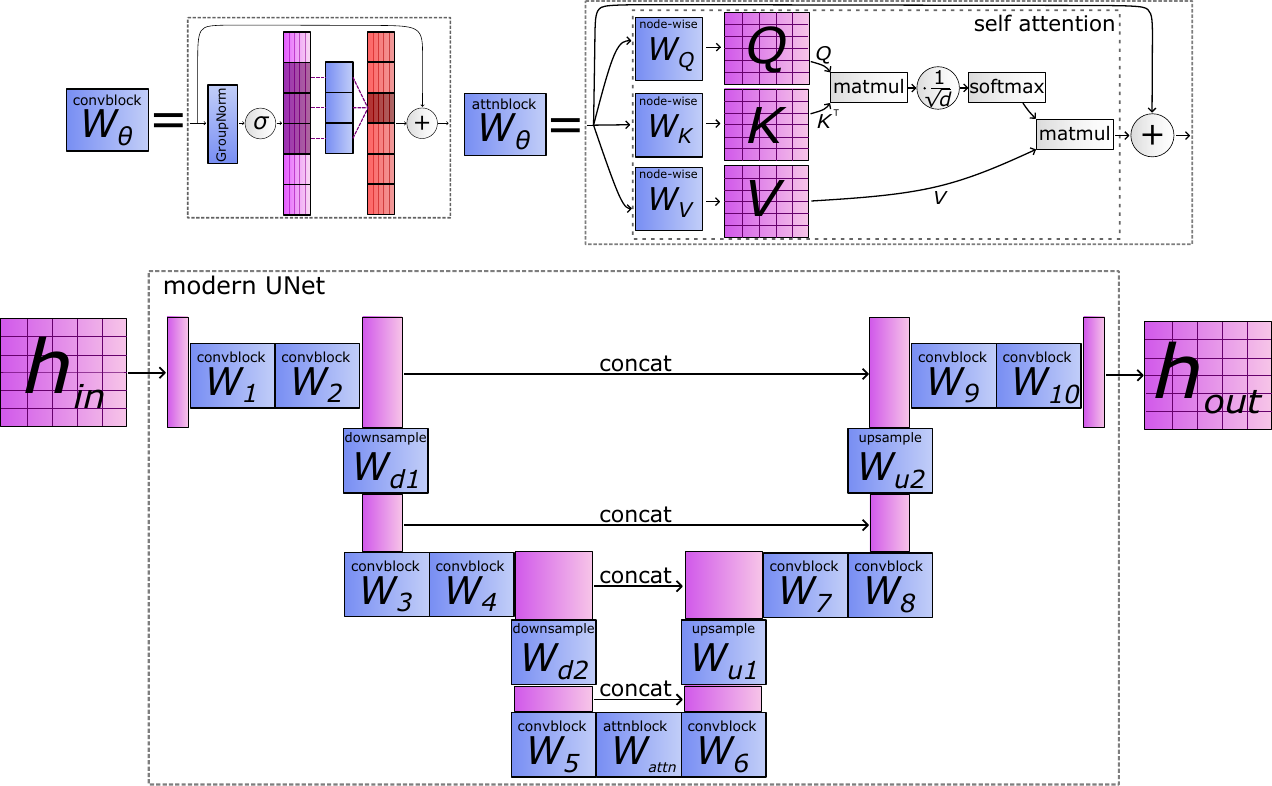}%
    \caption{Modern UNet~\cite{gupta2022}: The \textbf{UNet} applies convolutions on multiple spatial scales, with connections on the same scales.}
    \label{fig:unet}
\end{figure}

\begin{figure}[H]
    \centering
    \includegraphics[width=0.9\linewidth]{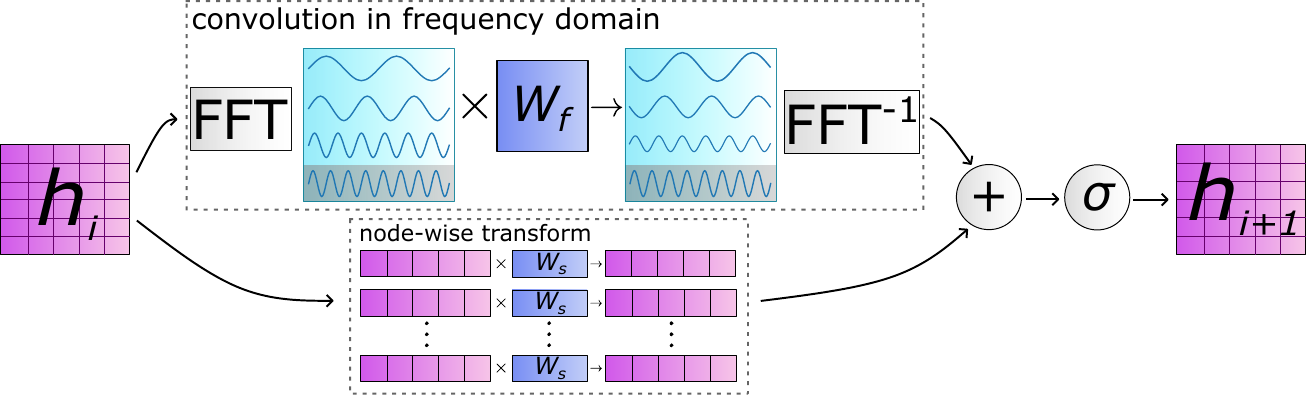}%
    \caption{Fourier Neural Operator~\cite{li2021}: The \textbf{FNO} parametrizes a convolution operator in Fourier space.}
    \label{fig:fno}
\end{figure}

\begin{figure}[H]
    \centering
    \includegraphics[width=0.9\linewidth]{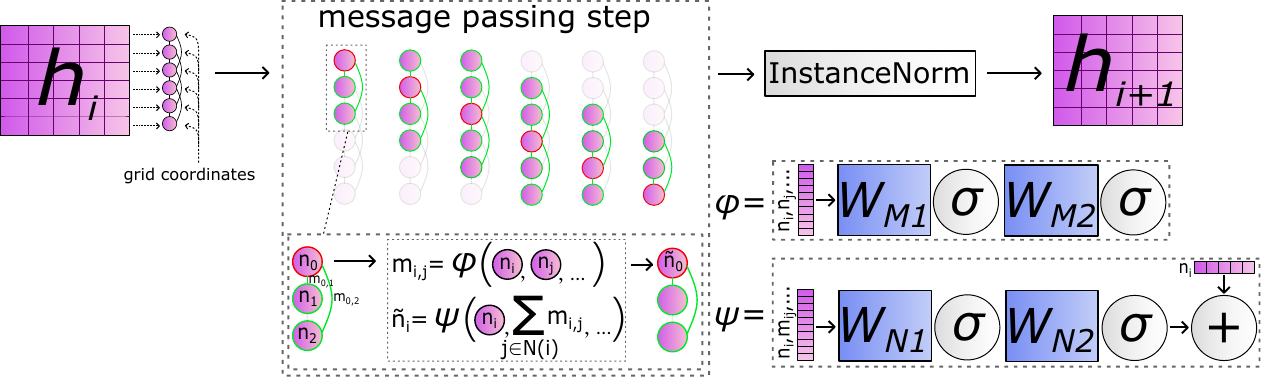}%
    \caption{Message Passing PDE Solver~\cite{brandstetter2022}: The \textbf{MP-PDE} solver updates a point's representation using information from itself and its neighbors.}
    \label{fig:gnn}
\end{figure}

\begin{figure}[H]
    \centering
    \includegraphics[width=0.9\linewidth]{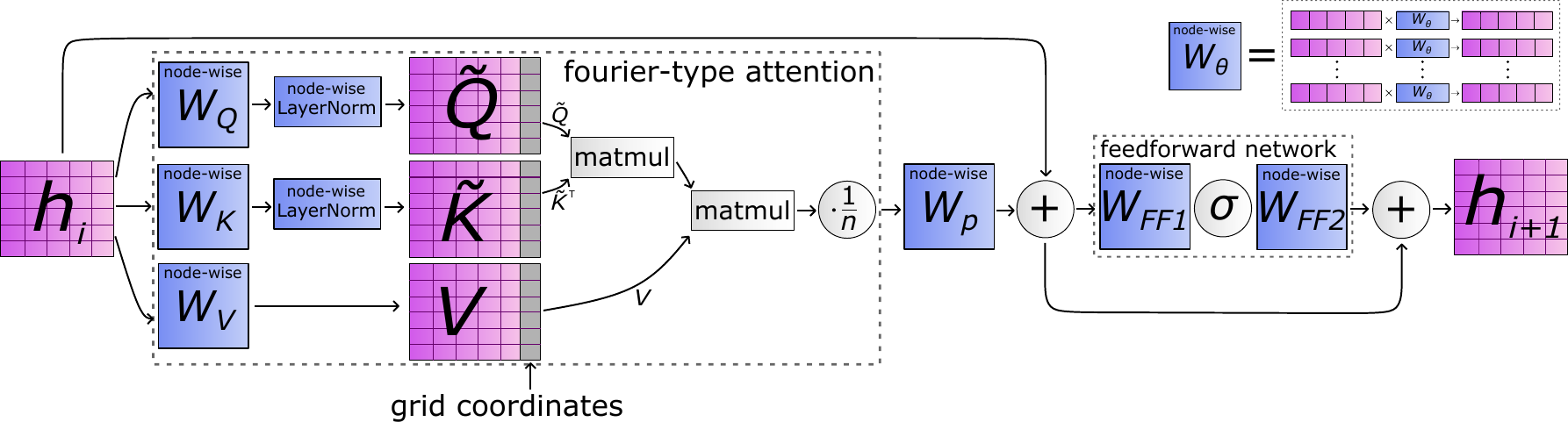}%
    \caption{Transformer with Fourier-type Attention~\cite{cao2021}: The \textbf{FT} updates a point's representation with the product of values $\mathbf{V}$ and attention score $\widetilde{\mathbf{Q}}\widetilde{\mathbf{K}}^\top$.}
    \label{fig:ft}
\end{figure}

This appendix contains more detailed illustrations of the architectures described in Section~\ref{ss:modelarch}. Figures~\ref{fig:drn}-\ref{fig:ft} contain illustrations for DRN~\cite{stachenfeld2022}, UNet~\cite{gupta2022}, FNO~\cite{li2021}, MP-PDE~\cite{brandstetter2022} and FT~\cite{cao2021}, respectively. All but the UNet illustrate a single hidden block that is repeated, the UNet illustration depicts the entire architecture.


\section{Time-Adjusted Batch Sampling}\label{ap:probs}
In this appendix we elaborate on the time-adjusted batch sampling strategy as introduced in Subsection~\ref{ss:model_div1d} and provide evaluations of its utility. The best settings are used to train the set of models found in Section~\ref{sec:results}.

For training, we sample batches of datapoints by first sampling the unrolling time $t$ from $p(t)$, and subsequently sample batch items $x$ with adjusted probabilities given this unrolling time from $p(x|t)$. The goal is to both adhere to the unrolling time distribution and to sample simulations uniformly, see also Equation~\ref{eq:pxsampling}.

In short, we compute $p(x|t)$ using intervals of simulation time windows \textit{Nt}, by placing simulations into time-based bins and adjusting sample probabilities based on the probability of sampling in an interval given $p(t)$. We start at the bin with the shortest simulations, and simply assign these shortest simulations the adjusted probability %
such that the marginalized probability is $\frac{1}{\text{N}}$. These items are no longer considered as they cannot be used when unrolling more steps in `longer' bins. The process is repeated with the now-shortest bin and its simulations while adjusting for the previously set probabilities, which is again repeated until all items are parsed.

The method for computing $p(x|t)$ is described in Pseudocode~\ref{pc:binprobs} in more detail. Distribution $p(x|t)$ is computed using a list of simulation times \var{times}, the block length \var{w} and the cumulative distribution function (CDF) of $p(t)$. Items are first placed in bins (line 2). These bins are computed by the number of unrollings one can do, i.e., by rounding the $Nt$ of each simulation down to a multiple of block length $w$. Each bin is defined by the items falling in this bin (\var{bins}$_i$.\var{items}) and the number of timesteps $t$ it can be unrolled (\var{bins}$_i$.\var{time}, a multiple of $w$), and is sorted on ascending time. Lines 6-15 describe the main procedure of assigning probabilities to items in bins. We start at the smallest bin \var{bins}$_1$, as items in this bin can only be used when unrolling for at most \var{bins}$_1$.\var{time}. Item probabilities are computed in line 8: Given the unassigned probability mass that is available to this item (\var{bin\_prob} -- from this bin, and \var{res\_prob} -- from smaller bins, which is 0 for the first iteration), the expectation of sampling it will be $\frac{1}{N}$. This procedure is repeated for all bins up until the last one. For the last bin we simply assign the remaining probability mass (line 16). Finally, in lines 17-21, the output object is post-processed to distribute the residual probability mass up to chain. We return the bins and probability assignments within the bin, defining $p(x|t)$.

\newcommand\commfont[1]{\scriptsize\ttfamily\textcolor{blue}{#1}}
\SetCommentSty{commfont}%

\begin{algorithm*}[h]
\DontPrintSemicolon
\SetKwFunction{FMain}{compute\_probabilities}
\SetKwProg{Fn}{Function}{:}{}
\Fn{\FMain{\var{times}, \var{w}, CDF}}{
$\var{bins} \gets$ $sort(create\_bins$(\var{times}, \var{w})) \tcp*{Binned by time $Nt$, with variables $\var{bins}_i.\var{time}$ and $\var{bins}_i.\var{items}$}
$N \gets$ $|\var{bins}|$\;
$\var{expected} \gets \frac{1}{|\var{times}|}$\tcp*{All simulations should be sampled uniformly}
$\var{res\_prob} \gets 0$\tcp*{Probability mass that is unassigned in already-parsed bins}
\For{$i \gets 1$ \KwTo $N-1$}{
$\var{bin\_prob} \gets CDF(\var{bins}_i.\var{time}) -  CDF(\var{bins}_{i-1}.\var{time})$\tcp*{Probability $t$ falls in this bin}
$\var{item\_prob} \gets \frac{\var{expected}}{\var{bin\_prob} + \var{res\_prob}}$\tcp*{Probability needed from this bin + previous residuals}
$\var{bin\_items} \gets$ $\{j: \var{item\_prob}\ |\  j \in \var{bins}_i.\var{items} \}$ \tcp*{Set probs for bin items}
$\var{other\_prob} \gets 1 - \sum\var{bin\_items}$\tcp*{Get bin residual}
\If(\tcp*[f]{Verify a valid assignment is possible}){$\var{other\_prob} < 0$}{
\textbf{error}: Bin residual $<$ 0, no valid assignment possible\;
}
$\var{bin\_items}[``\text{other}"] \gets \var{other\_prob}$\tcp*{Store bin residual}
$\var{bins}_i\var{.probs} \gets \var{bin\_items}$\tcp*{Save bin probabilities}
\tcp{Update \var{res\_prob}: Add residual from \var{bins}$_i$, remove residual consumed by \var{bins}$_i$.\var{items}}
$\var{res\_prob} \gets \var{res\_prob} + \var{bin\_prob} \cdot \var{other\_prob} - \var{res\_prob} \cdot (1 - \var{other\_prob})$\;
}
$\var{bins}_N\var{.probs} \gets \{j : \frac{1}{|\var{bins}_{N}.\var{items}|} |\ j \in \var{bins}_{N}.\var{items} \}$)\tcp*{Probabilities for final bin (no residual)}
\For(\tcp*[f]{Distribute residuals up the chain}){$i \gets N-1$ \KwTo$1$}{
$idx \gets$ $\cup_{j=i+1}^N \var{bins}_j.\var{items}$\tcp*{Indices of items in longer bins}
\ForEach{$j \in idx$}{
$\var{bins}_i.\var{probs}[j] \gets \var{bins}_i.\var{probs}[j] + \var{bins}_{i+1}.\var{probs}[j] \cdot \var{bins}_i.\var{probs}[``\text{other}"]$\;
}
\textbf{delete} $\var{bins}_i.\var{probs}[``\text{other}"]$\;
}
\KwRet \var{bins}\tcp*{Variables $\var{bins}_i.\var{time}$, $\var{bins}_i.\var{items}$ and $\var{bins}_i.\var{probs}$ (probablity mass assignment of $p(x|t)$)}
}
\caption{Compute $p(x|t)$ Algorithm}\label{pc:binprobs}
\end{algorithm*}

\begin{figure}[h]
    \centering
    \includegraphics[width=.4\linewidth]{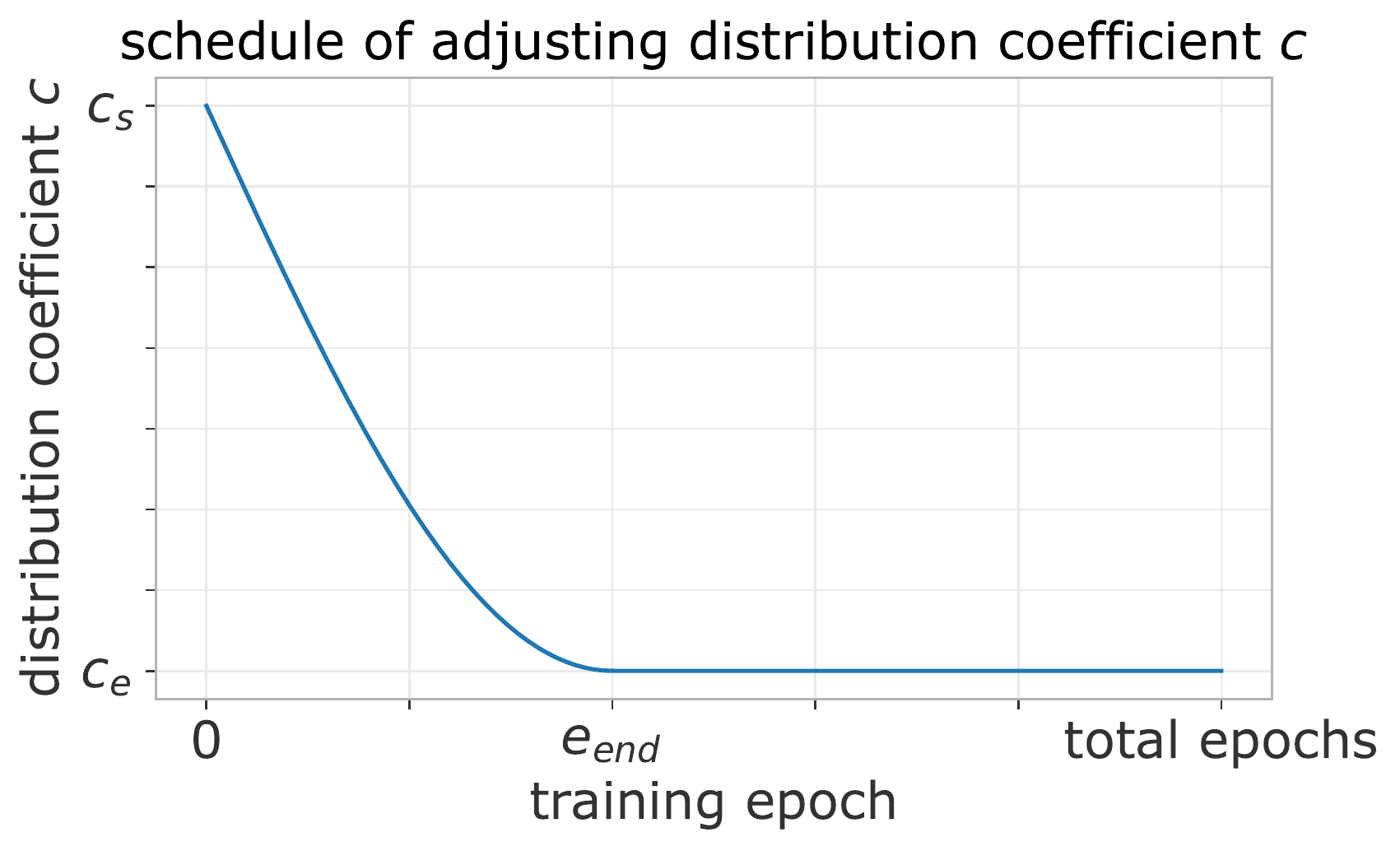}%
    \caption{Schedule of coefficient $c$ during training. $c$ is decayed from $c_s$ to $c_e$ through the bottom half of a sine wave until epoch $e_{end}$ (here at 40\% of training), after which it is kept fixed at $c_e$.}
    \label{fig:samplingschedule}
\end{figure}

To evaluate this approach we try different unrolling distributions $p(t)$ and do a comparison with other padding-free baselines (that also do not waste the majority of computation time, see Subsection~\ref{ss:model_div1d}). As probability mass function (PMF) of $p(t)$ we try the geometric distribution and a simple linear function. At the start of training, distribution parameters are chosen to make the PMFs steep, i.e., putting almost all weight towards short unrolling times. They are then made more shallow over time: As training goes on, we do longer and longer unrollings. We use the same schedule for adjusting the distribution coefficient, denoted as $c$, and try two coefficients per distribution family. A depiction of these distributions is provided in Figure~\ref{fig:distrillustrations}, with the coefficient schedule in Figure~\ref{fig:samplingschedule}.

\renewcommand{\rulesep}{\hfill}
\begin{figure*}[h]
    \centering
    \begin{subfigure}[b]{0.04\textwidth}
    \raisebox{1.87cm}{\rotatebox{90}{\textbf{Geometric}}}
    \end{subfigure}
    \begin{subfigure}[b]{0.31\textwidth}
    \begin{center}\includegraphics[width=\textwidth]{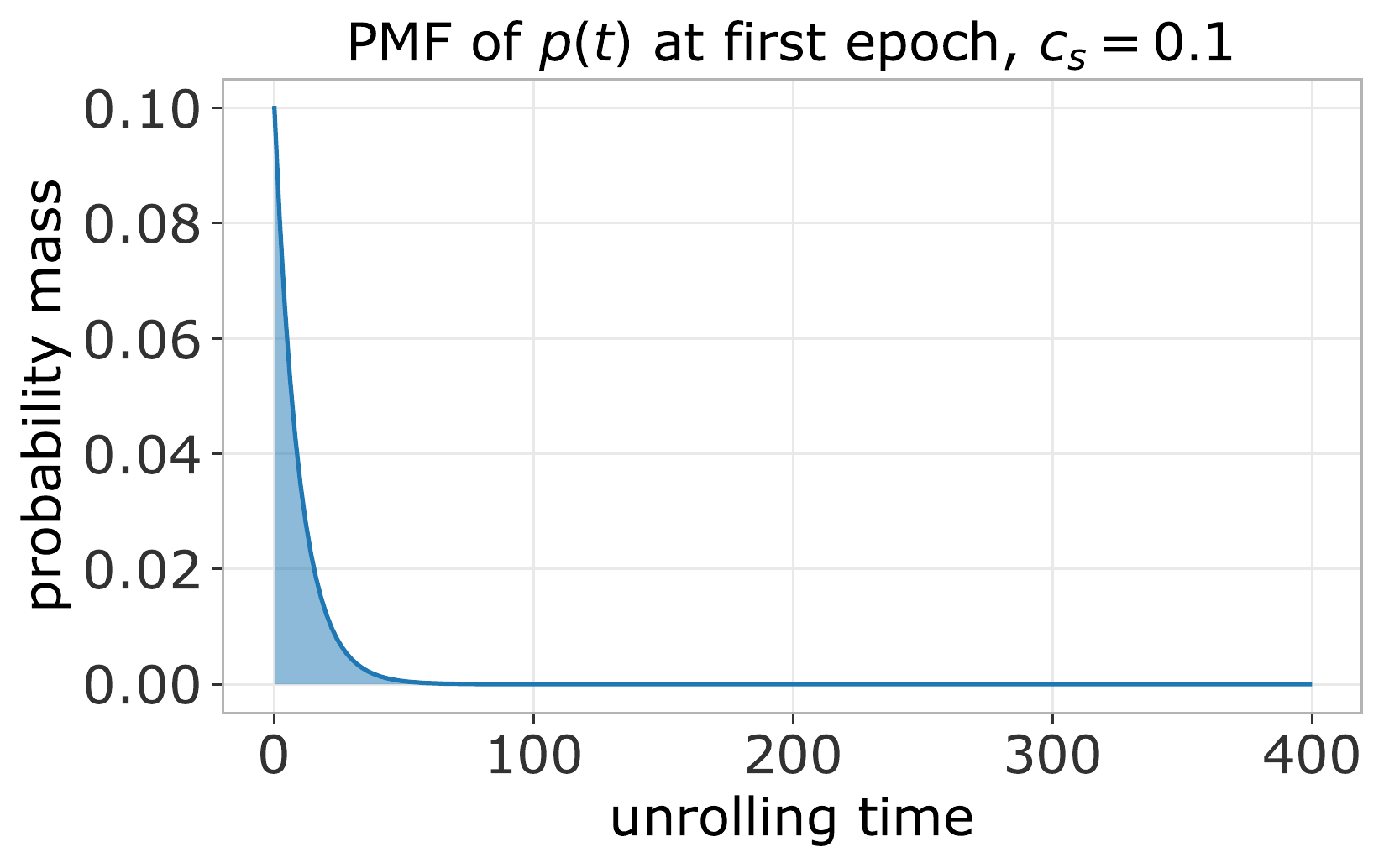}\end{center}
    \caption{At start epoch, $c_s=0.1$}{\label{fig:geom0}}
    \end{subfigure}
    \rulesep
    \begin{subfigure}[b]{0.31\textwidth}
    \begin{center}\includegraphics[width=\textwidth]{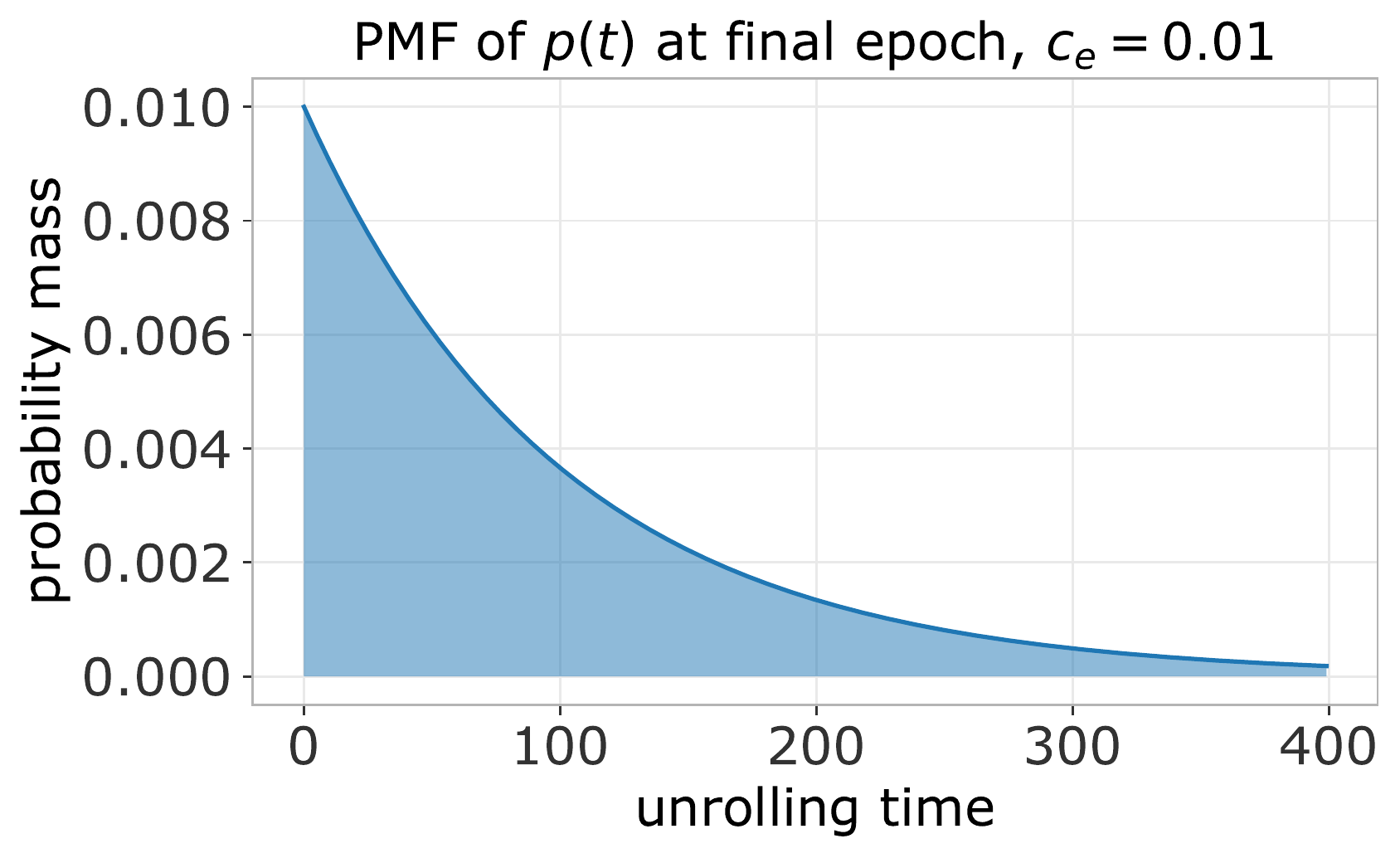}\end{center}
    \caption{At final epoch, $c_e=0.01$}{\label{fig:geom1a}}
    \end{subfigure}
    \rulesep
   \begin{subfigure}[b]{0.31\textwidth}
    \begin{center}\includegraphics[width=\textwidth]{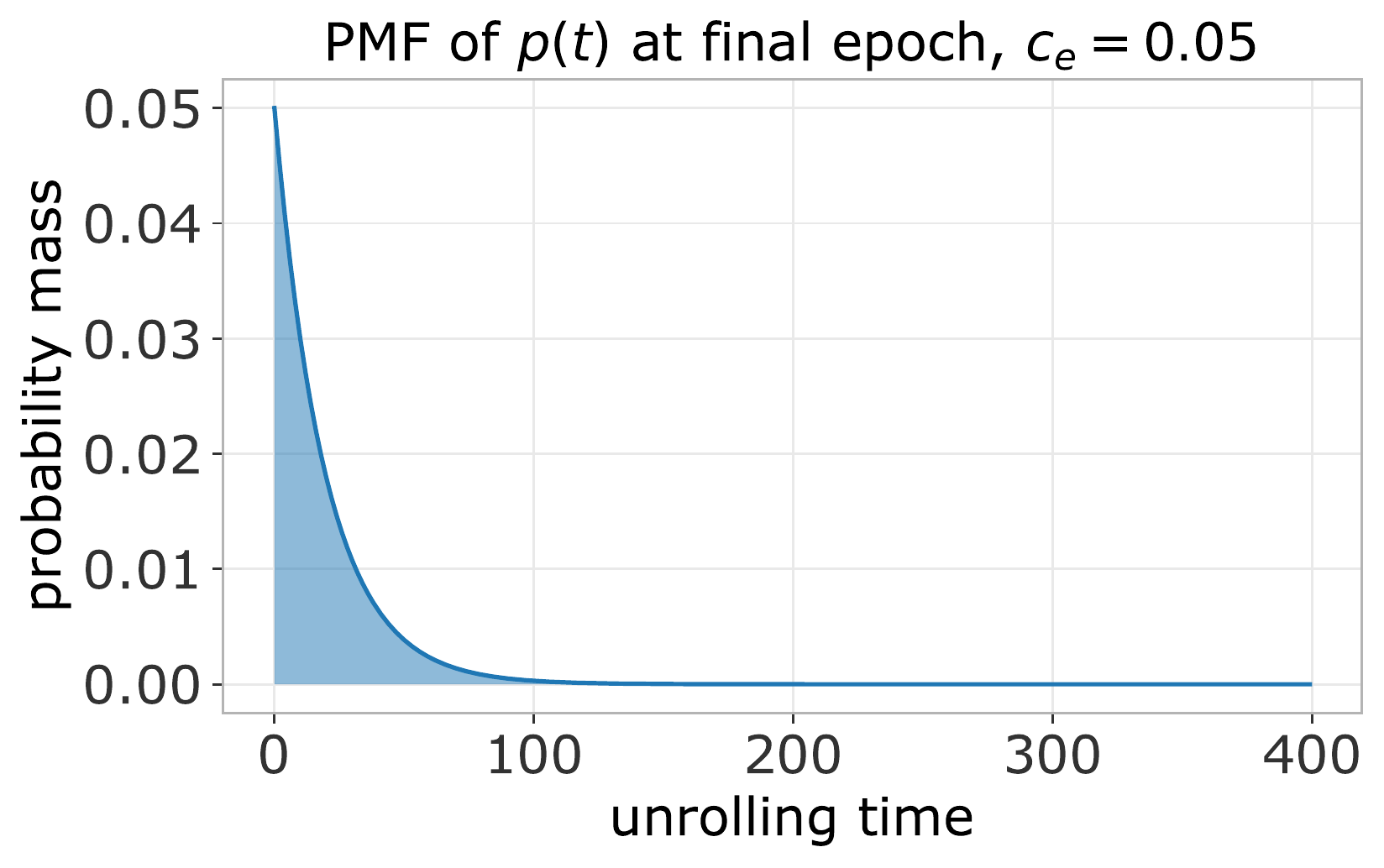}\end{center}
    \caption{At final epoch, $c_e=0.01$}{\label{fig:geom1b}}
    \end{subfigure} \\
    \begin{subfigure}[b]{0.04\textwidth}
    \raisebox{2.18cm}{\rotatebox{90}{\textbf{Linear}}}
    \end{subfigure}
    \begin{subfigure}[b]{0.31\textwidth}
    \begin{center}\includegraphics[width=\textwidth]{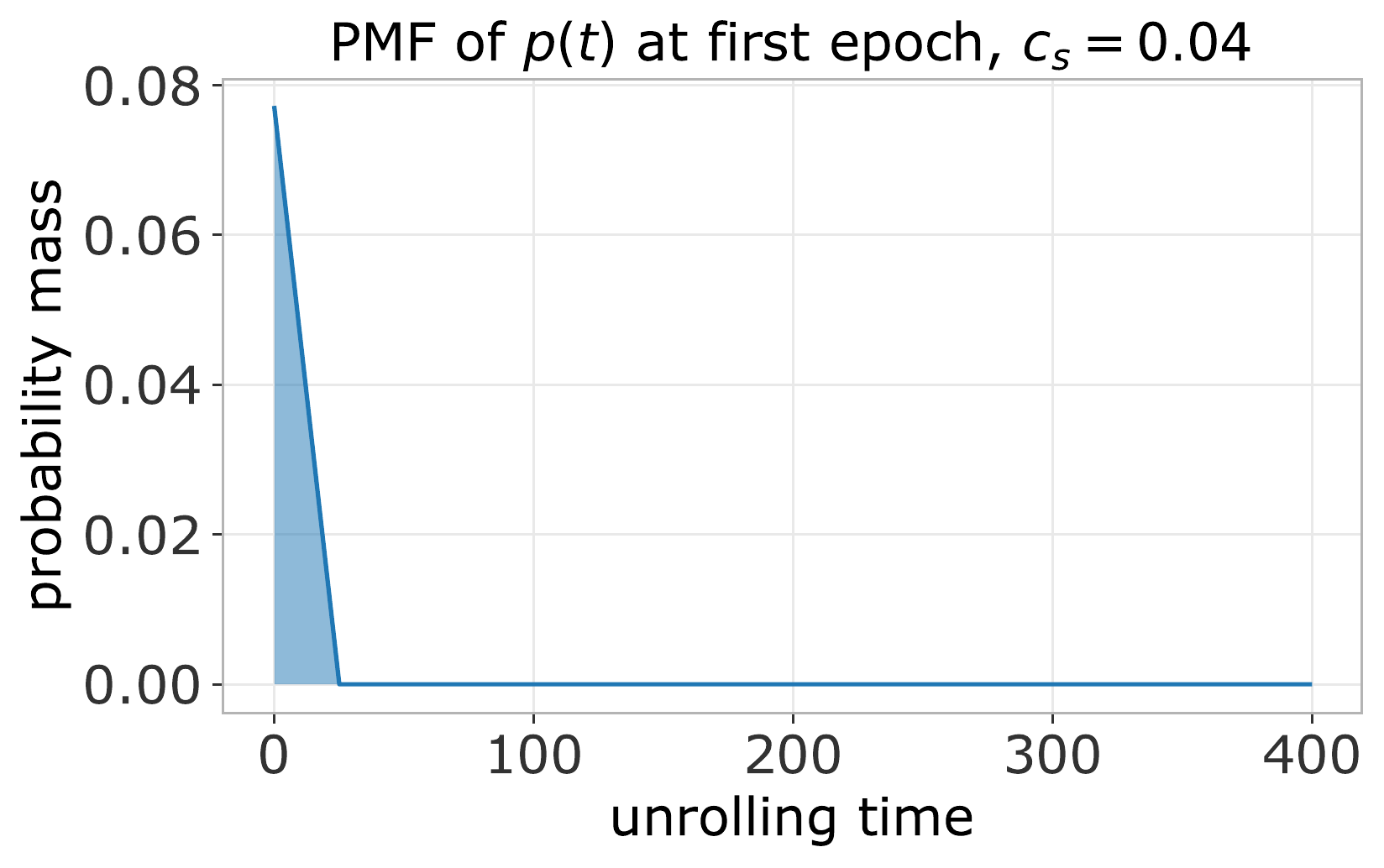}\end{center}
    \caption{At start epoch, $c_s=0.04$}{\label{fig:linear0}}
    \end{subfigure}
    \rulesep
    \begin{subfigure}[b]{0.31\textwidth}
    \begin{center}\includegraphics[width=\textwidth]{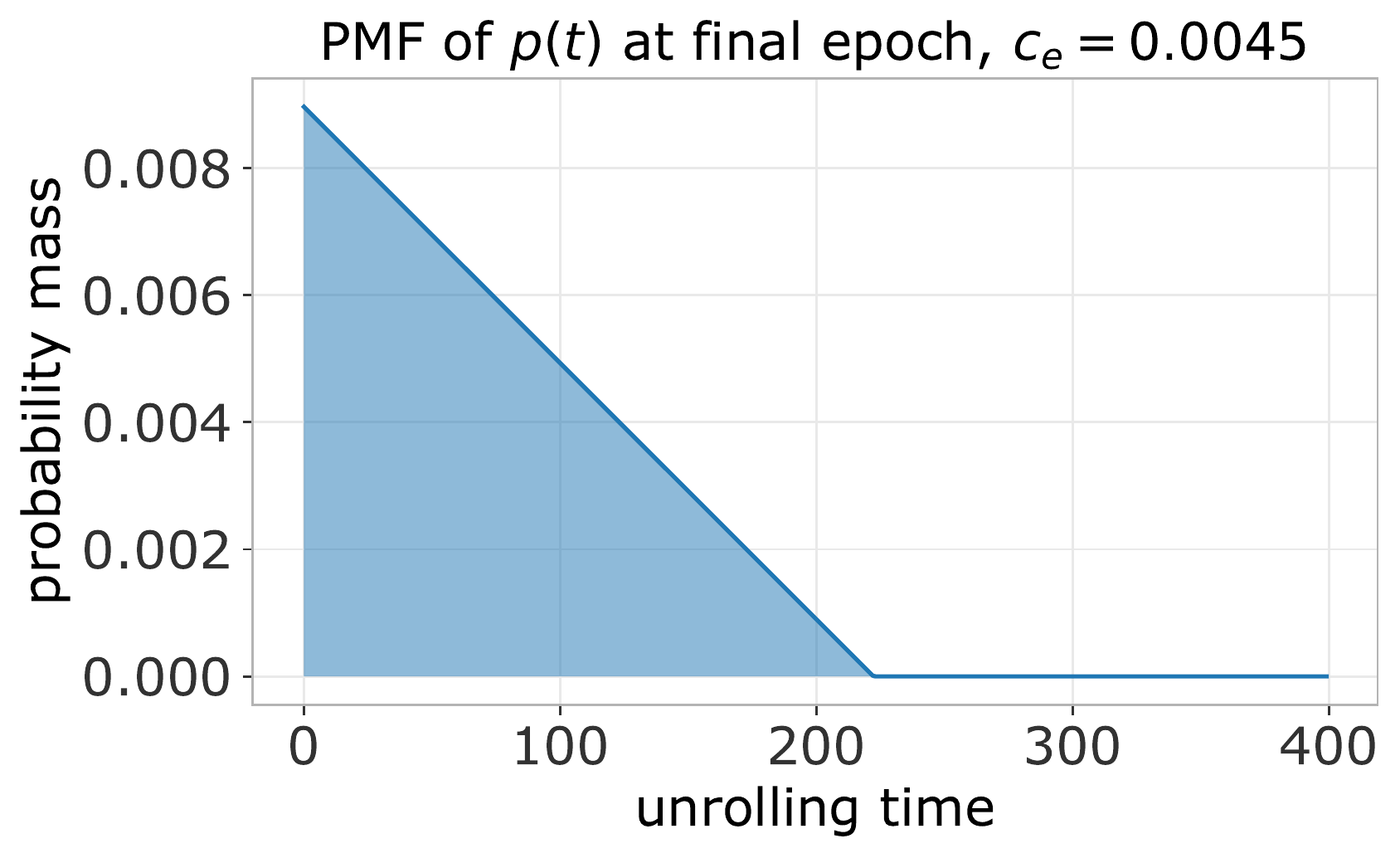}\end{center}
    \caption{At final epoch, $c_e=0.0045$}{\label{fig:linear1a}}
    \end{subfigure}
    \rulesep
   \begin{subfigure}[b]{0.31\textwidth}
    \begin{center}\includegraphics[width=\textwidth]{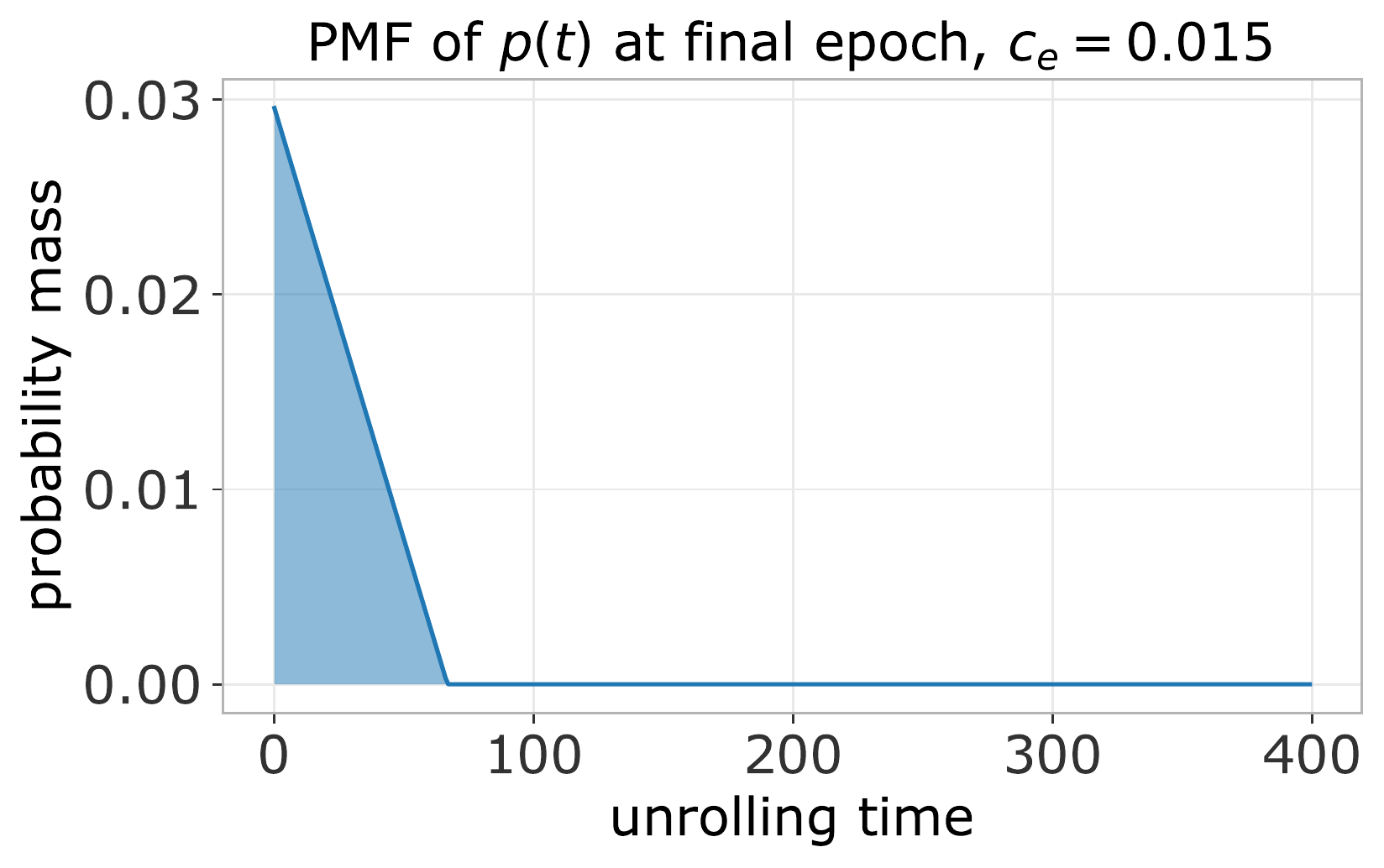}\end{center}
    \caption{At final epoch, $c_e=0.0015$}{\label{fig:linear1b}}
    \end{subfigure} \\
    \caption{PMFs we evaluate for sampling unrolling time $t \sim p(t)$: The geometric distribution and a linear function. Coefficient $c$ is decayed from $c_s$ to $c_e$, we try two end coefficients per distribution.}%
    \label{fig:distrillustrations}%
\end{figure*}

As baseline we consider two methods for sampling batches: (1) Uniformly sampling simulations in the minibatch and then sampling an unrolling time that fits given the batch items, and (2) sampling an unrolling time and then uniformly sampling simulations that fit this length. The downsides are that for (1) we have no guarantees on the sampling time distribution and for (2) we have no guarantees on sampling simulations uniformly. For both methods we uniformly sample unrolling times, either up until 160 timesteps or up until 400. This sampling window starts at 0 and is increased as training goes on, at 20 timesteps for each 5\% of training epochs (160 timesteps being reached at 40\%, 400 at the end of training).

All methods are tested by sampling batches in an epoch either with replacement or with no replacement. The latter setting, denoted with \textbf{-NR}, ensures we sample each item once per epoch, enforcing the desired expectation of $\frac{1}{N}$. However, this setting can bias the eligible unrolling windows to shorter windows.

The methods are evaluated before the final experiments (Section~\ref{sec:results}) in order to pick a setting that we choose for all models. We investigate using the density ramp dataset and use reference settings that we found to work reasonably well: Train for \num{10000} epochs with batch size 16 (where an epoch is one pass over each simulation, for a single random start point and its unrolling window). The model consists of an element-wise encoder, a Dilated Residual Network as processor (8 blocks; 512 hidden features; kernel size 5), and a time-wise convolution as decoder. We compute solutions in blocks of $w=20$ timesteps ($\text{\SI{2}{\milli\second}}$ real time).

The results are reported in Table~\ref{tab:samplingresults}. We choose the settings with the best results on the validation set for further experiments, the test set results are reported only for completeness. The adjusted sampling methods generally result in better models, with a `wide' linear distribution for the unrolling distribution $p(t)$ working best (AdjSampling-Linear$_{0.0045}$, Figure~\ref{fig:linear1a}).

\begin{table}[H]
\centering
\begin{tabular}{lcc}
\toprule
Sampling Strategy & Validation &Test \\
\bottomrule
AdjSampling-Geometric$_{0.01}$ & \cellcolor[RGB]{185,227,178} 0.00137 & \cellcolor[RGB]{184,226,177} 0.00291\\
AdjSampling-Geometric$_{0.01}$-NR & \cellcolor[RGB]{184,226,177} 0.00121 & \cellcolor[RGB]{187,228,181} 0.00336\\
AdjSampling-Geometric$_{0.05}$ & \cellcolor[RGB]{242,250,240} 0.00941 & \cellcolor[RGB]{238,248,235} 0.01026\\
AdjSampling-Geometric$_{0.05}$-NR & \cellcolor[RGB]{189,228,182} 0.00174 & \cellcolor[RGB]{187,228,181} 0.00336\\
\hline
AdjSampling-Linear$_{0.0045}$ & \cellcolor[RGB]{179,224,173} \textbf{0.00084} & \cellcolor[RGB]{179,224,173} \textbf{0.00254}\\
AdjSampling-Linear$_{0.0045}$-NR & \cellcolor[RGB]{199,233,192} 0.00280 & \cellcolor[RGB]{198,232,191} 0.00443\\
AdjSampling-Linear$_{0.015}$ & \cellcolor[RGB]{236,247,232} 0.00798 & \cellcolor[RGB]{255,255,255} 0.02169\\
AdjSampling-Linear$_{0.015}$-NR & \cellcolor[RGB]{255,255,255} 0.01938 & \cellcolor[RGB]{223,242,217} 0.00829\\
\hline
SimFirst-Unrolling$_{160}$ & \cellcolor[RGB]{241,249,238} 0.00899 & \cellcolor[RGB]{252,254,251} 0.01536\\
SimFirst-Unrolling$_{160}$-NR & \cellcolor[RGB]{255,255,255} 0.07273 & \cellcolor[RGB]{255,255,255} 0.55632\\
SimFirst-Unrolling$_{400}$ & \cellcolor[RGB]{214,239,208} 0.00504 & \cellcolor[RGB]{249,252,248} 0.01351\\
SimFirst-Unrolling$_{400}$-NR & \cellcolor[RGB]{236,247,232} 0.00786 & \cellcolor[RGB]{228,244,223} 0.00874\\
\hline
TimeFirst-Unrolling$_{160}$ & \cellcolor[RGB]{221,241,215} 0.00627 & \cellcolor[RGB]{254,254,254} 0.01761\\
TimeFirst-Unrolling$_{160}$-NR & \cellcolor[RGB]{255,255,255} 0.09000 & \cellcolor[RGB]{255,255,255} 0.01975\\
TimeFirst-Unrolling$_{400}$ & \cellcolor[RGB]{211,238,205} 0.00472 & \cellcolor[RGB]{255,255,255} 0.02893\\
TimeFirst-Unrolling$_{400}$-NR & \cellcolor[RGB]{254,254,254} 0.01572 & \cellcolor[RGB]{217,240,211} 0.00736\\
\toprule
\\
\end{tabular}
\caption{MSE on standardized data, full simulation rollouts on the validation and test set. Cells are colored by the error, with green being better. The best scores are marked in bold.}
\label{tab:samplingresults}
\end{table}

\section{Tables: Density Ramp Results}\label{ap:results}
In this appendix we provide full results for the evaluated neural PDE surrogate architectures for the density ramp data. All architectures follow the encode-process-decode structure, where $\textit{encoder} \colon \mathbf{ub}^{i-w:i} \in \mathbb{R}^{\textit{Nx}\times w \times 4} \to h_{in} \in \mathbb{R}^{\textit{Nx} \times D}$ maps the input block to a hidden representation, $\textit{processor} \colon h_{in} \in \mathbb{R}^{\textit{Nx} \times D} \to h_{out} \in \mathbb{R}^{\textit{Nx} \times D}$ transforms this hidden representation, and $\textit{decoder} \colon h_{out} \in \mathbb{R}^{\textit{Nx} \times D} \to {\widetilde{\mathbf{u}}\mathbf{b}^{i:i+w} \in \mathbb{R}^{\textit{Nx} \times w \times 4}}$ maps this representation back to the next block, all conditioned by~$(\mathbf{b}_{\text{s}}, \mathbf{b}_{\text{d}}^{\mathbf{t}_{i:i+w}}, \mathbf{c})$; see also Sections~\ref{ss:modelarch} and~\ref{ss:model_div1d}. 

Architectures are found by iteratively reducing the size of an architecture of `maximum width' and `maximum depth', where the former emphasises using many hidden dimensions, and the latter emphasises using many layers. For each architecture a set of hyperparameters is iteratively updated (up to some minimum values) until an architecture is found with an inference time below $\{0.5, 1, 2, 4, 6, 8, 10\}\text{\SI{}{\milli\second}}$\footnote{Note that not all inference times are reached for all models, due to the smallest architecture still exceeding the smallest desired inference times in most cases.}, computed as a forward pass of batch size 1 on an NVIDIA A100 40GB GPU with an Intel Xeon Platinum 8360Y CPU. The tested hyperparameters and other settings are as follows.

For all architectures the number of hidden dimensions is varied, which is denoted with $\mathit{D}$. All architectures besides the UNet consist of repeating layers (with identical architectures but non-shared parameters), where the number of layers is denoted with $\mathit{L}$. For all models, three combinations of an encoder/decoder/dynamic BC encoder are considered, denoted with \textbf{EncDec}. These are as follows:
\begin{itemize}
    \item \textbf{$\textrm{EncDec}_{\textrm{B}1}$.} \textit{Encoder}: Point-wise non-linear map to $\mathit{D}$ dimensions as in~\cite{brandstetter2022}, with Swish activations~\cite{ramachandran2018} and 1 hidden layer. \textit{Decoder}: Feature-wise non-linear convolution mapping back to a time-block of solutions as in~\cite{brandstetter2022}, with Swish activations and 1 hidden layer. \textit{BC encoder}: Non-linear convolution over the timesteps in the block, flattened and mapped to an embedding of 8 dimensions, using 1 hidden layer with GELU activations~\cite{hendrycks2016}.
    \item \textbf{$\textrm{EncDec}_{\textrm{B}2}$.} \textit{Encoder}: Point-wise non-linear map to $\mathit{D}$ dimensions with GELU activations and 1 hidden layer. \textit{Decoder}: Point-wise mapping from $\mathit{D} \to 12 \times w$ features, followed by a feature-wise non-linear convolution as in $\textrm{EncDec}_{\textrm{B}1}$, but with GELU activations instead. \textit{BC encoder}: Same as $\textrm{EncDec}_{\textrm{B}1}$.
    \item \textbf{$\textrm{EncDec}_{\textrm{S}}$.} \textit{Encoder}: Grid-wise linear convolution with kernel size 3. \textit{Decoder}: Grid-wise linear convolution with kernel size 3. \textit{BC encoder}: Non-linear convolution over the timesteps in the block, flattened and mapped to an embedding of 4 dimensions, with no hidden layers and GELU activations.
\end{itemize}
\textbf{B1} is used for all model sizes, whereas \textbf{B2} is used to re-evaluate the six best configurations of each model. We evaluate \textbf{B2} because convolving directly over hidden features in the decoder as done in \textbf{B1} can lead to unintentional sparsity when $\mathit{D} \gg w$ since the kernel will not slide over each hidden feature; in \textbf{B2} a learned downsampling is added. In practice we did not find this artifact to have a significant impact in the majority of configurations. \textbf{S} is used in the smallest configurations to avoid the encoder/decoder components being the bottleneck for the inference cost.

In all processor architectures GELU activations~\cite{hendrycks2016} are used where applicable. The hyperparameters for each specific architecture are as follows:
\begin{itemize}
\item \textbf{DRN.} We use convolutional layers with kernel size 5 and the dilation pattern from~\cite{stachenfeld2022}: Dilation rates of (1,~2,~4,~8,~4,~2,~1). For the smallest models (with $\textrm{EncDec}_{\textrm{S}}$) we use a kernel size of 3. We vary the number of blocks $\mathit{L}$ and the number of hidden dimensions $\mathit{D}$.
\item \textbf{UNet.} We use the modern UNet implementation from~\cite{gupta2022}. Each residual block consists of two convolutional layers with kernel size 3, a shortcut connection and group normalization~\cite{wu2019}. At the most subsampled layer, self attention~\cite{vaswani2017} is used. We vary the number of hidden dimensions $\mathit{D}$, alongside the number of downsampling (and corresponding upsampling) steps, denoted with $\mathit{Depth}$. The number of hidden dimensions $\mathit{D}$ is fixed throughout the network.
\item \textbf{FNO.} We use the implementation from~\cite{li2021}, and vary the number of blocks $\mathit{L}$, the number of hidden dimensions $\mathit{D}$, and the number of fourier modes $\mathit{M}$.
\item \textbf{MP-PDE.} We use the implementation from~\cite{brandstetter2022}, and connect each node to its 4 nearest neighbors (corresponding to a convolution of kernel size 5). We vary the number of blocks $\mathit{L}$ and the number of hidden dimensions $\mathit{D}$.
\item \textbf{FT.} We use the `transformer with fourier-type attention' implementation from~\cite{cao2021} and repeat a set of transformer encoder blocks. We vary the number of blocks $\mathit{L}$, the number of hidden dimensions $\mathit{D}$, the number of dimensions in the MLP following self attention denoted as $\mathit{MLP}$, and the number of attention heads $\mathit{NH}$\footnote{Since attention heads only split the total number of hidden dimensions in our case, the effect on inference time of varying $\mathit{NH}$ is rather small; we scale it as larger models often also use more attention heads.}. 
\item \textbf{FT-FNO.} We use a series of $\mathit{L}_{\mathit{FT}}$ FT layers followed by a series of $\mathit{L}_{\mathit{FNO}}$ FNO layers, with the remaining applicable hyperparameters as described in the corresponding methods (dimensions $\mathit{D}$; fully connected size $\mathit{MLP}$; fourier modes $\mathit{M}$).
\end{itemize}

In Table~\ref{tab:fullresult} the results for all models on the density ramp data are given. The table denotes the model configuration, the MSE on full test simulations (Figure~\ref{fig:wpall}), the MSE on the first 500 steps (Figure~\ref{fig:wp500}), the MSE on single-block predictions (Figure~\ref{fig:wpss}), and the compute cost (in milliseconds) of 1 block of $\text{\SI{2}{\milli\second}}$ on the GPU (NVIDIA A100 40GB; Figure~\ref{fig:wpdiv1dgpu}) and the CPU (AMD Rome 7H12, 1 core; Figure~\ref{fig:wpdiv1dcpu}). As a comparison, the DIV1D results when scaling the internal grid are provided in Table~\ref{tab:div1dresult}.

\setlength\tabcolsep{5.3pt}
\begingroup
\centering
\begin{longtable}{clccccc}
& & \multicolumn{3}{c}{Error metrics} & \multicolumn{2}{c}{Compute Time ($\text{\SI{}{\milli\second}}$) (per $\text{\SI{2}{\milli\second}}$)} \\
\cmidrule[\heavyrulewidth]{3-7}
& Model$_\textrm{configuration}$ & MSE & MSE$_{t < 500}$ & MSE$_\textrm{single}$ & \hphantom{xx.} GPU \hphantom{xx.} & \hphantom{x} CPU$_{\textrm{1-core}}$ \hphantom{x}\\
\cmidrule[\heavyrulewidth]{2-7}
\addlinespace[-\belowrulesep]
\multirow{3}{*}{\rotatebox{90}{depth}} \hspace{-0.3cm} & DRN$_{\textrm{L}:8,\textrm{D}:832,\textrm{EncDec}_{\textrm{B}1}}$ & \cellcolor[RGB]{183,226,176} 0.00110 & \cellcolor[RGB]{183,226,176} 0.00078 & \cellcolor[RGB]{186,227,180} 0.00031 & \cellcolor[RGB]{252,172,144} 9.23771 & \cellcolor[RGB]{252,166,137} 521.604 \\
 & DRN$_{\textrm{L}:6,\textrm{D}:832,\textrm{EncDec}_{\textrm{B}2}}$ & \cellcolor[RGB]{183,226,176} 0.00107 & \cellcolor[RGB]{182,225,175} 0.00065 & \cellcolor[RGB]{182,225,175} 0.00023 & \cellcolor[RGB]{252,194,171} 7.18971 & \cellcolor[RGB]{252,166,137} 395.983 \\
 & DRN$_{\textrm{L}:6,\textrm{D}:832,\textrm{EncDec}_{\textrm{B}1}}$ & \cellcolor[RGB]{199,233,192} 0.00299 & \cellcolor[RGB]{182,225,175} 0.00056 & \cellcolor[RGB]{184,226,177} 0.00027 & \cellcolor[RGB]{252,195,172} 7.13724 & \cellcolor[RGB]{252,166,137} 394.271 \\
 & DRN$_{\textrm{L}:7,\textrm{D}:768,\textrm{EncDec}_{\textrm{B}1}}$ & \cellcolor[RGB]{196,231,189} 0.00255 & \cellcolor[RGB]{184,226,177} 0.00092 & \cellcolor[RGB]{186,227,180} 0.00032 & \cellcolor[RGB]{253,216,200} 5.16661 & \cellcolor[RGB]{252,166,137} 398.770 \\
 & DRN$_{\textrm{L}:4,\textrm{D}:704,\textrm{EncDec}_{\textrm{B}2}}$ & \cellcolor[RGB]{180,225,174} 0.00086 & \cellcolor[RGB]{180,225,174} 0.00046 & \cellcolor[RGB]{179,224,173} 0.00019 & \cellcolor[RGB]{254,229,217} 3.55201 & \cellcolor[RGB]{252,166,137} 190.693 \\
 & DRN$_{\textrm{L}:4,\textrm{D}:704,\textrm{EncDec}_{\textrm{B}1}}$ & \cellcolor[RGB]{185,227,178} 0.00133 & \cellcolor[RGB]{182,225,175} 0.00061 & \cellcolor[RGB]{202,234,195} 0.00069 & \cellcolor[RGB]{254,229,218} 3.42041 & \cellcolor[RGB]{252,166,137} 188.738 \\
 & DRN$_{\textrm{L}:2,\textrm{D}:128,\textrm{EncDec}_{\textrm{S}}}$ & \cellcolor[RGB]{255,255,255} 2.28718 & \cellcolor[RGB]{230,245,225} 0.01285 & \cellcolor[RGB]{205,235,198} 0.00076 & \cellcolor[RGB]{254,239,232} 1.78569 & \cellcolor[RGB]{254,241,235} 10.2031 \\
 & DRN$_{\textrm{L}:1,\textrm{D}:64,\textrm{EncDec}_{\textrm{B}2}}$ & \cellcolor[RGB]{255,255,255} 0.09358 & \cellcolor[RGB]{246,251,244} 0.04252 & \cellcolor[RGB]{255,255,255} 0.00754 & \cellcolor[RGB]{254,241,235} 1.43229 & \cellcolor[RGB]{255,255,255} 3.39129 \\
 & DRN$_{\textrm{L}:1,\textrm{D}:64,\textrm{EncDec}_{\textrm{B}1}}$ & \cellcolor[RGB]{255,255,255} -  & \cellcolor[RGB]{255,255,255} 0.06412 & \cellcolor[RGB]{244,250,241} 0.00419 & \cellcolor[RGB]{254,241,235} 1.40984 & \cellcolor[RGB]{255,255,255} 3.22418 \\
 & DRN$_{\textrm{L}:1,\textrm{D}:64,\textrm{EncDec}_{\textrm{B}1}}$ & \cellcolor[RGB]{255,255,255} 0.22127 & \cellcolor[RGB]{255,255,255} 0.05735 & \cellcolor[RGB]{243,250,240} 0.00378 & \cellcolor[RGB]{254,241,235} 1.40591 & \cellcolor[RGB]{255,255,255} 3.22675 \\
 & DRN$_{\textrm{L}:1,\textrm{D}:128,\textrm{EncDec}_{\textrm{S}}}$ & \cellcolor[RGB]{255,255,255} -  & \cellcolor[RGB]{244,250,241} 0.03124 & \cellcolor[RGB]{213,238,207} 0.00104 & \cellcolor[RGB]{254,247,244} 1.18895 & \cellcolor[RGB]{254,250,248} 5.55577 \\
 & DRN$_{\textrm{L}:1,\textrm{D}:64,\textrm{EncDec}_{\textrm{S}}}$ & \cellcolor[RGB]{255,255,255} -  & \cellcolor[RGB]{255,255,255} 0.07665 & \cellcolor[RGB]{240,249,237} 0.00320 & \cellcolor[RGB]{254,250,248} 1.10879 & \cellcolor[RGB]{255,255,255} 2.26616 \\
\addlinespace[-2.32pt]
\cmidrule[\heavyrulewidth]{2-7}
\addlinespace[-\belowrulesep]
\multirow{3}{*}{\rotatebox{90}{width}} \hspace{-0.3cm} & DRN$_{\textrm{L}:6,\textrm{D}:1408,\textrm{EncDec}_{\textrm{B}1}}$ & \cellcolor[RGB]{209,237,202} 0.00464 & \cellcolor[RGB]{182,225,175} 0.00070 & \cellcolor[RGB]{186,227,180} 0.00032 & \cellcolor[RGB]{252,168,139} 9.59635 & \cellcolor[RGB]{252,166,137} 1214.41 \\
 & DRN$_{\textrm{L}:4,\textrm{D}:1536,\textrm{EncDec}_{\textrm{B}1}}$ & \cellcolor[RGB]{185,227,178} 0.00139 & \cellcolor[RGB]{182,225,175} 0.00056 & \cellcolor[RGB]{182,225,175} 0.00023 & \cellcolor[RGB]{252,191,166} 7.52427 & \cellcolor[RGB]{252,166,137} 941.731 \\
 & DRN$_{\textrm{L}:3,\textrm{D}:1536,\textrm{EncDec}_{\textrm{B}1}}$ & \cellcolor[RGB]{184,226,177} 0.00118 & \cellcolor[RGB]{184,226,177} 0.00093 & \cellcolor[RGB]{198,232,191} 0.00056 & \cellcolor[RGB]{253,208,189} 5.84262 & \cellcolor[RGB]{252,166,137} 700.517 \\
 & DRN$_{\textrm{L}:3,\textrm{D}:1152,\textrm{EncDec}_{\textrm{B}2}}$ & \cellcolor[RGB]{183,226,176} 0.00112 & \cellcolor[RGB]{182,225,175} 0.00058 & \cellcolor[RGB]{184,226,177} 0.00027 & \cellcolor[RGB]{254,227,214} 3.94719 & \cellcolor[RGB]{252,166,137} 394.983 \\
 & DRN$_{\textrm{L}:3,\textrm{D}:1152,\textrm{EncDec}_{\textrm{B}1}}$ & \cellcolor[RGB]{179,224,173} 0.00079 & \cellcolor[RGB]{179,224,173} 0.00037 & \cellcolor[RGB]{180,225,174} 0.00020 & \cellcolor[RGB]{254,227,214} 3.89898 & \cellcolor[RGB]{252,166,137} 399.986 \\
 & DRN$_{\textrm{L}:1,\textrm{D}:1152,\textrm{EncDec}_{\textrm{B}2}}$ & \cellcolor[RGB]{195,231,188} 0.00237 & \cellcolor[RGB]{180,225,174} 0.00043 & \cellcolor[RGB]{180,225,174} 0.00021 & \cellcolor[RGB]{254,239,231} 1.85389 & \cellcolor[RGB]{252,166,137} 143.139 \\
 & DRN$_{\textrm{L}:1,\textrm{D}:1152,\textrm{EncDec}_{\textrm{B}1}}$ & \cellcolor[RGB]{190,229,183} 0.00192 & \cellcolor[RGB]{180,225,174} 0.00041 & \cellcolor[RGB]{182,225,175} 0.00022 & \cellcolor[RGB]{254,239,232} 1.80736 & \cellcolor[RGB]{252,166,137} 140.160 \\
 & DRN$_{\textrm{L}:2,\textrm{D}:128,\textrm{EncDec}_{\textrm{S}}}$ & \cellcolor[RGB]{255,255,255} -  & \cellcolor[RGB]{239,248,235} 0.02125 & \cellcolor[RGB]{207,236,200} 0.00083 & \cellcolor[RGB]{254,239,232} 1.79184 & \cellcolor[RGB]{254,241,235} 10.2170 \\
 & DRN$_{\textrm{L}:1,\textrm{D}:128,\textrm{EncDec}_{\textrm{B}2}}$ & \cellcolor[RGB]{245,251,243} 0.03162 & \cellcolor[RGB]{218,240,212} 0.00835 & \cellcolor[RGB]{228,244,223} 0.00169 & \cellcolor[RGB]{254,240,233} 1.66056 & \cellcolor[RGB]{254,244,239} 8.40828 \\
 & DRN$_{\textrm{L}:1,\textrm{D}:128,\textrm{EncDec}_{\textrm{B}1}}$ & \cellcolor[RGB]{255,255,255} 0.04881 & \cellcolor[RGB]{236,248,233} 0.01862 & \cellcolor[RGB]{231,246,227} 0.00195 & \cellcolor[RGB]{254,241,234} 1.57086 & \cellcolor[RGB]{254,244,239} 8.09900 \\
 & DRN$_{\textrm{L}:1,\textrm{D}:128,\textrm{EncDec}_{\textrm{S}}}$ & \cellcolor[RGB]{255,255,255} -  & \cellcolor[RGB]{255,255,255} 0.05837 & \cellcolor[RGB]{216,240,210} 0.00113 & \cellcolor[RGB]{254,247,244} 1.19550 & \cellcolor[RGB]{254,250,248} 5.57275 \\
 & DRN$_{\textrm{L}:1,\textrm{D}:128,\textrm{EncDec}_{\textrm{S}}}$ & \cellcolor[RGB]{255,255,255} 0.38269 & \cellcolor[RGB]{244,251,242} 0.03310 & \cellcolor[RGB]{215,239,209} 0.00111 & \cellcolor[RGB]{254,247,244} 1.19353 & \cellcolor[RGB]{254,250,248} 5.56622 \\
\addlinespace[-2.32pt]
\cmidrule[\heavyrulewidth]{2-7}
\addlinespace[-\belowrulesep]
\multirow{3}{*}{\rotatebox{90}{depth}} \hspace{-0.3cm} & UNet$_{\textrm{D}:40,\textrm{Depth}:3,\textrm{EncDec}_{\textrm{B}2}}$ & \cellcolor[RGB]{232,246,227} 0.01126 & \cellcolor[RGB]{219,241,213} 0.00841 & \cellcolor[RGB]{255,255,255} 0.01552 & \cellcolor[RGB]{252,172,144} 9.26728 & \cellcolor[RGB]{254,247,244} 7.44607 \\
 & UNet$_{\textrm{D}:40,\textrm{Depth}:3,\textrm{EncDec}_{\textrm{B}1}}$ & \cellcolor[RGB]{255,255,255} 0.23186 & \cellcolor[RGB]{255,255,255} 0.13211 & \cellcolor[RGB]{255,255,255} 0.01521 & \cellcolor[RGB]{252,173,145} 9.13928 & \cellcolor[RGB]{254,247,244} 7.31869 \\
 & UNet$_{\textrm{D}:48,\textrm{Depth}:2,\textrm{EncDec}_{\textrm{B}2}}$ & \cellcolor[RGB]{237,248,234} 0.01562 & \cellcolor[RGB]{213,238,207} 0.00694 & \cellcolor[RGB]{246,251,244} 0.00589 & \cellcolor[RGB]{252,197,174} 7.01186 & \cellcolor[RGB]{254,250,248} 6.60978 \\
 & UNet$_{\textrm{D}:48,\textrm{Depth}:2,\textrm{EncDec}_{\textrm{B}1}}$ & \cellcolor[RGB]{255,255,255} 0.11321 & \cellcolor[RGB]{244,251,242} 0.03437 & \cellcolor[RGB]{255,255,255} 0.01592 & \cellcolor[RGB]{252,198,175} 6.94739 & \cellcolor[RGB]{254,250,248} 6.49898 \\
 & UNet$_{\textrm{D}:80,\textrm{Depth}:1,\textrm{EncDec}_{\textrm{B}2}}$ & \cellcolor[RGB]{198,232,191} 0.00283 & \cellcolor[RGB]{198,232,191} 0.00316 & \cellcolor[RGB]{245,251,243} 0.00547 & \cellcolor[RGB]{253,221,206} 4.65121 & \cellcolor[RGB]{254,247,244} 7.68850 \\
 & UNet$_{\textrm{D}:80,\textrm{Depth}:1,\textrm{EncDec}_{\textrm{B}1}}$ & \cellcolor[RGB]{227,244,221} 0.00887 & \cellcolor[RGB]{213,238,207} 0.00686 & \cellcolor[RGB]{244,250,241} 0.00439 & \cellcolor[RGB]{253,221,206} 4.62307 & \cellcolor[RGB]{254,247,244} 7.49870 \\
 & UNet$_{\textrm{D}:32,\textrm{Depth}:1,\textrm{EncDec}_{\textrm{B}1}}$ & \cellcolor[RGB]{255,255,255} 0.49890 & \cellcolor[RGB]{255,255,255} 0.22082 & \cellcolor[RGB]{255,255,255} 0.01546 & \cellcolor[RGB]{254,225,212} 4.08293 & \cellcolor[RGB]{255,255,255} 3.63259 \\
 & UNet$_{\textrm{D}:32,\textrm{Depth}:1,\textrm{EncDec}_{\textrm{S}}}$ & \cellcolor[RGB]{255,255,255} 0.04101 & \cellcolor[RGB]{231,246,227} 0.01389 & \cellcolor[RGB]{255,255,255} 0.00862 & \cellcolor[RGB]{254,227,214} 3.84332 & \cellcolor[RGB]{255,255,255} 3.26867 \\
\addlinespace[-2.32pt]
\cmidrule[\heavyrulewidth]{2-7}
\addlinespace[-\belowrulesep]
\multirow{3}{*}{\rotatebox{90}{width}} \hspace{-0.3cm} & UNet$_{\textrm{D}:40,\textrm{Depth}:3,\textrm{EncDec}_{\textrm{B}2}}$ & \cellcolor[RGB]{227,244,221} 0.00873 & \cellcolor[RGB]{217,240,211} 0.00791 & \cellcolor[RGB]{255,255,255} 0.01743 & \cellcolor[RGB]{252,172,144} 9.21076 & \cellcolor[RGB]{254,247,244} 7.42898 \\
 & UNet$_{\textrm{D}:40,\textrm{Depth}:3,\textrm{EncDec}_{\textrm{B}1}}$ & \cellcolor[RGB]{255,255,255} 0.27884 & \cellcolor[RGB]{255,255,255} 0.10758 & \cellcolor[RGB]{255,255,255} 0.01552 & \cellcolor[RGB]{252,173,145} 9.12114 & \cellcolor[RGB]{254,247,244} 7.32025 \\
 & UNet$_{\textrm{D}:48,\textrm{Depth}:2,\textrm{EncDec}_{\textrm{B}2}}$ & \cellcolor[RGB]{233,246,229} 0.01197 & \cellcolor[RGB]{210,237,203} 0.00592 & \cellcolor[RGB]{255,255,255} 0.00715 & \cellcolor[RGB]{252,197,174} 6.97971 & \cellcolor[RGB]{254,250,248} 6.59795 \\
 & UNet$_{\textrm{D}:48,\textrm{Depth}:2,\textrm{EncDec}_{\textrm{B}1}}$ & \cellcolor[RGB]{255,255,255} 0.09892 & \cellcolor[RGB]{245,251,243} 0.03511 & \cellcolor[RGB]{255,255,255} 0.01639 & \cellcolor[RGB]{252,198,175} 6.92277 & \cellcolor[RGB]{254,250,248} 6.50537 \\
 & UNet$_{\textrm{D}:128,\textrm{Depth}:1,\textrm{EncDec}_{\textrm{B}2}}$ & \cellcolor[RGB]{185,227,178} 0.00136 & \cellcolor[RGB]{189,228,182} 0.00167 & \cellcolor[RGB]{219,241,213} 0.00127 & \cellcolor[RGB]{253,219,203} 4.84794 & \cellcolor[RGB]{254,239,232} 12.5765 \\
 & UNet$_{\textrm{D}:128,\textrm{Depth}:1,\textrm{EncDec}_{\textrm{B}1}}$ & \cellcolor[RGB]{183,226,176} 0.00114 & \cellcolor[RGB]{186,227,180} 0.00137 & \cellcolor[RGB]{255,255,255} 0.00819 & \cellcolor[RGB]{253,220,205} 4.74599 & \cellcolor[RGB]{254,240,233} 12.2107 \\
 & UNet$_{\textrm{D}:32,\textrm{Depth}:1,\textrm{EncDec}_{\textrm{B}1}}$ & \cellcolor[RGB]{255,255,255} 0.43770 & \cellcolor[RGB]{255,255,255} 0.19121 & \cellcolor[RGB]{255,255,255} 0.01519 & \cellcolor[RGB]{254,225,212} 4.07527 & \cellcolor[RGB]{255,255,255} 3.62707 \\
 & UNet$_{\textrm{D}:32,\textrm{Depth}:1,\textrm{EncDec}_{\textrm{S}}}$ & \cellcolor[RGB]{241,249,238} 0.02002 & \cellcolor[RGB]{229,245,224} 0.01230 & \cellcolor[RGB]{255,255,255} 0.00846 & \cellcolor[RGB]{254,227,214} 3.84160 & \cellcolor[RGB]{255,255,255} 3.26604 \\
\addlinespace[-2.32pt]
\cmidrule[\heavyrulewidth]{2-7}
\addlinespace[-\belowrulesep]
\multirow{3}{*}{\rotatebox{90}{depth}} \hspace{-0.3cm} & FNO$_{\textrm{L}:12,\textrm{D}:512,\textrm{M}:44,\textrm{EncDec}_{\textrm{B}1}}$ & \cellcolor[RGB]{220,241,214} 0.00707 & \cellcolor[RGB]{184,226,177} 0.00104 & \cellcolor[RGB]{190,229,183} 0.00039 & \cellcolor[RGB]{252,171,142} 9.37148 & \cellcolor[RGB]{252,166,137} 729.302 \\
 & FNO$_{\textrm{L}:10,\textrm{D}:512,\textrm{M}:42,\textrm{EncDec}_{\textrm{B}2}}$ & \cellcolor[RGB]{196,231,189} 0.00253 & \cellcolor[RGB]{183,226,176} 0.00086 & \cellcolor[RGB]{195,231,188} 0.00049 & \cellcolor[RGB]{252,190,165} 7.67463 & \cellcolor[RGB]{252,166,137} 590.495 \\
 & FNO$_{\textrm{L}:10,\textrm{D}:512,\textrm{M}:42,\textrm{EncDec}_{\textrm{B}1}}$ & \cellcolor[RGB]{210,237,203} 0.00475 & \cellcolor[RGB]{183,226,176} 0.00088 & \cellcolor[RGB]{184,226,177} 0.00027 & \cellcolor[RGB]{252,190,165} 7.63494 & \cellcolor[RGB]{252,166,137} 591.633 \\
 & FNO$_{\textrm{L}:10,\textrm{D}:480,\textrm{M}:42,\textrm{EncDec}_{\textrm{B}2}}$ & \cellcolor[RGB]{225,243,219} 0.00816 & \cellcolor[RGB]{184,226,177} 0.00093 & \cellcolor[RGB]{184,226,177} 0.00027 & \cellcolor[RGB]{253,211,192} 5.71019 & \cellcolor[RGB]{252,166,137} 496.134 \\
 & FNO$_{\textrm{L}:10,\textrm{D}:480,\textrm{M}:42,\textrm{EncDec}_{\textrm{B}1}}$ & \cellcolor[RGB]{211,238,205} 0.00524 & \cellcolor[RGB]{185,227,178} 0.00109 & \cellcolor[RGB]{190,229,183} 0.00038 & \cellcolor[RGB]{253,211,192} 5.66940 & \cellcolor[RGB]{252,166,137} 493.843 \\
 & FNO$_{\textrm{L}:9,\textrm{D}:416,\textrm{M}:38,\textrm{EncDec}_{\textrm{B}2}}$ & \cellcolor[RGB]{230,245,225} 0.00992 & \cellcolor[RGB]{186,227,180} 0.00132 & \cellcolor[RGB]{184,226,177} 0.00028 & \cellcolor[RGB]{254,227,215} 3.81325 & \cellcolor[RGB]{252,166,137} 247.019 \\
 & FNO$_{\textrm{L}:9,\textrm{D}:416,\textrm{M}:38,\textrm{EncDec}_{\textrm{B}1}}$ & \cellcolor[RGB]{255,255,255} 0.05983 & \cellcolor[RGB]{186,227,180} 0.00130 & \cellcolor[RGB]{185,227,178} 0.00028 & \cellcolor[RGB]{254,228,216} 3.71339 & \cellcolor[RGB]{252,166,137} 246.963 \\
 & FNO$_{\textrm{L}:3,\textrm{D}:160,\textrm{M}:16,\textrm{EncDec}_{\textrm{B}1}}$ & \cellcolor[RGB]{255,255,255} -  & \cellcolor[RGB]{244,250,241} 0.03178 & \cellcolor[RGB]{239,249,236} 0.00308 & \cellcolor[RGB]{254,239,231} 1.85602 & \cellcolor[RGB]{254,250,248} 5.61218 \\
 & FNO$_{\textrm{L}:4,\textrm{D}:64,\textrm{M}:8,\textrm{EncDec}_{\textrm{S}}}$ & \cellcolor[RGB]{255,255,255} -  & \cellcolor[RGB]{255,255,255} 0.16240 & \cellcolor[RGB]{236,248,233} 0.00254 & \cellcolor[RGB]{254,239,232} 1.72311 & \cellcolor[RGB]{255,255,255} 1.90821 \\
 & FNO$_{\textrm{L}:2,\textrm{D}:96,\textrm{M}:10,\textrm{EncDec}_{\textrm{S}}}$ & \cellcolor[RGB]{255,255,255} 3.22016 & \cellcolor[RGB]{255,255,255} 0.24484 & \cellcolor[RGB]{242,250,239} 0.00373 & \cellcolor[RGB]{254,247,244} 1.21733 & \cellcolor[RGB]{255,255,255} 1.84118 \\
 & FNO$_{\textrm{L}:1,\textrm{D}:32,\textrm{M}:8,\textrm{EncDec}_{\textrm{B}1}}$ & \cellcolor[RGB]{255,255,255} 2.11400 & \cellcolor[RGB]{255,255,255} 1.34209 & \cellcolor[RGB]{255,255,255} 0.01761 & \cellcolor[RGB]{254,252,251} 1.04509 & \cellcolor[RGB]{255,255,255} 1.05574 \\
 & FNO$_{\textrm{L}:1,\textrm{D}:64,\textrm{M}:8,\textrm{EncDec}_{\textrm{S}}}$ & \cellcolor[RGB]{255,255,255} 1.43678 & \cellcolor[RGB]{255,255,255} 1.35991 & \cellcolor[RGB]{255,255,255} 0.01547 & \cellcolor[RGB]{255,255,255} 0.85336 & \cellcolor[RGB]{255,255,255} 0.94193 \\
 & FNO$_{\textrm{L}:1,\textrm{D}:32,\textrm{M}:8,\textrm{EncDec}_{\textrm{S}}}$ & \cellcolor[RGB]{255,255,255} -  & \cellcolor[RGB]{255,255,255} 0.41643 & \cellcolor[RGB]{255,255,255} 0.01855 & \cellcolor[RGB]{255,255,255} 0.82932 & \cellcolor[RGB]{255,255,255} 0.72637 \\
\addlinespace[-2.32pt]
\cmidrule[\heavyrulewidth]{2-7}
\addlinespace[-\belowrulesep]
\multirow{3}{*}{\rotatebox{90}{width}} \hspace{-0.3cm} & FNO$_{\textrm{L}:5,\textrm{D}:768,\textrm{M}:42,\textrm{EncDec}_{\textrm{B}1}}$ & \cellcolor[RGB]{255,255,255} 1.79074 & \cellcolor[RGB]{186,227,180} 0.00129 & \cellcolor[RGB]{199,233,192} 0.00059 & \cellcolor[RGB]{252,166,137} 9.73918 & \cellcolor[RGB]{252,166,137} 703.420 \\
 & FNO$_{\textrm{L}:5,\textrm{D}:704,\textrm{M}:40,\textrm{EncDec}_{\textrm{B}2}}$ & \cellcolor[RGB]{255,255,255} 0.48578 & \cellcolor[RGB]{187,228,181} 0.00144 & \cellcolor[RGB]{185,227,178} 0.00029 & \cellcolor[RGB]{252,187,162} 7.89942 & \cellcolor[RGB]{252,166,137} 564.436 \\
 & FNO$_{\textrm{L}:5,\textrm{D}:704,\textrm{M}:40,\textrm{EncDec}_{\textrm{B}1}}$ & \cellcolor[RGB]{255,255,255} 0.07451 & \cellcolor[RGB]{187,228,181} 0.00149 & \cellcolor[RGB]{186,227,180} 0.00031 & \cellcolor[RGB]{252,187,162} 7.85474 & \cellcolor[RGB]{252,166,137} 561.600 \\
 & FNO$_{\textrm{L}:5,\textrm{D}:640,\textrm{M}:34,\textrm{EncDec}_{\textrm{B}1}}$ & \cellcolor[RGB]{255,255,255} 3.02045 & \cellcolor[RGB]{210,237,203} 0.00582 & \cellcolor[RGB]{237,248,234} 0.00273 & \cellcolor[RGB]{253,209,191} 5.74161 & \cellcolor[RGB]{252,166,137} 397.831 \\
 & FNO$_{\textrm{L}:4,\textrm{D}:576,\textrm{M}:30,\textrm{EncDec}_{\textrm{B}2}}$ & \cellcolor[RGB]{255,255,255} 0.30959 & \cellcolor[RGB]{196,231,189} 0.00282 & \cellcolor[RGB]{189,228,182} 0.00037 & \cellcolor[RGB]{254,228,216} 3.63196 & \cellcolor[RGB]{252,166,137} 229.023 \\
 & FNO$_{\textrm{L}:4,\textrm{D}:576,\textrm{M}:30,\textrm{EncDec}_{\textrm{B}1}}$ & \cellcolor[RGB]{255,255,255} -  & \cellcolor[RGB]{196,231,189} 0.00275 & \cellcolor[RGB]{200,233,193} 0.00062 & \cellcolor[RGB]{254,229,217} 3.59567 & \cellcolor[RGB]{252,166,137} 227.535 \\
 & FNO$_{\textrm{L}:3,\textrm{D}:384,\textrm{M}:20,\textrm{EncDec}_{\textrm{B}1}}$ & \cellcolor[RGB]{255,255,255} 1.19278 & \cellcolor[RGB]{232,246,227} 0.01421 & \cellcolor[RGB]{214,239,208} 0.00107 & \cellcolor[RGB]{254,239,231} 1.86434 & \cellcolor[RGB]{252,166,137} 54.7761 \\
 & FNO$_{\textrm{L}:3,\textrm{D}:512,\textrm{M}:22,\textrm{EncDec}_{\textrm{S}}}$ & \cellcolor[RGB]{255,255,255} -  & \cellcolor[RGB]{216,240,210} 0.00772 & \cellcolor[RGB]{198,232,191} 0.00056 & \cellcolor[RGB]{254,240,233} 1.63385 & \cellcolor[RGB]{252,166,137} 110.777 \\
 & FNO$_{\textrm{L}:2,\textrm{D}:192,\textrm{M}:18,\textrm{EncDec}_{\textrm{S}}}$ & \cellcolor[RGB]{255,255,255} -  & \cellcolor[RGB]{255,255,255} 0.53540 & \cellcolor[RGB]{232,246,228} 0.00207 & \cellcolor[RGB]{254,247,244} 1.23171 & \cellcolor[RGB]{254,250,248} 5.78802 \\
 & FNO$_{\textrm{L}:1,\textrm{D}:64,\textrm{M}:8,\textrm{EncDec}_{\textrm{B}2}}$ & \cellcolor[RGB]{255,255,255} 1.00942 & \cellcolor[RGB]{255,255,255} 0.39667 & \cellcolor[RGB]{255,255,255} 0.01002 & \cellcolor[RGB]{254,247,244} 1.18821 & \cellcolor[RGB]{255,255,255} 1.36239 \\
 & FNO$_{\textrm{L}:1,\textrm{D}:64,\textrm{M}:8,\textrm{EncDec}_{\textrm{B}1}}$ & \cellcolor[RGB]{255,255,255} 2.79691 & \cellcolor[RGB]{255,255,255} 0.80250 & \cellcolor[RGB]{255,255,255} 0.01356 & \cellcolor[RGB]{254,250,248} 1.15519 & \cellcolor[RGB]{255,255,255} 1.20475 \\
 & FNO$_{\textrm{L}:1,\textrm{D}:64,\textrm{M}:8,\textrm{EncDec}_{\textrm{S}}}$ & \cellcolor[RGB]{255,255,255} 1.43678 & \cellcolor[RGB]{255,255,255} 1.35991 & \cellcolor[RGB]{255,255,255} 0.01547 & \cellcolor[RGB]{255,255,255} 0.85205 & \cellcolor[RGB]{255,255,255} 0.95389 \\
 & FNO$_{\textrm{L}:1,\textrm{D}:64,\textrm{M}:8,\textrm{EncDec}_{\textrm{S}}}$ & \cellcolor[RGB]{255,255,255} 1.43678 & \cellcolor[RGB]{255,255,255} 1.35991 & \cellcolor[RGB]{255,255,255} 0.01547 & \cellcolor[RGB]{255,255,255} 0.85094 & \cellcolor[RGB]{255,255,255} 0.94833 \\
\addlinespace[-2.32pt]
\cmidrule[\heavyrulewidth]{2-7}
\addlinespace[-\belowrulesep]
\multirow{3}{*}{\rotatebox{90}{depth}} \hspace{-0.3cm} & MP-PDE$_{\textrm{L}:9,\textrm{D}:640,\textrm{EncDec}_{\textrm{B}1}}$ & \cellcolor[RGB]{255,255,255} 1.38904 & \cellcolor[RGB]{255,255,255} 0.76555 & \cellcolor[RGB]{255,255,255} 0.01350 & \cellcolor[RGB]{252,172,144} 9.25938 & \cellcolor[RGB]{252,166,137} 848.036 \\
 & MP-PDE$_{\textrm{L}:7,\textrm{D}:640,\textrm{EncDec}_{\textrm{B}1}}$ & \cellcolor[RGB]{255,255,255} 1.68327 & \cellcolor[RGB]{255,255,255} 0.56863 & \cellcolor[RGB]{255,255,255} 0.01428 & \cellcolor[RGB]{252,192,168} 7.48872 & \cellcolor[RGB]{252,166,137} 660.928 \\
 & MP-PDE$_{\textrm{L}:5,\textrm{D}:128,\textrm{EncDec}_{\textrm{B}2}}$ & \cellcolor[RGB]{255,255,255} 0.16393 & \cellcolor[RGB]{255,255,255} 0.16624 & \cellcolor[RGB]{255,255,255} 0.01652 & \cellcolor[RGB]{253,209,191} 5.82885 & \cellcolor[RGB]{253,223,209} 30.3772 \\
 & MP-PDE$_{\textrm{L}:5,\textrm{D}:128,\textrm{EncDec}_{\textrm{B}1}}$ & \cellcolor[RGB]{255,255,255} 0.45055 & \cellcolor[RGB]{255,255,255} 0.33469 & \cellcolor[RGB]{255,255,255} 0.01319 & \cellcolor[RGB]{253,211,192} 5.71171 & \cellcolor[RGB]{253,223,209} 30.1035 \\
 & MP-PDE$_{\textrm{L}:3,\textrm{D}:64,\textrm{EncDec}_{\textrm{B}2}}$ & \cellcolor[RGB]{255,255,255} 4.18210 & \cellcolor[RGB]{255,255,255} 1.19208 & \cellcolor[RGB]{255,255,255} 0.01560 & \cellcolor[RGB]{254,229,217} 3.52891 & \cellcolor[RGB]{254,244,239} 8.32650 \\
 & MP-PDE$_{\textrm{L}:3,\textrm{D}:64,\textrm{EncDec}_{\textrm{B}1}}$ & \cellcolor[RGB]{255,255,255} 2.08942 & \cellcolor[RGB]{255,255,255} 1.40994 & \cellcolor[RGB]{255,255,255} 0.01580 & \cellcolor[RGB]{254,229,218} 3.43175 & \cellcolor[RGB]{254,244,239} 8.16214 \\
 & MP-PDE$_{\textrm{L}:1,\textrm{D}:64,\textrm{EncDec}_{\textrm{B}2}}$ & \cellcolor[RGB]{255,255,255} -  & \cellcolor[RGB]{255,255,255} 0.96522 & \cellcolor[RGB]{255,255,255} 0.01389 & \cellcolor[RGB]{254,238,230} 1.98164 & \cellcolor[RGB]{255,255,255} 3.53774 \\
 & MP-PDE$_{\textrm{L}:1,\textrm{D}:64,\textrm{EncDec}_{\textrm{B}1}}$ & \cellcolor[RGB]{255,255,255} 1.95242 & \cellcolor[RGB]{255,255,255} 1.33502 & \cellcolor[RGB]{255,255,255} 0.01736 & \cellcolor[RGB]{254,239,231} 1.89047 & \cellcolor[RGB]{255,255,255} 3.36327 \\
 & MP-PDE$_{\textrm{L}:1,\textrm{D}:64,\textrm{EncDec}_{\textrm{S}}}$ & \cellcolor[RGB]{255,255,255} 2.20677 & \cellcolor[RGB]{255,255,255} 1.32638 & \cellcolor[RGB]{255,255,255} 0.02348 & \cellcolor[RGB]{254,241,234} 1.60772 & \cellcolor[RGB]{255,255,255} 3.08230 \\
 & MP-PDE$_{\textrm{L}:1,\textrm{D}:64,\textrm{EncDec}_{\textrm{S}}}$ & \cellcolor[RGB]{255,255,255} 2.20676 & \cellcolor[RGB]{255,255,255} 1.32657 & \cellcolor[RGB]{255,255,255} 0.02350 & \cellcolor[RGB]{254,241,234} 1.57680 & \cellcolor[RGB]{255,255,255} 3.08918 \\
\addlinespace[-2.32pt]
\cmidrule[\heavyrulewidth]{2-7}
\addlinespace[-\belowrulesep]
\multirow{3}{*}{\rotatebox{90}{width}} \hspace{-0.3cm} & MP-PDE$_{\textrm{L}:7,\textrm{D}:2432,\textrm{EncDec}_{\textrm{B}1}}$ & \cellcolor[RGB]{255,255,255} 0.05416 & \cellcolor[RGB]{255,255,255} 0.08421 & \cellcolor[RGB]{255,255,255} 0.01263 & \cellcolor[RGB]{252,169,141} 9.40597 & \cellcolor[RGB]{252,166,137} 8866.54 \\
 & MP-PDE$_{\textrm{L}:5,\textrm{D}:2432,\textrm{EncDec}_{\textrm{B}1}}$ & \cellcolor[RGB]{255,255,255} 0.09611 & \cellcolor[RGB]{255,255,255} 0.13401 & \cellcolor[RGB]{255,255,255} 0.01341 & \cellcolor[RGB]{252,195,172} 7.13519 & \cellcolor[RGB]{252,166,137} 6334.37 \\
 & MP-PDE$_{\textrm{L}:5,\textrm{D}:1536,\textrm{EncDec}_{\textrm{B}2}}$ & \cellcolor[RGB]{255,255,255} 0.04448 & \cellcolor[RGB]{201,234,194} 0.00376 & \cellcolor[RGB]{243,250,240} 0.00379 & \cellcolor[RGB]{253,208,189} 5.85646 & \cellcolor[RGB]{252,166,137} 2575.59 \\
 & MP-PDE$_{\textrm{L}:5,\textrm{D}:1536,\textrm{EncDec}_{\textrm{B}1}}$ & \cellcolor[RGB]{242,250,239} 0.02180 & \cellcolor[RGB]{245,251,243} 0.04098 & \cellcolor[RGB]{255,255,255} 0.01231 & \cellcolor[RGB]{253,211,192} 5.69930 & \cellcolor[RGB]{252,166,137} 2573.31 \\
 & MP-PDE$_{\textrm{L}:2,\textrm{D}:1920,\textrm{EncDec}_{\textrm{B}2}}$ & \cellcolor[RGB]{183,226,176} 0.00109 & \cellcolor[RGB]{185,227,178} 0.00115 & \cellcolor[RGB]{200,233,193} 0.00062 & \cellcolor[RGB]{254,230,219} 3.29961 & \cellcolor[RGB]{252,166,137} 1623.96 \\
 & MP-PDE$_{\textrm{L}:2,\textrm{D}:1920,\textrm{EncDec}_{\textrm{B}1}}$ & \cellcolor[RGB]{255,255,255} 0.20460 & \cellcolor[RGB]{255,255,255} 0.24257 & \cellcolor[RGB]{255,255,255} 0.01241 & \cellcolor[RGB]{254,231,220} 3.18353 & \cellcolor[RGB]{252,166,137} 1622.10 \\
 & MP-PDE$_{\textrm{L}:1,\textrm{D}:128,\textrm{EncDec}_{\textrm{B}2}}$ & \cellcolor[RGB]{255,255,255} 2.29486 & \cellcolor[RGB]{255,255,255} 0.10277 & \cellcolor[RGB]{255,255,255} 0.01534 & \cellcolor[RGB]{254,237,228} 2.19144 & \cellcolor[RGB]{254,247,244} 7.27126 \\
 & MP-PDE$_{\textrm{L}:1,\textrm{D}:128,\textrm{EncDec}_{\textrm{B}1}}$ & \cellcolor[RGB]{255,255,255} 1.63561 & \cellcolor[RGB]{255,255,255} 1.05139 & \cellcolor[RGB]{255,255,255} 0.01595 & \cellcolor[RGB]{254,238,230} 2.05083 & \cellcolor[RGB]{254,247,244} 6.93801 \\
 & MP-PDE$_{\textrm{L}:1,\textrm{D}:128,\textrm{EncDec}_{\textrm{S}}}$ & \cellcolor[RGB]{255,255,255} 2.29374 & \cellcolor[RGB]{255,255,255} 1.35059 & \cellcolor[RGB]{255,255,255} 0.02195 & \cellcolor[RGB]{254,239,232} 1.75542 & \cellcolor[RGB]{254,250,248} 6.73953 \\
 & MP-PDE$_{\textrm{L}:1,\textrm{D}:128,\textrm{EncDec}_{\textrm{S}}}$ & \cellcolor[RGB]{255,255,255} 2.29385 & \cellcolor[RGB]{255,255,255} 1.35036 & \cellcolor[RGB]{255,255,255} 0.02194 & \cellcolor[RGB]{254,239,232} 1.75485 & \cellcolor[RGB]{254,250,248} 6.73666 \\
\addlinespace[-2.32pt]
\cmidrule[\heavyrulewidth]{2-7}
\addlinespace[-\belowrulesep]
\multirow{3}{*}{\rotatebox{90}{depth}} \hspace{-0.3cm} & FT$_{\textrm{L}:9,\textrm{D}:512,\textrm{MLP}:640,\textrm{NH}:3,\textrm{EncDec}_{\textrm{B}1}}$ & \cellcolor[RGB]{197,232,190} 0.00264 & \cellcolor[RGB]{185,227,178} 0.00114 & \cellcolor[RGB]{182,225,175} 0.00023 & \cellcolor[RGB]{252,172,144} 9.24877 & \cellcolor[RGB]{252,166,137} 158.036 \\
 & FT$_{\textrm{L}:7,\textrm{D}:256,\textrm{MLP}:512,\textrm{NH}:2,\textrm{EncDec}_{\textrm{B}2}}$ & \cellcolor[RGB]{217,240,211} 0.00634 & \cellcolor[RGB]{202,234,195} 0.00412 & \cellcolor[RGB]{193,230,187} 0.00046 & \cellcolor[RGB]{252,197,174} 7.03963 & \cellcolor[RGB]{252,189,163} 46.7115 \\
 & FT$_{\textrm{L}:7,\textrm{D}:256,\textrm{MLP}:512,\textrm{NH}:2,\textrm{EncDec}_{\textrm{B}1}}$ & \cellcolor[RGB]{209,237,202} 0.00468 & \cellcolor[RGB]{198,232,191} 0.00326 & \cellcolor[RGB]{189,228,182} 0.00037 & \cellcolor[RGB]{252,198,175} 6.89447 & \cellcolor[RGB]{252,190,165} 46.1182 \\
 & FT$_{\textrm{L}:6,\textrm{D}:128,\textrm{MLP}:512,\textrm{NH}:2,\textrm{EncDec}_{\textrm{B}2}}$ & \cellcolor[RGB]{232,246,227} 0.01120 & \cellcolor[RGB]{214,239,208} 0.00704 & \cellcolor[RGB]{203,234,196} 0.00069 & \cellcolor[RGB]{253,213,195} 5.44432 & \cellcolor[RGB]{254,234,224} 19.9295 \\
 & FT$_{\textrm{L}:6,\textrm{D}:128,\textrm{MLP}:512,\textrm{NH}:2,\textrm{EncDec}_{\textrm{B}1}}$ & \cellcolor[RGB]{244,251,242} 0.02755 & \cellcolor[RGB]{229,245,224} 0.01242 & \cellcolor[RGB]{211,238,205} 0.00098 & \cellcolor[RGB]{253,214,197} 5.29166 & \cellcolor[RGB]{254,234,224} 19.5117 \\
 & FT$_{\textrm{L}:4,\textrm{D}:64,\textrm{MLP}:320,\textrm{NH}:2,\textrm{EncDec}_{\textrm{B}1}}$ & \cellcolor[RGB]{255,255,255} 0.70294 & \cellcolor[RGB]{246,251,244} 0.04705 & \cellcolor[RGB]{237,248,233} 0.00263 & \cellcolor[RGB]{254,229,217} 3.54636 & \cellcolor[RGB]{254,247,244} 7.39704 \\
 & FT$_{\textrm{L}:1,\textrm{D}:128,\textrm{MLP}:192,\textrm{NH}:2,\textrm{EncDec}_{\textrm{B}2}}$ & \cellcolor[RGB]{255,255,255} 0.68101 & \cellcolor[RGB]{255,255,255} 0.12418 & \cellcolor[RGB]{245,251,243} 0.00521 & \cellcolor[RGB]{254,239,232} 1.78545 & \cellcolor[RGB]{254,252,251} 3.77128 \\
 & FT$_{\textrm{L}:1,\textrm{D}:128,\textrm{MLP}:192,\textrm{NH}:2,\textrm{EncDec}_{\textrm{B}1}}$ & \cellcolor[RGB]{255,255,255} 0.13445 & \cellcolor[RGB]{255,255,255} 0.11040 & \cellcolor[RGB]{245,251,243} 0.00553 & \cellcolor[RGB]{254,240,233} 1.64835 & \cellcolor[RGB]{255,255,255} 3.42118 \\
 & FT$_{\textrm{L}:1,\textrm{D}:448,\textrm{MLP}:192,\textrm{NH}:3,\textrm{EncDec}_{\textrm{S}}}$ & \cellcolor[RGB]{255,255,255} -  & \cellcolor[RGB]{255,255,255} 0.12445 & \cellcolor[RGB]{231,245,226} 0.00193 & \cellcolor[RGB]{254,240,233} 1.62812 & \cellcolor[RGB]{254,239,232} 12.4719 \\
 & FT$_{\textrm{L}:1,\textrm{D}:64,\textrm{MLP}:64,\textrm{NH}:1,\textrm{EncDec}_{\textrm{B}1}}$ & \cellcolor[RGB]{255,255,255} 0.42559 & \cellcolor[RGB]{255,255,255} 0.32091 & \cellcolor[RGB]{255,255,255} 0.01931 & \cellcolor[RGB]{254,241,235} 1.44765 & \cellcolor[RGB]{255,255,255} 1.94380 \\
 & FT$_{\textrm{L}:1,\textrm{D}:64,\textrm{MLP}:64,\textrm{NH}:2,\textrm{EncDec}_{\textrm{S}}}$ & \cellcolor[RGB]{255,255,255} 3.63296 & \cellcolor[RGB]{255,255,255} 0.17905 & \cellcolor[RGB]{255,255,255} 0.00950 & \cellcolor[RGB]{254,247,244} 1.26231 & \cellcolor[RGB]{255,255,255} 1.98313 \\
 & FT$_{\textrm{L}:1,\textrm{D}:64,\textrm{MLP}:64,\textrm{NH}:1,\textrm{EncDec}_{\textrm{S}}}$ & \cellcolor[RGB]{255,255,255} 1.15685 & \cellcolor[RGB]{255,255,255} 0.67446 & \cellcolor[RGB]{255,255,255} 0.01413 & \cellcolor[RGB]{254,247,244} 1.16531 & \cellcolor[RGB]{255,255,255} 1.66528 \\
\addlinespace[-2.32pt]
\cmidrule[\heavyrulewidth]{2-7}
\addlinespace[-\belowrulesep]
\multirow{3}{*}{\rotatebox{90}{width}} \hspace{-0.3cm} & FT$_{\textrm{L}:7,\textrm{D}:3840,\textrm{MLP}:4096,\textrm{NH}:6,\textrm{EncDec}_{\textrm{B}1}}$ & \cellcolor[RGB]{255,255,255} -  & \cellcolor[RGB]{255,255,255} -  & \cellcolor[RGB]{255,255,255} -  & \cellcolor[RGB]{252,173,145} 9.18213 & \cellcolor[RGB]{252,166,137} 5513.29 \\
 & FT$_{\textrm{L}:7,\textrm{D}:3328,\textrm{MLP}:3840,\textrm{NH}:5,\textrm{EncDec}_{\textrm{B}1}}$ & \cellcolor[RGB]{255,255,255} 2.60587 & \cellcolor[RGB]{255,255,255} 1.67681 & \cellcolor[RGB]{255,255,255} 0.03231 & \cellcolor[RGB]{252,191,166} 7.58866 & \cellcolor[RGB]{252,166,137} 4234.88 \\
 & FT$_{\textrm{L}:5,\textrm{D}:1792,\textrm{MLP}:2048,\textrm{NH}:4,\textrm{EncDec}_{\textrm{B}2}}$ & \cellcolor[RGB]{255,255,255} 2.64014 & \cellcolor[RGB]{255,255,255} 1.67762 & \cellcolor[RGB]{255,255,255} 0.03231 & \cellcolor[RGB]{253,208,189} 5.86695 & \cellcolor[RGB]{252,166,137} 910.374 \\
 & FT$_{\textrm{L}:5,\textrm{D}:1792,\textrm{MLP}:2048,\textrm{NH}:4,\textrm{EncDec}_{\textrm{B}1}}$ & \cellcolor[RGB]{182,225,175} 0.00098 & \cellcolor[RGB]{183,226,176} 0.00075 & \cellcolor[RGB]{182,225,175} 0.00022 & \cellcolor[RGB]{253,211,192} 5.69377 & \cellcolor[RGB]{252,166,137} 906.339 \\
 & FT$_{\textrm{L}:3,\textrm{D}:256,\textrm{MLP}:1280,\textrm{NH}:3,\textrm{EncDec}_{\textrm{B}2}}$ & \cellcolor[RGB]{244,251,242} 0.02808 & \cellcolor[RGB]{218,240,212} 0.00826 & \cellcolor[RGB]{200,233,193} 0.00061 & \cellcolor[RGB]{254,227,215} 3.83484 & \cellcolor[RGB]{253,220,205} 32.3728 \\
 & FT$_{\textrm{L}:3,\textrm{D}:256,\textrm{MLP}:1280,\textrm{NH}:3,\textrm{EncDec}_{\textrm{B}1}}$ & \cellcolor[RGB]{255,255,255} 0.11371 & \cellcolor[RGB]{230,245,225} 0.01287 & \cellcolor[RGB]{203,234,196} 0.00070 & \cellcolor[RGB]{254,228,216} 3.71192 & \cellcolor[RGB]{253,221,206} 31.7293 \\
 & FT$_{\textrm{L}:1,\textrm{D}:256,\textrm{MLP}:256,\textrm{NH}:1,\textrm{EncDec}_{\textrm{B}2}}$ & \cellcolor[RGB]{255,255,255} 0.34787 & \cellcolor[RGB]{255,255,255} 0.06518 & \cellcolor[RGB]{255,255,255} 0.01229 & \cellcolor[RGB]{254,239,231} 1.92979 & \cellcolor[RGB]{254,247,244} 7.07917 \\
 & FT$_{\textrm{L}:1,\textrm{D}:256,\textrm{MLP}:256,\textrm{NH}:1,\textrm{EncDec}_{\textrm{B}1}}$ & \cellcolor[RGB]{245,251,243} 0.03365 & \cellcolor[RGB]{244,250,241} 0.03264 & \cellcolor[RGB]{245,251,243} 0.00554 & \cellcolor[RGB]{254,239,232} 1.79323 & \cellcolor[RGB]{254,250,248} 6.43597 \\
 & FT$_{\textrm{L}:1,\textrm{D}:256,\textrm{MLP}:256,\textrm{NH}:1,\textrm{EncDec}_{\textrm{B}1}}$ & \cellcolor[RGB]{255,255,255} 0.05331 & \cellcolor[RGB]{245,251,243} 0.03498 & \cellcolor[RGB]{255,255,255} 0.00700 & \cellcolor[RGB]{254,239,232} 1.78549 & \cellcolor[RGB]{254,250,248} 6.43803 \\
 & FT$_{\textrm{L}:1,\textrm{D}:2816,\textrm{MLP}:3840,\textrm{NH}:2,\textrm{EncDec}_{\textrm{S}}}$ & \cellcolor[RGB]{255,255,255} 0.08461 & \cellcolor[RGB]{226,243,220} 0.01072 & \cellcolor[RGB]{205,235,198} 0.00075 & \cellcolor[RGB]{254,241,234} 1.53436 & \cellcolor[RGB]{252,166,137} 458.610 \\
 & FT$_{\textrm{L}:1,\textrm{D}:256,\textrm{MLP}:256,\textrm{NH}:1,\textrm{EncDec}_{\textrm{S}}}$ & \cellcolor[RGB]{255,255,255} 1.68240 & \cellcolor[RGB]{255,255,255} 0.66721 & \cellcolor[RGB]{255,255,255} 0.00673 & \cellcolor[RGB]{254,241,235} 1.46035 & \cellcolor[RGB]{254,250,248} 5.85290 \\
\addlinespace[-2.32pt]
\cmidrule[\heavyrulewidth]{2-7}
\addlinespace[-\belowrulesep]
\multirow{3}{*}{\rotatebox{90}{depth}} \hspace{-0.3cm} & FT-FNO$_{\textrm{L}_\textrm{FT}:10,\textrm{L}_\textrm{FNO}:3,\textrm{D}:512,\textrm{MLP}:960,\textrm{M}:16,\textrm{EncDec}_{\textrm{B}2}}$ & \cellcolor[RGB]{255,255,255} 1.88162 & \cellcolor[RGB]{255,255,255} 1.68076 & \cellcolor[RGB]{255,255,255} 0.03125 & \cellcolor[RGB]{252,173,145} 9.17574 & \cellcolor[RGB]{252,166,137} 275.703 \\
 & FT-FNO$_{\textrm{L}_\textrm{FT}:10,\textrm{L}_\textrm{FNO}:3,\textrm{D}:512,\textrm{MLP}:960,\textrm{M}:16,\textrm{EncDec}_{\textrm{B}1}}$ & \cellcolor[RGB]{192,230,185} 0.00211 & \cellcolor[RGB]{186,227,180} 0.00134 & \cellcolor[RGB]{187,228,181} 0.00033 & \cellcolor[RGB]{252,175,147} 9.05003 & \cellcolor[RGB]{252,166,137} 275.295 \\
 & FT-FNO$_{\textrm{L}_\textrm{FT}:7,\textrm{L}_\textrm{FNO}:3,\textrm{D}:512,\textrm{MLP}:832,\textrm{M}:16,\textrm{EncDec}_{\textrm{B}1}}$ & \cellcolor[RGB]{255,255,255} 1.48946 & \cellcolor[RGB]{255,255,255} 0.99874 & \cellcolor[RGB]{255,255,255} 0.02365 & \cellcolor[RGB]{252,197,174} 7.00912 & \cellcolor[RGB]{252,166,137} 209.876 \\
 & FT-FNO$_{\textrm{L}_\textrm{FT}:5,\textrm{L}_\textrm{FNO}:3,\textrm{D}:352,\textrm{MLP}:704,\textrm{M}:16,\textrm{EncDec}_{\textrm{B}2}}$ & \cellcolor[RGB]{211,237,204} 0.00490 & \cellcolor[RGB]{195,231,188} 0.00259 & \cellcolor[RGB]{186,227,180} 0.00032 & \cellcolor[RGB]{253,211,192} 5.71605 & \cellcolor[RGB]{252,166,137} 86.1326 \\
 & FT-FNO$_{\textrm{L}_\textrm{FT}:5,\textrm{L}_\textrm{FNO}:3,\textrm{D}:352,\textrm{MLP}:704,\textrm{M}:16,\textrm{EncDec}_{\textrm{B}1}}$ & \cellcolor[RGB]{208,236,201} 0.00449 & \cellcolor[RGB]{200,233,193} 0.00367 & \cellcolor[RGB]{202,234,195} 0.00066 & \cellcolor[RGB]{253,212,194} 5.61349 & \cellcolor[RGB]{252,166,137} 86.0099 \\
 & FT-FNO$_{\textrm{L}_\textrm{FT}:2,\textrm{L}_\textrm{FNO}:3,\textrm{D}:512,\textrm{MLP}:896,\textrm{M}:15,\textrm{EncDec}_{\textrm{B}2}}$ & \cellcolor[RGB]{235,247,231} 0.01344 & \cellcolor[RGB]{199,233,192} 0.00345 & \cellcolor[RGB]{189,228,182} 0.00037 & \cellcolor[RGB]{254,228,216} 3.64507 & \cellcolor[RGB]{252,166,137} 118.348 \\
 & FT-FNO$_{\textrm{L}_\textrm{FT}:2,\textrm{L}_\textrm{FNO}:3,\textrm{D}:512,\textrm{MLP}:896,\textrm{M}:15,\textrm{EncDec}_{\textrm{B}1}}$ & \cellcolor[RGB]{238,248,235} 0.01597 & \cellcolor[RGB]{200,233,193} 0.00372 & \cellcolor[RGB]{195,231,188} 0.00048 & \cellcolor[RGB]{254,229,217} 3.51748 & \cellcolor[RGB]{252,166,137} 118.280 \\
 & FT-FNO$_{\textrm{L}_\textrm{FT}:1,\textrm{L}_\textrm{FNO}:1,\textrm{D}:64,\textrm{MLP}:64,\textrm{M}:9,\textrm{EncDec}_{\textrm{B}1}}$ & \cellcolor[RGB]{255,255,255} -  & \cellcolor[RGB]{255,255,255} 0.33748 & \cellcolor[RGB]{255,255,255} 0.01043 & \cellcolor[RGB]{254,239,232} 1.79479 & \cellcolor[RGB]{255,255,255} 2.34146 \\
 & FT-FNO$_{\textrm{L}_\textrm{FT}:1,\textrm{L}_\textrm{FNO}:2,\textrm{D}:32,\textrm{MLP}:64,\textrm{M}:8,\textrm{EncDec}_{\textrm{S}}}$ & \cellcolor[RGB]{255,255,255} -  & \cellcolor[RGB]{255,255,255} 0.99186 & \cellcolor[RGB]{255,255,255} 0.00816 & \cellcolor[RGB]{254,239,232} 1.72589 & \cellcolor[RGB]{255,255,255} 1.84310 \\
 & FT-FNO$_{\textrm{L}_\textrm{FT}:1,\textrm{L}_\textrm{FNO}:1,\textrm{D}:32,\textrm{MLP}:64,\textrm{M}:8,\textrm{EncDec}_{\textrm{B}1}}$ & \cellcolor[RGB]{255,255,255} 1.77179 & \cellcolor[RGB]{255,255,255} 0.86157 & \cellcolor[RGB]{255,255,255} 0.01556 & \cellcolor[RGB]{254,240,233} 1.66973 & \cellcolor[RGB]{255,255,255} 1.95149 \\
 & FT-FNO$_{\textrm{L}_\textrm{FT}:1,\textrm{L}_\textrm{FNO}:1,\textrm{D}:32,\textrm{MLP}:64,\textrm{M}:8,\textrm{EncDec}_{\textrm{S}}}$ & \cellcolor[RGB]{255,255,255} -  & \cellcolor[RGB]{255,255,255} 1.26927 & \cellcolor[RGB]{255,255,255} 0.01005 & \cellcolor[RGB]{254,241,235} 1.45551 & \cellcolor[RGB]{255,255,255} 1.59881 \\
\addlinespace[-2.32pt]
\cmidrule[\heavyrulewidth]{2-7}
\addlinespace[-\belowrulesep]
\multirow{3}{*}{\rotatebox{90}{width}} \hspace{-0.3cm} & FT-FNO$_{\textrm{L}_\textrm{FT}:6,\textrm{L}_\textrm{FNO}:2,\textrm{D}:896,\textrm{MLP}:1792,\textrm{M}:40,\textrm{EncDec}_{\textrm{B}2}}$ & \cellcolor[RGB]{255,255,255} 2.61245 & \cellcolor[RGB]{255,255,255} 1.67738 & \cellcolor[RGB]{255,255,255} 0.03231 & \cellcolor[RGB]{252,167,138} 9.67148 & \cellcolor[RGB]{252,166,137} 713.672 \\
 & FT-FNO$_{\textrm{L}_\textrm{FT}:6,\textrm{L}_\textrm{FNO}:2,\textrm{D}:896,\textrm{MLP}:1792,\textrm{M}:40,\textrm{EncDec}_{\textrm{B}1}}$ & \cellcolor[RGB]{203,234,196} 0.00363 & \cellcolor[RGB]{186,227,180} 0.00132 & \cellcolor[RGB]{190,229,183} 0.00038 & \cellcolor[RGB]{252,168,139} 9.54798 & \cellcolor[RGB]{252,166,137} 710.717 \\
 & FT-FNO$_{\textrm{L}_\textrm{FT}:5,\textrm{L}_\textrm{FNO}:2,\textrm{D}:832,\textrm{MLP}:1664,\textrm{M}:34,\textrm{EncDec}_{\textrm{B}2}}$ & \cellcolor[RGB]{202,234,195} 0.00347 & \cellcolor[RGB]{191,229,184} 0.00214 & \cellcolor[RGB]{185,227,178} 0.00030 & \cellcolor[RGB]{252,190,165} 7.65755 & \cellcolor[RGB]{252,166,137} 524.350 \\
 & FT-FNO$_{\textrm{L}_\textrm{FT}:5,\textrm{L}_\textrm{FNO}:2,\textrm{D}:832,\textrm{MLP}:1664,\textrm{M}:34,\textrm{EncDec}_{\textrm{B}1}}$ & \cellcolor[RGB]{226,243,220} 0.00843 & \cellcolor[RGB]{191,229,184} 0.00209 & \cellcolor[RGB]{201,234,194} 0.00064 & \cellcolor[RGB]{252,191,166} 7.59206 & \cellcolor[RGB]{252,166,137} 521.618 \\
 & FT-FNO$_{\textrm{L}_\textrm{FT}:5,\textrm{L}_\textrm{FNO}:2,\textrm{D}:576,\textrm{MLP}:1408,\textrm{M}:26,\textrm{EncDec}_{\textrm{B}2}}$ & \cellcolor[RGB]{209,237,202} 0.00464 & \cellcolor[RGB]{191,229,184} 0.00204 & \cellcolor[RGB]{189,228,182} 0.00036 & \cellcolor[RGB]{253,211,192} 5.65682 & \cellcolor[RGB]{252,166,137} 240.784 \\
 & FT-FNO$_{\textrm{L}_\textrm{FT}:5,\textrm{L}_\textrm{FNO}:2,\textrm{D}:576,\textrm{MLP}:1408,\textrm{M}:26,\textrm{EncDec}_{\textrm{B}1}}$ & \cellcolor[RGB]{229,245,224} 0.00966 & \cellcolor[RGB]{189,228,182} 0.00162 & \cellcolor[RGB]{192,230,185} 0.00043 & \cellcolor[RGB]{253,212,194} 5.60030 & \cellcolor[RGB]{252,166,137} 239.314 \\
 & FT-FNO$_{\textrm{L}_\textrm{FT}:4,\textrm{L}_\textrm{FNO}:2,\textrm{D}:64,\textrm{MLP}:256,\textrm{M}:8,\textrm{EncDec}_{\textrm{B}1}}$ & \cellcolor[RGB]{255,255,255} 0.55685 & \cellcolor[RGB]{255,255,255} 0.32376 & \cellcolor[RGB]{255,255,255} 0.00870 & \cellcolor[RGB]{254,228,216} 3.64401 & \cellcolor[RGB]{254,250,248} 6.65231 \\
 & FT-FNO$_{\textrm{L}_\textrm{FT}:1,\textrm{L}_\textrm{FNO}:1,\textrm{D}:192,\textrm{MLP}:640,\textrm{M}:10,\textrm{EncDec}_{\textrm{S}}}$ & \cellcolor[RGB]{255,255,255} -  & \cellcolor[RGB]{255,255,255} 0.42749 & \cellcolor[RGB]{240,249,237} 0.00315 & \cellcolor[RGB]{254,239,232} 1.81580 & \cellcolor[RGB]{254,247,244} 7.05801 \\
 & FT-FNO$_{\textrm{L}_\textrm{FT}:1,\textrm{L}_\textrm{FNO}:1,\textrm{D}:64,\textrm{MLP}:128,\textrm{M}:8,\textrm{EncDec}_{\textrm{B}1}}$ & \cellcolor[RGB]{255,255,255} 3.65552 & \cellcolor[RGB]{255,255,255} 0.57266 & \cellcolor[RGB]{255,255,255} 0.01047 & \cellcolor[RGB]{254,239,232} 1.77369 & \cellcolor[RGB]{255,255,255} 2.38160 \\
 & FT-FNO$_{\textrm{L}_\textrm{FT}:1,\textrm{L}_\textrm{FNO}:1,\textrm{D}:64,\textrm{MLP}:128,\textrm{M}:8,\textrm{EncDec}_{\textrm{B}1}}$ & \cellcolor[RGB]{255,255,255} 2.07186 & \cellcolor[RGB]{255,255,255} 1.24912 & \cellcolor[RGB]{255,255,255} 0.00986 & \cellcolor[RGB]{254,239,232} 1.77189 & \cellcolor[RGB]{255,255,255} 2.38057 \\
 & FT-FNO$_{\textrm{L}_\textrm{FT}:1,\textrm{L}_\textrm{FNO}:1,\textrm{D}:64,\textrm{MLP}:128,\textrm{M}:8,\textrm{EncDec}_{\textrm{S}}}$ & \cellcolor[RGB]{255,255,255} -  & \cellcolor[RGB]{255,255,255} 1.40775 & \cellcolor[RGB]{255,255,255} 0.00663 & \cellcolor[RGB]{254,241,235} 1.47542 & \cellcolor[RGB]{255,255,255} 2.08789 \\
\addlinespace[-2.32pt]
\cmidrule[\heavyrulewidth]{2-7}
\addlinespace[-\belowrulesep]

\\
\caption{Complete results on test data for the density ramp dataset. Architectures are found by optimizing hyperparameters to a set of inference times between $\text{\SI{0.5}{}}$ to $\text{\SI{10}{\milli\second}}$. Evaluations are provided for three error metrics, where green indicates a lower error, alongside the inference speed for two types of hardware, where red indicates a larger inference cost. A dash indicates that the model converged to noise, filtered as the error being greater than 5.}\label{tab:fullresult}
\end{longtable}
\endgroup
\setlength\tabcolsep{6pt}

\setlength\tabcolsep{4.5pt}
\begin{table}[H]
\centering
\begin{tabular}{lcc}
&  & Compute Time ($\text{\SI{}{\milli\second}}$) \\
&  &  (per $\text{\SI{2}{\milli\second}}$) \\
\addlinespace[-2.32pt]
\cmidrule[\heavyrulewidth]{3-3}
\addlinespace[-\belowrulesep]
Setting \hphantom{xxxxxxxx} & \hphantom{xxx} MSE \hphantom{xxx} & CPU$_{\textrm{1-core}}$\\
\cmidrule[\heavyrulewidth]{1-3}
\addlinespace[-\belowrulesep]
DIV1D-Nx500 & \cellcolor[RGB]{179,224,173} 0.00000 & \cellcolor[RGB]{252,166,137} 642882 \\
DIV1D-Nx450 & \cellcolor[RGB]{182,225,175} 0.00094 & \cellcolor[RGB]{252,166,137} 422383 \\
DIV1D-Nx400 & \cellcolor[RGB]{217,240,211} 0.00491 & \cellcolor[RGB]{252,166,137} 361780 \\
DIV1D-Nx300 & \cellcolor[RGB]{255,255,255} 0.03963 & \cellcolor[RGB]{252,166,137} 301293 \\
DIV1D-Nx200 & \cellcolor[RGB]{255,255,255} 0.22677 & \cellcolor[RGB]{252,166,137} 89738.1 \\
DIV1D-Nx100 & \cellcolor[RGB]{255,255,255} 2.59551 & \cellcolor[RGB]{252,166,137} 9861.76 \\
\addlinespace[-2.32pt]
\cmidrule[\heavyrulewidth]{1-3}
\addlinespace[-\belowrulesep]
DIV1D$_{\textit{fast}}$-Nx500 & \cellcolor[RGB]{179,224,173} 0.00000 & \cellcolor[RGB]{252,166,137} 22839.0 \\
DIV1D$_{\textit{fast}}$-Nx450 & \cellcolor[RGB]{183,226,176} 0.00098 & \cellcolor[RGB]{252,166,137} 15565.1 \\
DIV1D$_{\textit{fast}}$-Nx400 & \cellcolor[RGB]{220,241,214} 0.00534 & \cellcolor[RGB]{252,166,137} 8788.88 \\
DIV1D$_{\textit{fast}}$-Nx300 & \cellcolor[RGB]{255,255,255} 0.04417 & \cellcolor[RGB]{252,166,137} 4508.63 \\
DIV1D$_{\textit{fast}}$-Nx200 & \cellcolor[RGB]{255,255,255} 0.28126 & \cellcolor[RGB]{252,166,137} 8664.64 \\
DIV1D$_{\textit{fast}}$-Nx100 & \cellcolor[RGB]{255,255,255} 2.75022 & \cellcolor[RGB]{252,166,137} 415.797 \\
\addlinespace[-2.32pt]
\cmidrule[\heavyrulewidth]{1-3}
\addlinespace[-\belowrulesep]
\\
\end{tabular}
\caption{The trade-off between error and speed when decreasing DIV1D's spatial grid size for density ramps. Note that all grids are non-equidistant following the scaling from~\cite{derks2022}, only the NN surrogates use equidistant grids. For both DIV1D and DIV1D$_\textit{fast}$ the error measurements are taken with respect to their own reference solutions at 500 gridpoints. Cells are colored using the same scale as Table~\ref{tab:fullresult}.}\label{tab:div1dresult}
\end{table}
\setlength\tabcolsep{6pt}

\section{Tables: Fast Transients Results}\label{ap:transients}
In this appendix extra results for the fast transient data evaluation are provided. We provide the scaling of DIV1D's performance with respect to its internal spatial grid to contextualize the error quantification of DIV1D-NN. Additionally, we provide the confusion matrix for attachment/detachment predictions.

As a reference for DIV1D-NN's MSE, we provide DIV1D's MSE as it scales when decreasing the internal grid. In this setting, DIV1D$_\textit{fast}$ was used to generate the dataset. The scaling of DIV1D$_\textit{fast}$ is provided in Table~\ref{tab:div1dtransient}. For the density ramp data the best NN surrogates were comparable in MSE to running DIV1D at $\approx$450 gridpoints, whereas for the fast transient data the NN surrogate had an error residing somewhere between running DIV1D$_\textit{fast}$ at 300 and 400 gridpoints (0.02298 for DIV1D-NN, compared to 0.06775 and 0.01569 for DIV1D$_\textit{fast}$ using 300 and 400 gridpoints, respectively).

\setlength\tabcolsep{4.5pt}
\begin{table}[H]
\centering
\begin{tabular}{lcc}
&  & Compute Time ($\text{\SI{}{\milli\second}}$) \\
&  & (per $\text{\SI{2}{\milli\second}}$) \\
\addlinespace[-2.32pt]
\cmidrule[\heavyrulewidth]{3-3}
\addlinespace[-\belowrulesep]
Setting \hphantom{xxxxxxxx} & \hphantom{xxx} MSE \hphantom{xxx} & CPU$_{\textrm{1-core}}$\\
\cmidrule[\heavyrulewidth]{1-3}
\addlinespace[-\belowrulesep]
DIV1D$_{\textit{fast}}$-Nx500 & \cellcolor[RGB]{179,224,173} 0.00000 & \cellcolor[RGB]{252,166,137} 66104.1 \\
DIV1D$_{\textit{fast}}$-Nx450 & \cellcolor[RGB]{179,224,173} 0.00648 & \cellcolor[RGB]{252,166,137} 45520.9 \\
DIV1D$_{\textit{fast}}$-Nx400 & \cellcolor[RGB]{184,226,177} 0.01569 & \cellcolor[RGB]{252,166,137} 30600.1 \\
DIV1D$_{\textit{fast}}$-Nx300 & \cellcolor[RGB]{208,236,201} 0.06775 & \cellcolor[RGB]{252,166,137} 13403.2 \\
DIV1D$_{\textit{fast}}$-Nx200 & \cellcolor[RGB]{255,255,255} 0.19863 & \cellcolor[RGB]{252,166,137} 12964.4 \\
DIV1D$_{\textit{fast}}$-Nx100 & \cellcolor[RGB]{255,255,255} 0.36371 & \cellcolor[RGB]{252,166,137} 7259.12 \\
\addlinespace[-2.32pt]
\cmidrule[\heavyrulewidth]{1-3}
\addlinespace[-\belowrulesep]
\\
\end{tabular}
\caption{The trade-off between error and speed when decreasing DIV1D's spatial grid size for fast transients. The error measurements are taken with respect to 
reference solutions at 500 gridpoints. Error cells are colored such that green indicates better results, scaled to the DIV1D-NN error on this dataset. Compute time cells are colored with the same scale as in Table~\ref{tab:div1dresult}.}\label{tab:div1dtransient}
\end{table}
\setlength\tabcolsep{6pt}
\hphantom{xx} %

For predicting whether the plasma is in an attached or detached state (defined as the target temperature being above or below $\text{\SI{7}{\electronvolt}}$, see also Section~\ref{ss:casestudies}), DIV1D-NN's accuracy with respect to DIV1D's reference solutions is 98.61\%. For more detail, the confusion matrix is provided in Table~\ref{tab:attachmatrixtransient}.

\begin{table}[h]
\centering
\begin{tabular}{cccc}
& & \multicolumn{2}{c}{DIV1D$_\textit{fast}$} \\
& \multicolumn{1}{c?{\heavyrulewidth}}{} & Detached & Attached \\
\addlinespace[-2.32pt]
\cmidrule[\heavyrulewidth]{2-4}
\addlinespace[-\belowrulesep]
\multirow{2}{*}{DIV1D-NN} &\multicolumn{1}{c?{\heavyrulewidth}}{Detached} & \cellcolor[RGB]{151,198,223} 325565 & \cellcolor[RGB]{255,255,255} 2123 \\
&\multicolumn{1}{c?{\heavyrulewidth}}{Attached} & \cellcolor[RGB]{245,249,254} 4170 & \cellcolor[RGB]{211,227,243} 120142 \\
\addlinespace[-2.32pt]
\cmidrule[\heavyrulewidth]{2-4}
\end{tabular}
\caption{Confusion matrix of attachment predictions for the fast transient data, where attachment is defined as the target temperature being below $\text{\SI{7}{\electronvolt}}$. The accuracy of predicting attachment is 98.61\% over the entire test set.}
\label{tab:attachmatrixtransient}
\end{table}

\section{Tables: Inter- and Extrapolation Results}\label{ap:interp}
In this appendix we provide more results for the inter- and extrapolation evaluation of DIV1D-NN on the density ramp data. To evaluate the ability of DIV1D-NN to interpolate and extrapolate within the parameter space, we train separate models where all simulations containing either a specific value for the upstream heat flux ($q_{\|\mathrm{u}}$) or for the impurity fraction ($\xi_{\mathrm{C}}$) are left out from the training and validation data. These models are then tested on simulations with these parameter values, to evaluate the quality when simulating unseen parameters. In Table~\ref{tab:leaveout}, the results for all values of $q_{\|\mathrm{u}}$ and $\xi_{\mathrm{C}}$ are given.

Additionally, we consider linear interpolation of surrounding parameter values as a baseline for this experiment. New simulations are generated by linearly inter- or extrapolating simulations of identical density ramps with the surrounding parameter values for $q_{\|\mathrm{u}}$ and $\xi_{\mathrm{C}}$, and are then compared to the reference simulations with these parameter values. These results are provided in Table~\ref{tab:leaveoutlinear}. The ratios of errors between DIV1D-NN and linear interpolation are provided in Table~\ref{tab:leaveoutratio}. In general, DIV1D-NN performed much better, as also demonstrated in Section~\ref{ss:interp}: The NN-based approach manages to capture non-linear dependencies in the dynamics. However, it is still not advisable to use the NN-based surrogate for extrapolation, as indicated by the higher errors (and worse ratios compared to linear interpolation) on the boundaries of the parameter space.

\begin{table}[H]
\centering
\begin{tabular}{cccccc}
\diagbox[linewidth=\heavyrulewidth,innerwidth=.8cm]{$\xi_{\mathrm{C}}$}{$q_{\|\mathrm{u}}$} & 10 & 15 & 20 & 25 & 30 \\
\addlinespace[-2.32pt]
\cmidrule[\heavyrulewidth]{2-6}
\addlinespace[-\belowrulesep]
\multicolumn{1}{c?{\heavyrulewidth}}{0.01} & \cellcolor[RGB]{255,255,255} 0.0239 & \cellcolor[RGB]{255,255,255} 0.0321 & \cellcolor[RGB]{216,240,210} 0.0090 & \cellcolor[RGB]{207,236,200} 0.0073 & \cellcolor[RGB]{255,255,255} 0.0694\\
\multicolumn{1}{c?{\heavyrulewidth}}{0.02} & \cellcolor[RGB]{224,243,218} 0.0104 & \cellcolor[RGB]{161,217,155} 0.0015 & \cellcolor[RGB]{156,214,151} 0.0011 & \cellcolor[RGB]{166,219,160} 0.0021 & \cellcolor[RGB]{255,255,255} 0.0254\\
\multicolumn{1}{c?{\heavyrulewidth}}{0.03} & \cellcolor[RGB]{235,247,231} 0.0141 & \cellcolor[RGB]{166,219,160} 0.0020 & \cellcolor[RGB]{154,214,149} 0.0010 & \cellcolor[RGB]{162,217,156} 0.0017 & \cellcolor[RGB]{255,255,255} 0.0892\\
\multicolumn{1}{c?{\heavyrulewidth}}{0.04} & \cellcolor[RGB]{208,236,201} 0.0074 & \cellcolor[RGB]{161,217,155} 0.0015 & \cellcolor[RGB]{161,217,155} 0.0015 & \cellcolor[RGB]{170,220,163} 0.0025 & \cellcolor[RGB]{232,246,228} 0.0130\\
\multicolumn{1}{c?{\heavyrulewidth}}{0.05} & \cellcolor[RGB]{239,248,235} 0.0160 & \cellcolor[RGB]{199,233,192} 0.0059 & \cellcolor[RGB]{183,226,176} 0.0039 & \cellcolor[RGB]{190,229,183} 0.0048 & \cellcolor[RGB]{255,255,255} 0.2290\\
\toprule
\end{tabular}
\caption{Validating the inter- and extrapolation capabilities of DIV1D-NN. We train different models from scratch where we leave out \textit{all} simulations using a given upstream heat flux ($q_{\|\mathrm{u}}$) \textit{or} impurity fraction ($\xi_{\mathrm{C}}$). The values indicate the MSE with these left-out simulations, when using the remaining simulations for training and validation (cells colored by MSE, greener is better). As expected, when leaving out data within the domain (middle cells), such that we are strictly interpolating in parameter space, the model performs a lot better.}
\label{tab:leaveout}
\end{table}

\begin{table}[H]
\centering
\begin{tabular}{cccccc}
\diagbox[linewidth=\heavyrulewidth,innerwidth=.8cm]{$\xi_{\mathrm{C}}$}{$q_{\|\mathrm{u}}$} & 10 & 15 & 20 & 25 & 30 \\
\addlinespace[-2.32pt]
\cmidrule[\heavyrulewidth]{2-6}
\addlinespace[-\belowrulesep]
\multicolumn{1}{c?{\heavyrulewidth}}{0.01} & \cellcolor[RGB]{255,255,255} 0.0354 & \cellcolor[RGB]{224,243,218} 0.0105 & \cellcolor[RGB]{222,242,216} 0.0100 & \cellcolor[RGB]{212,238,206} 0.0082 & \cellcolor[RGB]{255,255,255} 0.0313\\
\multicolumn{1}{c?{\heavyrulewidth}}{0.02} & \cellcolor[RGB]{255,255,255} 0.0313 & \cellcolor[RGB]{216,240,210} 0.0090 & \cellcolor[RGB]{214,239,208} 0.0085 & \cellcolor[RGB]{204,235,197} 0.0067 & \cellcolor[RGB]{246,251,244} 0.0224\\
\multicolumn{1}{c?{\heavyrulewidth}}{0.03} & \cellcolor[RGB]{255,255,255} 0.0311 & \cellcolor[RGB]{212,238,206} 0.0083 & \cellcolor[RGB]{211,237,204} 0.0079 & \cellcolor[RGB]{200,233,193} 0.0061 & \cellcolor[RGB]{246,251,244} 0.0221\\
\multicolumn{1}{c?{\heavyrulewidth}}{0.04} & \cellcolor[RGB]{255,255,255} 0.0312 & \cellcolor[RGB]{212,238,206} 0.0082 & \cellcolor[RGB]{210,237,203} 0.0077 & \cellcolor[RGB]{199,233,192} 0.0059 & \cellcolor[RGB]{246,251,244} 0.0220\\
\multicolumn{1}{c?{\heavyrulewidth}}{0.05} & \cellcolor[RGB]{255,255,255} 0.0349 & \cellcolor[RGB]{214,239,208} 0.0086 & \cellcolor[RGB]{211,238,205} 0.0080 & \cellcolor[RGB]{200,233,193} 0.0061 & \cellcolor[RGB]{245,251,243} 0.0216\\
\toprule
\end{tabular}
\caption{Evaluation of the inter- and extrapolation error when using linear combinations of simulations with surrounding values to compute a given simulation. For each setting of the upstream heat flux ($q_{\|\mathrm{u}}$) and impurity fraction ($\xi_{\mathrm{C}}$), the simulations are computed by using only linear interpolation (or extrapolation) of simulations without either of those values present; cells indicate the MSE with reference simulations for the given setting. Cells are colored by MSE, using the same scale as Table~\ref{tab:leaveout}.}
\label{tab:leaveoutlinear}
\end{table}

\begin{table}[H]
\centering
\begin{tabular}{cccccc}
\diagbox[linewidth=\heavyrulewidth,innerwidth=.8cm]{$\xi_{\mathrm{C}}$}{$q_{\|\mathrm{u}}$} & 10 & 15 & 20 & 25 & 30 \\
\addlinespace[-2.32pt]
\cmidrule[\heavyrulewidth]{2-6}
\addlinespace[-\belowrulesep]
\multicolumn{1}{c?{\heavyrulewidth}}{0.01} & \cellcolor[RGB]{232,230,241}0.68$\times$ & \cellcolor[RGB]{253,183,121}3.06$\times$ & \cellcolor[RGB]{243,242,247}0.90$\times$ & \cellcolor[RGB]{243,241,247}0.89$\times$ & \cellcolor[RGB]{253,204,156}2.22$\times$\\
\multicolumn{1}{c?{\heavyrulewidth}}{0.02} & \cellcolor[RGB]{197,197,224}0.33$\times$ & \cellcolor[RGB]{150,146,196}0.17$\times$ & \cellcolor[RGB]{130,127,187}0.13$\times$ & \cellcolor[RGB]{193,194,222}0.31$\times$ & \cellcolor[RGB]{254,234,214}1.13$\times$\\
\multicolumn{1}{c?{\heavyrulewidth}}{0.03} & \cellcolor[RGB]{215,215,233}0.45$\times$ & \cellcolor[RGB]{176,175,212}0.24$\times$ & \cellcolor[RGB]{127,124,185}0.13$\times$ & \cellcolor[RGB]{186,187,219}0.28$\times$ & \cellcolor[RGB]{253,163,91}4.04$\times$\\
\multicolumn{1}{c?{\heavyrulewidth}}{0.04} & \cellcolor[RGB]{175,174,211}0.24$\times$ & \cellcolor[RGB]{157,153,199}0.18$\times$ & \cellcolor[RGB]{162,158,202}0.19$\times$ & \cellcolor[RGB]{211,211,231}0.42$\times$ & \cellcolor[RGB]{227,226,239}0.59$\times$\\
\multicolumn{1}{c?{\heavyrulewidth}}{0.05} & \cellcolor[RGB]{215,215,233}0.46$\times$ & \cellcolor[RGB]{232,231,242}0.69$\times$ & \cellcolor[RGB]{219,219,235}0.49$\times$ & \cellcolor[RGB]{239,237,245}0.79$\times$ & \cellcolor[RGB]{240,104,18}10.60$\times$\\
\toprule
\end{tabular}
\caption{Ratio of MSEs between DIV1D-NN and linear interpolation, that is, between the values from Tables~\ref{tab:leaveout} and~\ref{tab:leaveoutlinear}. Purple indicates the NN-based interpolation of parameter values performed better, whereas orange indicates that linear interpolation performed better; magnitudes are scaled according to the relative error. In general, NN-based interpolation results in a significantly lower error (as also demonstrated in Section~\ref{ss:interp}), however, for extrapolation it does not necessarily perform better.}
\label{tab:leaveoutratio}
\end{table}

\end{document}